\documentclass[a4paper,11pt]{article}
\pdfoutput=1 

\usepackage{jheppub} 

\usepackage[T1]{fontenc} 
 
\usepackage{graphicx}
\usepackage{dcolumn}
\usepackage{bm}
\usepackage{ulem}
\usepackage{color}

\usepackage{float} 
\usepackage{lscape}

\definecolor{dG}{rgb}{0, 0.4, 0}
\definecolor{sRed}{rgb}{0.7,0.0,0.1} 

\def\GeV{\textrm{GeV}}
\def\MeV{\textrm{MeV}}
\def\Aprime{A^\prime}

\def\BF{\mathcal{B}}
\def\BFBAA{\mathcal{B}(B^0 \to \Aprime \Aprime)}
\def\prodee{\mathcal{B}(B^0 \to \Aprime \Aprime) \times \mathcal{B}(\Aprime \to e^+ e^-)^2}
\def\prodmm{\mathcal{B}(B^0 \to \Aprime \Aprime) \times \mathcal{B}(\Aprime \to \mu^+ \mu^-)^2}

\def\eeee{e^+ e^- e^+ e^-}
\def\mmmm{\mu^+ \mu^- \mu^+ \mu^-}
\def\eemm{e^+ e^- \mu^+ \mu^-}
\def\eepp{e^+ e^- \pi^+ \pi^-}
\def\mmpp{\mu^+ \mu^- \pi^+ \pi^-}
\def\mAbin{m_{\Aprime}^{\textrm{bin}}}
\def\epem{e^+ e^-}
\def\mpmm{\mu^+ \mu^-}
\def\ipim{\pi^+ \pi^-}

\begin{document}

\title{\boldmath Search for the dark photon in $B^0 \to \Aprime \Aprime$, $\Aprime \to e^+ e^-$, $\mu^+ \mu^-$, and $\pi^+ \pi^-$ decays at Belle}

\newcounter{AffiliationCounter}
\stepcounter{AffiliationCounter}\edef\instBilbao{\protect\theAffiliationCounter}
\stepcounter{AffiliationCounter}\edef\instBonn{\protect\theAffiliationCounter}
\stepcounter{AffiliationCounter}\edef\instBNL{\protect\theAffiliationCounter}
\stepcounter{AffiliationCounter}\edef\instBINP{\protect\theAffiliationCounter}
\stepcounter{AffiliationCounter}\edef\instCharles{\protect\theAffiliationCounter}
\stepcounter{AffiliationCounter}\edef\instChonnam{\protect\theAffiliationCounter}
\stepcounter{AffiliationCounter}\edef\instCincinnati{\protect\theAffiliationCounter}
\stepcounter{AffiliationCounter}\edef\instDESY{\protect\theAffiliationCounter}
\stepcounter{AffiliationCounter}\edef\instFlorida{\protect\theAffiliationCounter}
\stepcounter{AffiliationCounter}\edef\instFuJen{\protect\theAffiliationCounter}
\stepcounter{AffiliationCounter}\edef\instFudan{\protect\theAffiliationCounter}
\stepcounter{AffiliationCounter}\edef\instGifu{\protect\theAffiliationCounter}
\stepcounter{AffiliationCounter}\edef\instSokendai{\protect\theAffiliationCounter}
\stepcounter{AffiliationCounter}\edef\instGyeongsang{\protect\theAffiliationCounter}
\stepcounter{AffiliationCounter}\edef\instHanyang{\protect\theAffiliationCounter}
\stepcounter{AffiliationCounter}\edef\instHawaii{\protect\theAffiliationCounter}
\stepcounter{AffiliationCounter}\edef\instKEK{\protect\theAffiliationCounter}
\stepcounter{AffiliationCounter}\edef\instJPARC{\protect\theAffiliationCounter}
\stepcounter{AffiliationCounter}\edef\instHSE{\protect\theAffiliationCounter}
\stepcounter{AffiliationCounter}\edef\instJuelich{\protect\theAffiliationCounter}
\stepcounter{AffiliationCounter}\edef\instIKER{\protect\theAffiliationCounter}
\stepcounter{AffiliationCounter}\edef\instIISERM{\protect\theAffiliationCounter}
\stepcounter{AffiliationCounter}\edef\instIITG{\protect\theAffiliationCounter}
\stepcounter{AffiliationCounter}\edef\instIITH{\protect\theAffiliationCounter}
\stepcounter{AffiliationCounter}\edef\instIITM{\protect\theAffiliationCounter}
\stepcounter{AffiliationCounter}\edef\instIndiana{\protect\theAffiliationCounter}
\stepcounter{AffiliationCounter}\edef\instIHEP{\protect\theAffiliationCounter}
\stepcounter{AffiliationCounter}\edef\instProtvino{\protect\theAffiliationCounter}
\stepcounter{AffiliationCounter}\edef\instVienna{\protect\theAffiliationCounter}
\stepcounter{AffiliationCounter}\edef\instNapoli{\protect\theAffiliationCounter}
\stepcounter{AffiliationCounter}\edef\instTorino{\protect\theAffiliationCounter}
\stepcounter{AffiliationCounter}\edef\instJAEA{\protect\theAffiliationCounter}
\stepcounter{AffiliationCounter}\edef\instJSI{\protect\theAffiliationCounter}
\stepcounter{AffiliationCounter}\edef\instKarlsruhe{\protect\theAffiliationCounter}
\stepcounter{AffiliationCounter}\edef\instIPMU{\protect\theAffiliationCounter}
\stepcounter{AffiliationCounter}\edef\instKAU{\protect\theAffiliationCounter}
\stepcounter{AffiliationCounter}\edef\instKitasato{\protect\theAffiliationCounter}
\stepcounter{AffiliationCounter}\edef\instKISTI{\protect\theAffiliationCounter}
\stepcounter{AffiliationCounter}\edef\instKorea{\protect\theAffiliationCounter}
\stepcounter{AffiliationCounter}\edef\instKyotoSangyo{\protect\theAffiliationCounter}
\stepcounter{AffiliationCounter}\edef\instKyungpook{\protect\theAffiliationCounter}
\stepcounter{AffiliationCounter}\edef\instLAL{\protect\theAffiliationCounter}
\stepcounter{AffiliationCounter}\edef\instLebedev{\protect\theAffiliationCounter}
\stepcounter{AffiliationCounter}\edef\instLNNU{\protect\theAffiliationCounter}
\stepcounter{AffiliationCounter}\edef\instLjubljana{\protect\theAffiliationCounter}
\stepcounter{AffiliationCounter}\edef\instLMU{\protect\theAffiliationCounter}
\stepcounter{AffiliationCounter}\edef\instLuther{\protect\theAffiliationCounter}
\stepcounter{AffiliationCounter}\edef\instMNIT{\protect\theAffiliationCounter}
\stepcounter{AffiliationCounter}\edef\instMaribor{\protect\theAffiliationCounter}
\stepcounter{AffiliationCounter}\edef\instMPI{\protect\theAffiliationCounter}
\stepcounter{AffiliationCounter}\edef\instMelbourne{\protect\theAffiliationCounter}
\stepcounter{AffiliationCounter}\edef\instMississippi{\protect\theAffiliationCounter}
\stepcounter{AffiliationCounter}\edef\instMiyazaki{\protect\theAffiliationCounter}
\stepcounter{AffiliationCounter}\edef\instMEPhI{\protect\theAffiliationCounter}
\stepcounter{AffiliationCounter}\edef\instNagoya{\protect\theAffiliationCounter}
\stepcounter{AffiliationCounter}\edef\instNagoyaKMI{\protect\theAffiliationCounter}
\stepcounter{AffiliationCounter}\edef\instUNapoli{\protect\theAffiliationCounter}
\stepcounter{AffiliationCounter}\edef\instNara{\protect\theAffiliationCounter}
\stepcounter{AffiliationCounter}\edef\instNCU{\protect\theAffiliationCounter}
\stepcounter{AffiliationCounter}\edef\instNUU{\protect\theAffiliationCounter}
\stepcounter{AffiliationCounter}\edef\instTaiwan{\protect\theAffiliationCounter}
\stepcounter{AffiliationCounter}\edef\instKrakow{\protect\theAffiliationCounter}
\stepcounter{AffiliationCounter}\edef\instNihonDental{\protect\theAffiliationCounter}
\stepcounter{AffiliationCounter}\edef\instNiigata{\protect\theAffiliationCounter}
\stepcounter{AffiliationCounter}\edef\instNovaGorica{\protect\theAffiliationCounter}
\stepcounter{AffiliationCounter}\edef\instNovosibirsk{\protect\theAffiliationCounter}
\stepcounter{AffiliationCounter}\edef\instOsakaCity{\protect\theAffiliationCounter}
\stepcounter{AffiliationCounter}\edef\instPNNL{\protect\theAffiliationCounter}
\stepcounter{AffiliationCounter}\edef\instPanjab{\protect\theAffiliationCounter}
\stepcounter{AffiliationCounter}\edef\instPeking{\protect\theAffiliationCounter}
\stepcounter{AffiliationCounter}\edef\instPittsburgh{\protect\theAffiliationCounter}
\stepcounter{AffiliationCounter}\edef\instNPC{\protect\theAffiliationCounter}
\stepcounter{AffiliationCounter}\edef\instRIKENMSL{\protect\theAffiliationCounter}
\stepcounter{AffiliationCounter}\edef\instUSTC{\protect\theAffiliationCounter}
\stepcounter{AffiliationCounter}\edef\instSeoul{\protect\theAffiliationCounter}
\stepcounter{AffiliationCounter}\edef\instShoyaku{\protect\theAffiliationCounter}
\stepcounter{AffiliationCounter}\edef\instSoochow{\protect\theAffiliationCounter}
\stepcounter{AffiliationCounter}\edef\instSoongsil{\protect\theAffiliationCounter}
\stepcounter{AffiliationCounter}\edef\instSungkyunkwan{\protect\theAffiliationCounter}
\stepcounter{AffiliationCounter}\edef\instSydney{\protect\theAffiliationCounter}
\stepcounter{AffiliationCounter}\edef\instTabuk{\protect\theAffiliationCounter}
\stepcounter{AffiliationCounter}\edef\instTata{\protect\theAffiliationCounter}
\stepcounter{AffiliationCounter}\edef\instTUM{\protect\theAffiliationCounter}
\stepcounter{AffiliationCounter}\edef\instTelAviv{\protect\theAffiliationCounter}
\stepcounter{AffiliationCounter}\edef\instToho{\protect\theAffiliationCounter}
\stepcounter{AffiliationCounter}\edef\instTohoku{\protect\theAffiliationCounter}
\stepcounter{AffiliationCounter}\edef\instERI{\protect\theAffiliationCounter}
\stepcounter{AffiliationCounter}\edef\instTokyo{\protect\theAffiliationCounter}
\stepcounter{AffiliationCounter}\edef\instTIT{\protect\theAffiliationCounter}
\stepcounter{AffiliationCounter}\edef\instTMU{\protect\theAffiliationCounter}
\stepcounter{AffiliationCounter}\edef\instUtkal{\protect\theAffiliationCounter}
\stepcounter{AffiliationCounter}\edef\instVPI{\protect\theAffiliationCounter}
\stepcounter{AffiliationCounter}\edef\instWayneState{\protect\theAffiliationCounter}
\stepcounter{AffiliationCounter}\edef\instYamagata{\protect\theAffiliationCounter}
\stepcounter{AffiliationCounter}\edef\instYonsei{\protect\theAffiliationCounter}

\collaborationImg{\includegraphics[width=2cm]{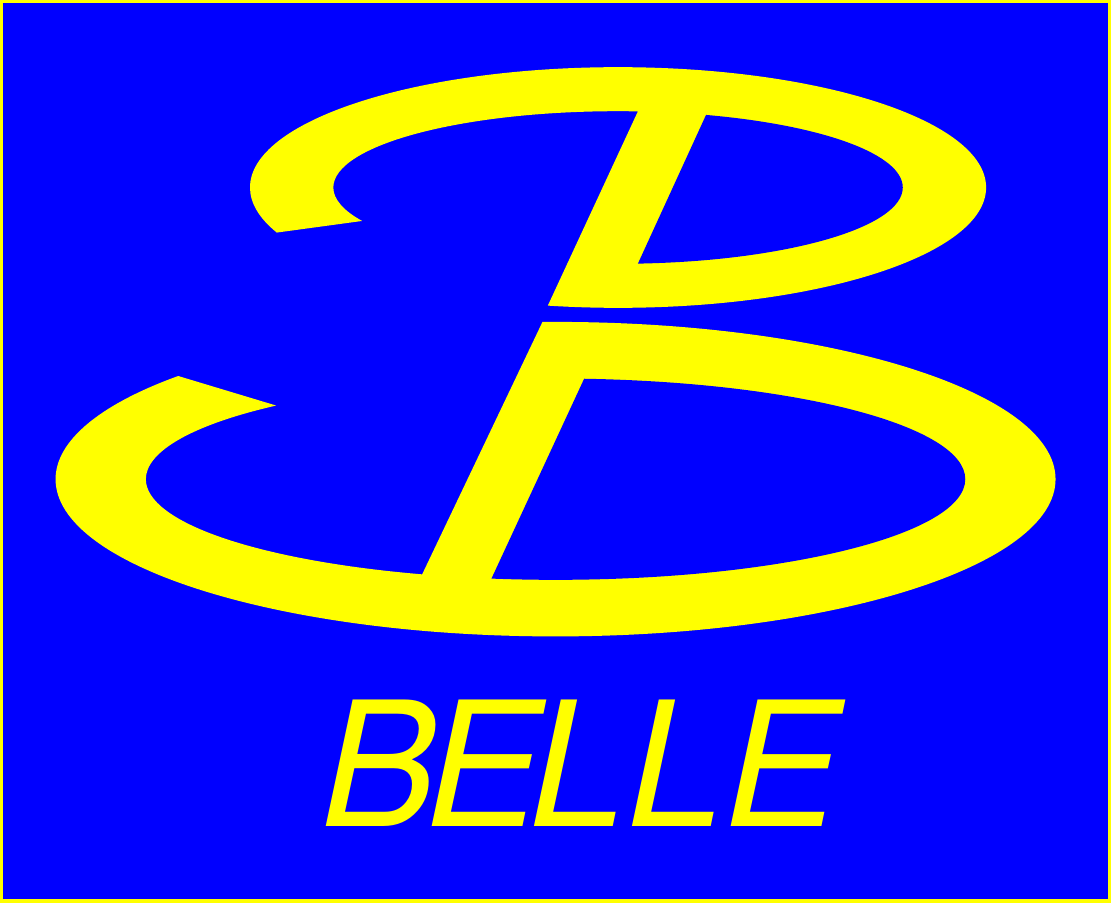}}
\collaboration{The Belle Collaboration}
  \author[\instYonsei]{S.-H.~Park,} 
  \author[\instYonsei]{Y.-J.~Kwon,} 
  \author[\instKEK,\instSokendai]{I.~Adachi,} 
  \author[\instTokyo]{H.~Aihara,} 
  \author[\instTabuk,\instKAU]{S.~Al~Said,} 
  \author[\instBNL]{D.~M.~Asner,} 
  \author[\instCincinnati]{H.~Atmacan,} 
  \author[\instHSE]{T.~Aushev,} 
  \author[\instTabuk]{R.~Ayad,} 
  \author[\instDESY]{V.~Babu,} 
% ...

\abstract{
We present a search for the dark photon $\Aprime$ in the $B^0 \to \Aprime \Aprime$ decays, where $\Aprime$ subsequently decays to $e^+ e^-$, $\mu^+ \mu^-$, and $\pi^+ \pi^-$.  
The search is performed by analyzing $772 \times 10^6$ $B\overline{B}$ events collected by the Belle detector at the KEKB $e^+ e^-$ energy-asymmetric collider at the $\Upsilon (4S)$ resonance.  
No signal is found in the dark photon mass range $0.01~\GeV/c^2 \le m_{\Aprime} \le 2.62~\GeV/c^2$, and we set upper limits of the branching fraction of $B^0 \to \Aprime \Aprime$ at the 90\% confidence level. 
The products of branching fractions, $\prodee$ and $\prodmm$, have limits of the order of $10^{-8}$ depending on the $\Aprime$ mass.
Furthermore, considering $\Aprime$ decay rate to each pair of charged particles, the upper limits of $\BFBAA$ are of the order of $10^{-8}$--$10^{-5}$.
From the upper limits of $\BFBAA$, we obtain the Higgs portal coupling for each assumed dark photon and dark Higgs mass.
The Higgs portal couplings are of the order of $10^{-2}$--$10^{-1}$ at $m_{h'} \simeq m_{B^0} \pm 40~\MeV/c^2$ and $10^{-1}$--$1$ at $m_{h'} \simeq m_{B^0} \pm 3~\GeV/c^2$.
}


\vspace*{0.2cm}
\preprint{\vbox{ \hbox{   }
\hbox{Belle Preprint 2020-19}
\hbox{KEK Preprint   2020-36}
}}

\maketitle
\flushbottom

\section{Introduction}

The validity of the Standard Model (SM) has been confirmed by various experimental measurements~\cite{PDG}, but it is also known that the SM is incomplete and cannot
explain several phenomena occurring in nature, e.g. neutrino oscillations~\cite{neutrino,neutrino2} 
and the baryon asymmetry~\cite{baryo}.
A possible way to explain the above problems while keeping the internal structure of the SM unaffected is to introduce a dark sector~\cite{PhysRevD.83.054005} that interacts with the SM particles only very weakly.
For example, a vector mediator of hypothetical $U'(1)$ gauge interaction in the dark sector, the so-called dark photon, may interact with matter through various portals with a small coupling strength \cite{HOLDOM1986196,FAYET1980285,FAYET1981184}.  
Such a model of the dark sector with portal interaction to the SM could explain the muon $g-2$ anomaly 
\cite{mug2,PhysRevD.80.095002,PhysRevD.83.101702,PhysRevLett.107.011803},
baryogenesis~\cite{baryodark}, and high energy positron fraction anomaly in cosmic rays~\cite{ATIC,PAMELA,FermiLAT,AMS,CHEN2009255}.

\begin{figure}[]
\centering
\includegraphics[width=0.60\textwidth]{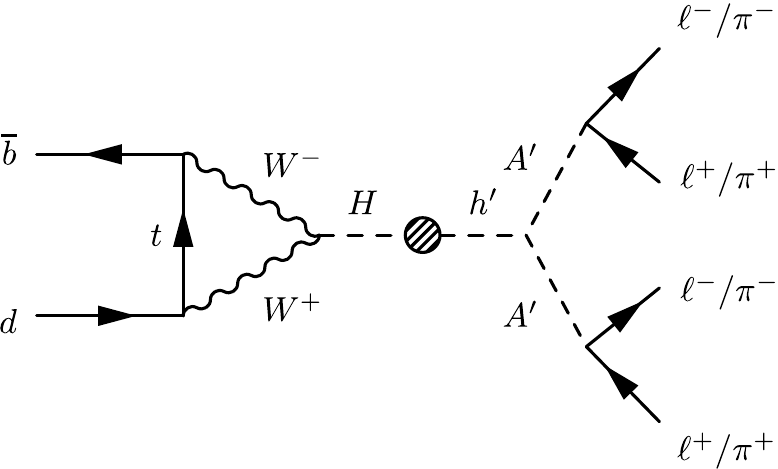}
  \caption{A possible diagram of $B^0 \to \Aprime \Aprime$ decay through off-shell Higgs--dark Higgs mixing indicated by the shaded circle.}
\label{Fig:feynman}
\end{figure}

In this paper, we report a search for the dark photon $\Aprime$, in the decays of $B^0$ mesons by analyzing the $e^+ e^-$ collision data from the Belle experiment. 
In particular, we study $B^0$ decays into a pair of dark photons, $B^0 \to \Aprime\Aprime$, which are mediated by an off-shell dark Higgs $h'$~\cite{PhysRevD.83.054005} (Fig.~\ref{Fig:feynman}), wherein we scan the $\Aprime$ mass range between 
$0.01~\GeV/c^2$ and $2.62~\GeV/c^2$ in $10~\MeV/c^2$ ($m_{\Aprime} < 1.1~\GeV/c^2$) and $20~\MeV/c^2$ ($m_{\Aprime} > 1.1~\GeV/c^2$) intervals. Throughout the paper, the charge-conjugate modes are always implied. 
In this paper, we restrict ourselves to the hypothesis that  
all dark-sector particles coupling to $\Aprime$ are heavier than $\Aprime$, therefore the latter can only decay to SM particles. Moreover, we assume that the $\Aprime$ decays promptly.
In the kinematic range of this analysis, the allowed $\Aprime$ decay are to $e^+ e^-$, $\mu^+ \mu^-$, or hadronic final states.
Lepton-flavor-violating decays~\cite{osti_1254817,PhysRevD.97.075022} $\Aprime \to e^\pm \mu^\mp$ are not considered in this analysis.

\subsection{Branching fraction of dark photon decay}

In order to obtain $\BF(B^0 \to \Aprime \Aprime)$ from the analysis of the decays into the final states considered, 
we need to know the branching fractions of $\Aprime$ to a particular final state.  
Below the $\tau^+ \tau^-$ threshold, the branching fraction of the dark photon that is consistent with our hypothesis is obtained as
\begin{equation} \label{eq:darkbr}
\begin{aligned}
  \mathcal{B}(\Aprime \to \ell^+ \ell^- / \pi^+ \pi^-) = \frac{\Gamma_{\Aprime \to \ell^+ \ell^- / \pi^+ \pi^- }}{\Gamma_{\Aprime \to e^+ e^-} + \Gamma_{\Aprime \to \mu^+ \mu^-} + \Gamma_{\Aprime \to \textrm{hadrons}}},
\end{aligned}
\end{equation}
where $\ell = e~\textrm{or}~\mu$.
Following Ref.~\cite{PhysRevD.79.115008}, we write down the partial widths to $\ell^+ \ell^-$ and hadrons as 
\begin{equation} \label{eq:llwidth}
\begin{aligned}
  \Gamma_{\Aprime \to \ell^+ \ell^-} =& \frac{1}{3} \alpha \varepsilon_{\textrm{mix}}^2 m_{\Aprime} \sqrt{1 - {4 m_\ell^2}/{m_{\Aprime}^2}}(1 + 2 m_\ell^2 / m_{\Aprime}^2),\\ 
  \Gamma_{\Aprime \to \textrm{hadrons}} =& \Gamma_{\Aprime \to \mu^+ \mu^-} \times R(s = m_{\Aprime}^2),
\end{aligned}
\end{equation}
with the square of the total center-of-mass (CM) frame energy $s$, 
the kinetic mixing parameter $\varepsilon_{\textrm{mix}}$, and
$R(s) = \sum_{e^+ e^- \to \textrm{hadrons}} / \sum_{e^+ e^- \to \mu^+ \mu^-}$ which is determined by various experiments \cite{PDG}.  
The branching fraction of $\Aprime \to \pi^+ \pi^-$ is then obtained as ~\cite{PhysRevLett.108.211801}:
\begin{equation}
\begin{aligned}
\mathcal{B}(\Aprime \to \pi^+ \pi^-) = \mathcal{B}(\Aprime \to \textrm{hadrons}) \times \sum(e^+ e^- \to \pi^+ \pi^-) / \sum(e^+ e^- \to \textrm{hadrons}).
\end{aligned}
\end{equation}

\subsection{The SM expectation of $B^0$ decays to four charged leptons}

The $B^0$-decay final states that we analyze are $e^+ e^- e^+ e^-$, $e^+ e^- \mu^+ \mu^-$, 
$\mu^+ \mu^- \mu^+ \mu^-$, $e^+ e^- \pi^+ \pi^-$, and $\mu^+ \mu^- \pi^+ \pi^-$. 
In the SM, branching fractions of $B^0$-meson decays to four-charged-lepton final states are expected to be $\mathcal{O}(10^{-12})$~\cite{Danilina2018}. 
Due to the low SM signal and background yields expected, these multileptonic $B$-meson decay channels can be a sensitive probe for dark sector bosons. 
The LHCb experiment has set an upper limit $\BF(B^0 \to \mu^+ \mu^- \mu^+ \mu^-) < 6.9 \times 10^{-10}$ at $95\%$ confidence level (C.L.)~\cite{Aaij2017} and measured $\mathcal{B} (B^0 \to \mu^+ \mu^- \pi^+ \pi^- ) = (2.1 \pm 0.5) \times 10^{-8}$~\cite{Aaij:2014lba}.

\section{The Belle detector}

Our analysis is based on the full $711~\textrm{fb}^{-1}$ integrated luminosity of 
the $\Upsilon(4S)$ data set from the Belle detector \cite{Belle1,Belle2} at KEKB $e^+ e^-$ energy-asymmetric collider \cite{KEKB1,KEKB2}.  
The Belle detector consists of seven subdetectors with $1.5~\textrm{T}$ magnetic field along the beam axis.
Inside the coil, there are the silicon vertex detector, the central drift chamber (CDC), 
the aerogel threshold Cherenkov counters (ACC), the time-of-flight scintillation counters (TOF), and
the electromagnetic calorimeter (ECL).
In the return yoke outside the coil, a $K_L^0$ meson and muon detector (KLM) is instrumented.

We perform a blind search in this analysis, for which we generate Monte Carlo (MC) simulation samples using EvtGen \cite{evtgen} for event generation and GEANT3 \cite{geant3} for detector simulation. 
Signal efficiencies are determined from the signal MC set, where one million events are generated for each signal mode and dark photon mass.  
The event shape and amount of the background events are studied
by using generic MC samples simulating $e^+ e^- \to \Upsilon(4S) \to B\overline{B}$ and $e^+ e^- \to q\bar{q}~(q=u,d,s,c)$ (`continuum') processes. The size of MC samples for  $\Upsilon(4S)$ and continuum simulation corresponds to 10 and 6 times that of real data, respectively.

\section{Signal event selection}

To select signal events, we retain tracks satisfying the following track reconstruction quality requirements.  
Because we assume prompt dark photon decays, all tracks are required to originate from near the interaction point (IP). 
In particular, each track should satisfy the following 
conditions on the impact parameters in the transverse and longitudinal directions, $dr < 0.2~\textrm{cm}$ and $|dz| < 3.0~\textrm{cm}$, respectively.
The impact parameters are calculated using the beam IP and track helix, and the $z$-axis is aligned opposite the direction of positron beam.
We also require a good track fit based upon $\chi^2$ per degree of freedom ($N_{\textrm{d.o.f.}}$) by accepting only the tracks with $\chi^2 / N_{\textrm{d.o.f.}} < 5$.

The species of the charged particles are identified by considering the likelihood ratios.  Muons are identified by requiring ${\mathcal{L}_{\mu}}/({\mathcal{L}_{\mu} + \mathcal{L}_{K} + \mathcal{L}_{\pi}}) > 0.9$, where the likelihood $\mathcal{L}_j\ (j=\mu, K, \pi)$~\cite{Abashian:2002bd} 
is constructed by the hit position and penetration in the KLM.  Electrons are required to meet ${\mathcal{L}_{e}}/({\mathcal{L}_{e} + \mathcal{L}_{\textrm{not-}e}}) > 0.9$ where the likelihood $\mathcal{L}_j\ (j = e, \textrm{not-}e)$~\cite{Hanagaki:2001fz} is determined by $dE/dx$ from the CDC, ratio of the ECL cluster energy to the matched track momentum, shower shape of the ECL cluster, and the ACC photoelectron response.
Charged pions and kaons are identified by the likelihood~\cite{Nakano:2002jw} using the $dE/dx$ from the CDC, the ACC photoelectron response, and the time-of-flight information from the TOF.
The tracks with ${\mathcal{L}_{\pi}}/({\mathcal{L}_{K} + \mathcal{L}_{\pi}}) > 0.4$ are identified as pions.

To recover energy losses by $e^\pm$ candidates due to bremsstrahlung, radiative photons are added to the electron momentum if they fall within a 0.05 radian cone around the $e^\pm$ direction.
We require these photons to exceed an energy threshold that depends on the ECL region: $E_{\gamma} > 50$ (barrel), $100$ (forward endcap), 
and $150$ (backward endcap) $\textrm{MeV}$.

The dark photon candidate is reconstructed in the following modes: $\Aprime \to e^+ e^-$, $\mu^+ \mu^-$, and $\pi^+ \pi^-$.  For $B^0 \to e^+ e^- e^+ e^-$ and $\mu^+ \mu^- \mu^+ \mu^-$ modes, we have an ambiguity between $(\ell_1^+ \ell_1^-)(\ell_2^+ \ell_2^-)$ and $(\ell_1^+ \ell_2^-)(\ell_2^+ \ell_1^-)$, where the lepton pair from a single $\Aprime$ decay is indicated in parentheses.  
To find a single dark photon combination per event, we choose that corresponding to the smallest invariant mass difference of dark photon candidates, $\Delta M_{\Aprime}$.

Finally, $B^0$ candidates are reconstructed from two dark photon candidates.  To extract signal events from data, we use the following five variables, defined 
in the CM frame: $M_{\textrm{bc}}$, $\Delta E$, $E_{\textrm{miss}}$, $\Delta M_{\Aprime}$, and $\sum \delta M_{\Aprime}$. 
$M_{\textrm{bc}} \equiv \sqrt{(\sqrt{s}/2)^2 - \vec{\textrm{p}}_B^2}$ is the beam-energy-constrained mass, 
where $\vec{\textrm{p}}_B$ is the momentum of the reconstructed $B^0$.  
$\Delta E \equiv E_{B^0} - (\sqrt{s}/2)$ is the difference between the $B^0$-candidate energy and the beam energy ($=\sqrt{s}/2$), and $E_{\textrm{miss}}$ is the missing energy, $E_{\textrm{miss}} \equiv \sqrt{s} - \sum_j E_j$ where the index $j$ is for all charged and neutral particles in the event.  
The missing energy is useful to reduce combinatorial background due to multiple semileptonic decays from 
$b \to c \ell^- \bar{\nu}_\ell$ and $c \to (s,d) \ell^+ \nu_\ell$ for both $B$ and $\overline{B}$.
For the two dark photon candidates in an event, we calculate 
$\Delta M_{\Aprime} \equiv |M_{\Aprime_1} - M_{\Aprime_2}|$ and $\sum \delta M_{\Aprime} \equiv |(M_{\Aprime_1} - \mAbin) + (M_{\Aprime_2} - \mAbin)|$, where $M_{\Aprime_{j}}$ is the reconstructed mass of $\Aprime_{j}\ (j=1,2)$ and $\mAbin$ is the nominal $\Aprime$ mass for a particular bin of $m_{\Aprime}$. 

For the signal event selection, we require
$M_{\textrm{bc}} > 5.27~\GeV/c^2$ and $E_{\textrm{miss}} < 3.5~\GeV$ for all modes. 
Considering the energy loss from $e^\pm$, $\Delta E$ requirements are chosen separately for different modes:  
$-0.2~\GeV < \Delta E < 0.05~\GeV$ for $B^0 \to e^+ e^- e^+ e^-$, 
$-0.1~\GeV < \Delta E < 0.04~\GeV$ for $B^0 \to e^+ e^- \mu^+ \mu^-$ and $e^+ e^- \pi^+ \pi^-$, and 
$-0.03~\GeV < \Delta E < 0.03~\GeV$ for $B^0 \to \mu^+ \mu^- \mu^+ \mu^-$ and $\mu^+ \mu^- \pi^+ \pi^-$. 
We use $\Delta M_{\Aprime}$ and $\sum \delta M_{\Aprime}$ to set the search window for each $\mAbin$ and the final-state mode. The requirements on these variables depend on both $\mAbin $ and the number of electrons in the final state. 
For $\mAbin > 0.1~\GeV/c^2$, we require $\Delta M_{\Aprime}(\sum \delta M_{\Aprime}) < 0.06 \times \mAbin + 0.03 ~\GeV/c^2$ for 
$B^0 \to e^+ e^- e^+ e^-$, $\Delta M_{\Aprime}(\sum \delta M_{\Aprime}) < 0.03 \times \mAbin + 0.01 ~\GeV/c^2$ for 
$B^0 \to e^+ e^- \mu^+ \mu^-$ and $e^+ e^- \pi^+ \pi^-$, and 
$\Delta M_{\Aprime}(\sum \delta M_{\Aprime}) < 0.01 \times \mAbin + 0.01 ~\GeV/c^2$ for $B^0 \to \mu^+ \mu^- \mu^+ \mu^-$ and $\mu^+ \mu^- \pi^+ \pi^-$. 
The above conditions are determined so that if we consider the distribution of $\Delta M_{\Aprime}$ the upper edge of the accepted region has a value of nearly $3$--$5\%$ of the peak value. In addition, we make use of the approximately linear increase of the $\Delta M_{\Aprime}$ width as a function of $\mAbin $.
We choose the same selection for $\sum \delta M_{\Aprime}$ since the distribution is almost the same as $\Delta M_{\Aprime}$.  
For $\mAbin \le 0.1~\GeV/c^2$, we apply slightly different selection conditions 
for $\Delta M_{\Aprime}$ and $\sum \delta M_{\Aprime}$, while requirements on $M_{\textrm{bc}}$ and $\Delta E$ remain the same as for $\mAbin > 0.1~\GeV/c^2$.  
We do not use  $E_{\textrm{miss}}$ for $\mAbin \le 0.1~\GeV/c^2$, because for such low-mass dark photons, little background is expected from generic $B$ decays. 
For $\mAbin \le 0.1~\GeV/c^2$, the resolutions of both $\Delta M_{\Aprime}$ and $\sum \delta M_{\Aprime}$ are nearly independent of $\mAbin$.
Therefore, we require $\Delta M_{\Aprime} < 0.02~\GeV/c^2$ and $\sum \delta M_{\Aprime} < 0.02~\GeV/c^2$ for all $m_{\Aprime} \le 0.1~\GeV/c^2$. 
From the MC study, our $\Delta M_{\Aprime}$ selections in $\Aprime \to \mpmm$ and $\ipim$ cover roughly $\pm 2.5$ times the mass resolution. 
In case of $\Aprime \to \epem$, the mass resolution is worse, and our selections correspond to $\pm (1.7-2.5)$ times the mass resolution, depending on $m_{\Aprime}$.  
For instance, the $M_{\Aprime}$ resolution of the $1.5~\mathrm{GeV}$ dark photon is about $5~\mathrm{MeV}$ for $\Aprime \to \mpmm$ or $\ipim$, while for $\Aprime \to \epem$ it is about $20~\mathrm{MeV}$.
The union of the search windows determined using $\Delta M_{\Aprime}$ and $\sum \delta M_{\Aprime}$ for all $\mAbin $ covers the entire dark photon mass range of our study without any gap.

The dominant SM background sources for $\ell^+ \ell^-$ pairs are photon conversion and charmonium meson decays, mostly $J/\psi$ and $\psi(2S)$.
To reduce the background events from photon conversion, $e^+ e^-$ pairs with $M_{e^+ e^-} < 0.1 ~\GeV/c^2$ are rejected when we search for $m_{\Aprime} > 0.1~\GeV/c^2$.  
On the other hand, this veto is not applied for the searches in the region $m_{\Aprime} \leq 0.1~\GeV/c^2$. 
To suppress the lepton pairs from charmonium decays such as $J/\psi$ or $\psi(2S) \to \ell^+ \ell^-$, we reject two regions: 
$3.00(3.05) ~\GeV/c^2 < M_{e^+ e^-(\mu^+ \mu^-)} < 3.15(3.13) ~\GeV/c^2$ for $J/\psi$ and 
$3.60(3.65) ~\GeV/c^2 < M_{e^+ e^-(\mu^+ \mu^-)} < 3.75(3.73) ~\GeV/c^2$ for $\psi(2S)$.

For the charged pion pairs, there is strong background from light mesons, such as $K_S^0$, $\rho^0$, and $f_0(980)$.  
Because of possible $K$--$\pi$ misidentification, $K^{*0}$, $\phi$ and so on are also a source of possible background.  Since production of such mesons is copious, especially that of $\rho^0$ mesons,
we reject the $0.45 ~\GeV/c^2 < M_{\pi^+ \pi^-} < 1.1 ~\GeV/c^2$.
Another source of pion pairs is $D^0$ meson.  Two decay channels, $D^0 \to \pi^+ \pi^-$ and $D^0 \to \pi^+ K^-$ are considered.  
A direct $D^0$ veto is applied by removing $\pi^+ \pi^-$ combinations which satisfy $1.85 ~\GeV/c^2 < M_{\pi^+ \pi^-} < 1.88 ~\GeV/c^2$.
The other decay channel, $D^0 \to \pi^+ K^-$, can mimic the signal via $K$--$\pi$ misidentification.
We reject these events by discarding the $1.85 ~\GeV/c^2 < M_{\pi^+ K^-} < 1.88 ~\GeV/c^2$ mass range.

After signal selection, most of the combinatorial background is in the 
$B^0 \to \ell^+ \ell^- \pi^+ \pi^-$ mode, coming from the continuum processes $e^+ e^- \to q\bar{q}$ ($q= u,d,s$ or $c$).  
In the four-lepton mode, on the other hand, there is almost no background left.
The continuum background is suppressed via multivariate analysis (MVA) using the Fisher discriminant~\cite{Fisher} method in the TMVA~\cite{Hocker:2007ht} package. 
We make use of 16 event shape variables: 
the cosine of angle between the beam axis and $B^0$ momentum ($\cos{\theta_{B}}$), 
the cosine of angle between the thrust axis of the $B^0$ daughters and that of the rest of the event ($\cos{\theta_{\textrm{T}}}$), 
and the Fisher discriminant components of modified Fox-Wolfram moments \cite{PhysRevLett.91.261801}. 
The MVA training is performed for the $\ell^+ \ell^- \pi^+ \pi^-$ final state for each $\mAbin$, using the signal and continuum MC.  
We apply MVA selection creteria to retain from 75\% to 90\% of signal and from 10\% to 30\% of continuum background, depending on $m_{\Aprime}$ and final state.

\section{Systematic uncertainties}

\begin{figure*}
\centering
\includegraphics[width=.46\textwidth]{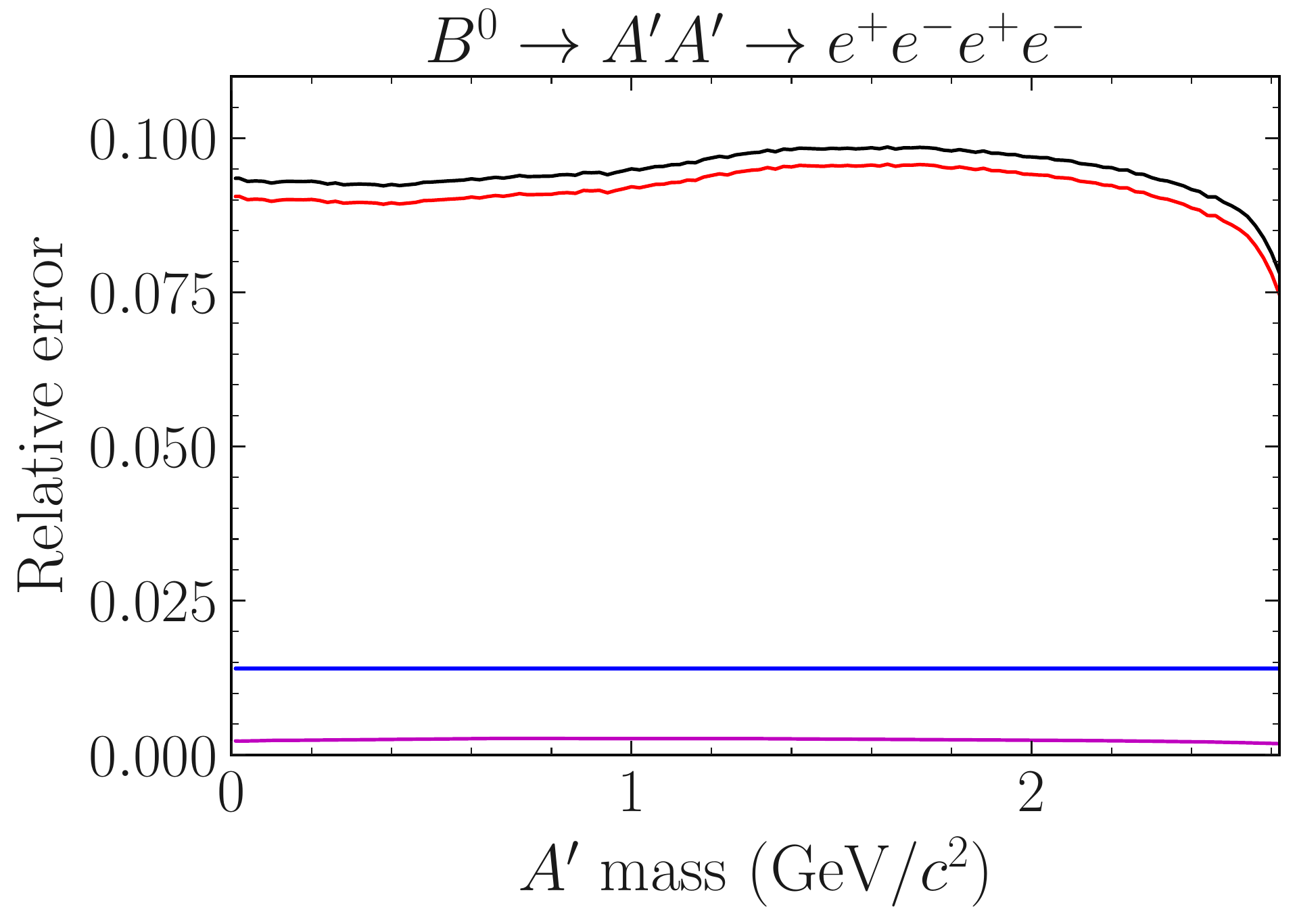}\quad
\includegraphics[width=.46\textwidth]{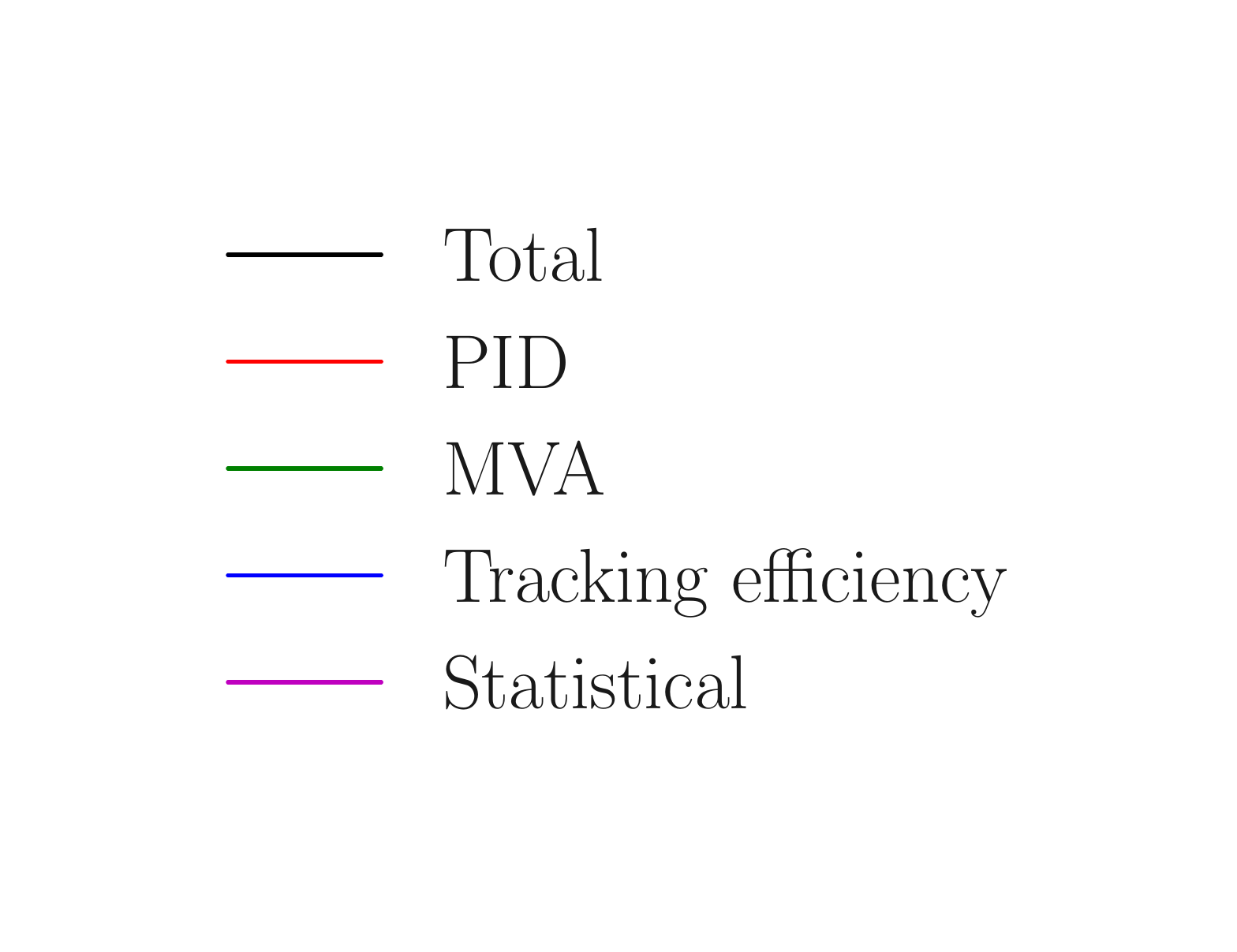}

\medskip

\includegraphics[width=.46\textwidth]{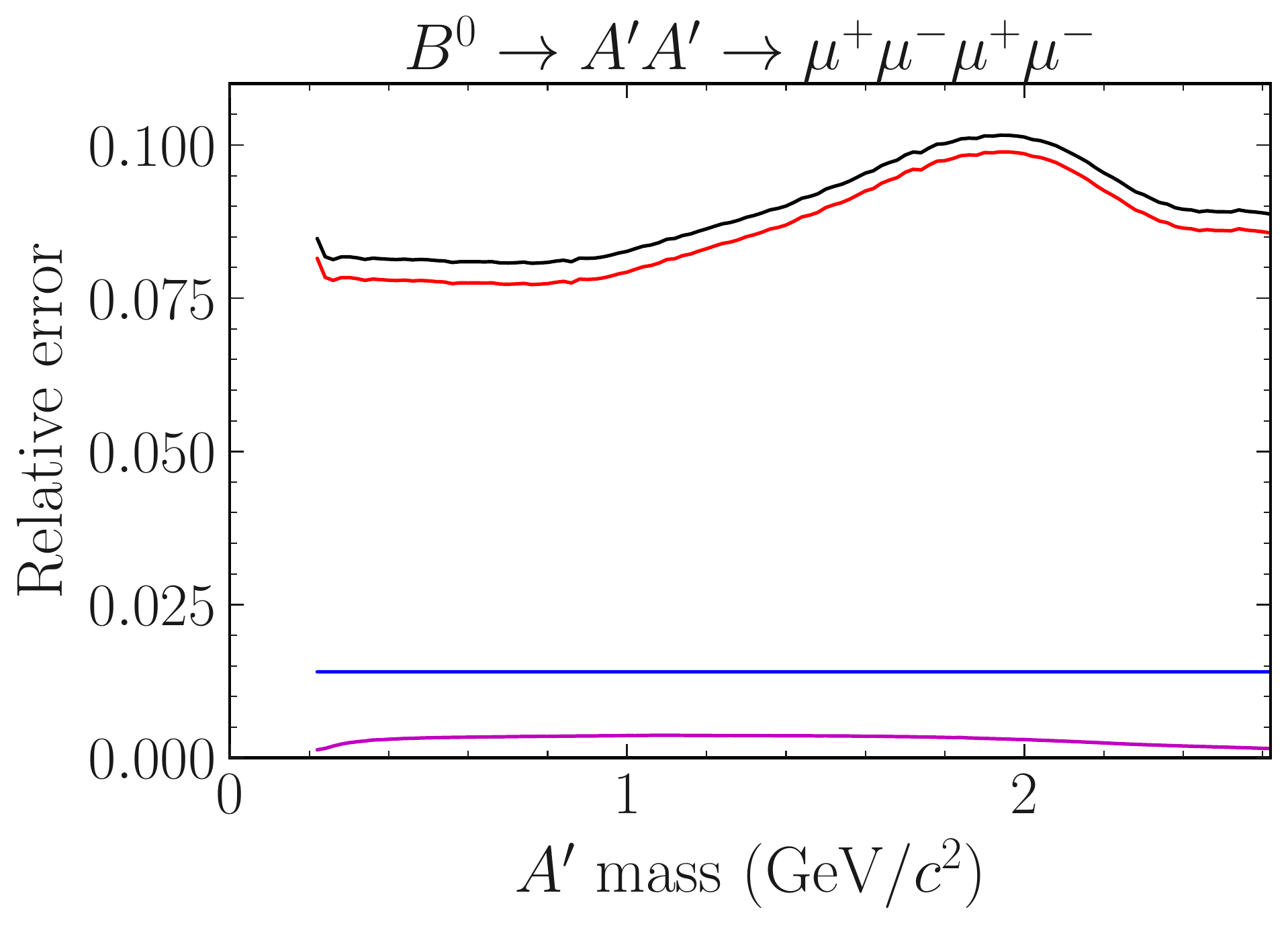}\quad
\includegraphics[width=.46\textwidth]{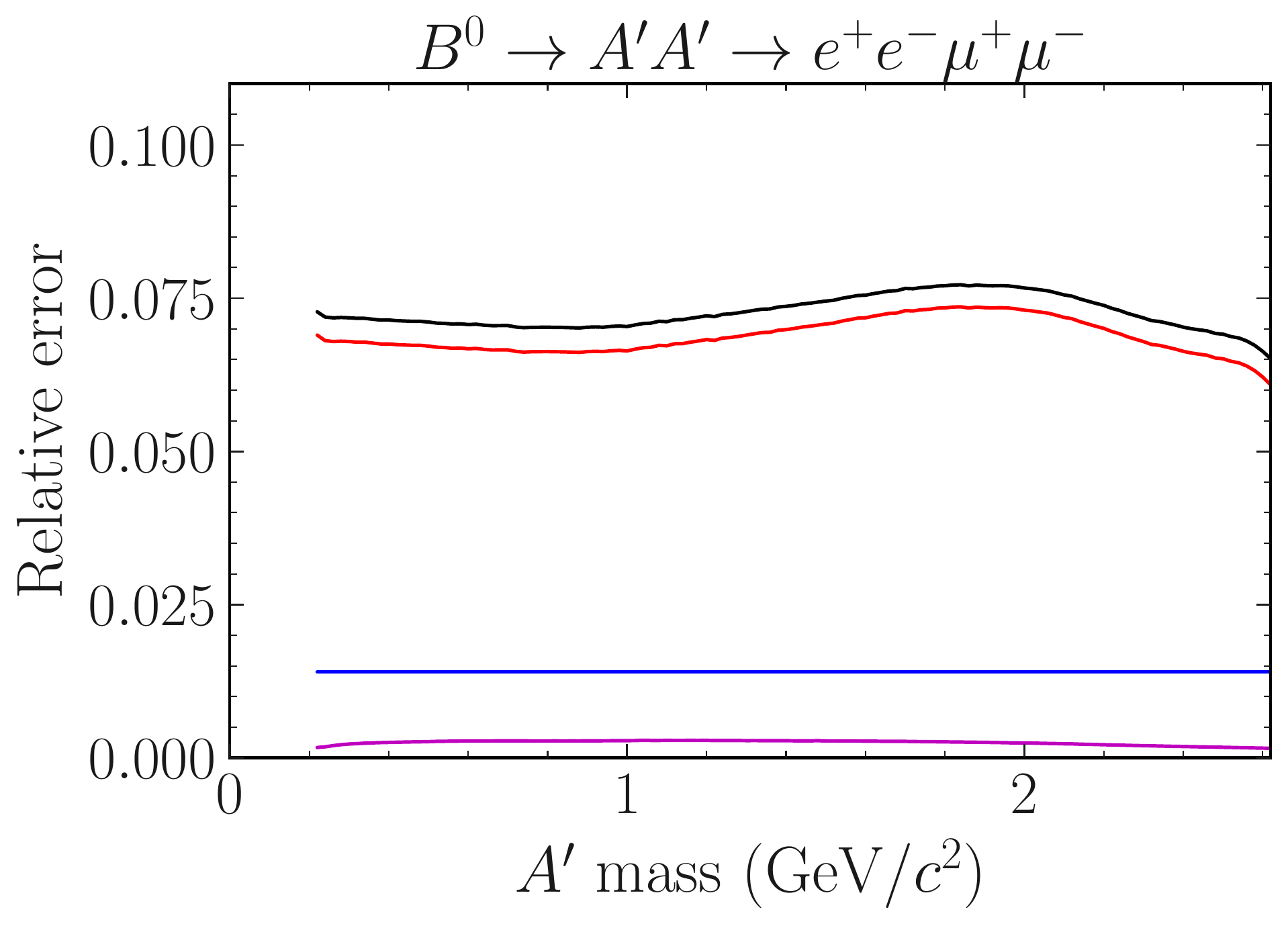}

\medskip

\includegraphics[width=.46\textwidth]{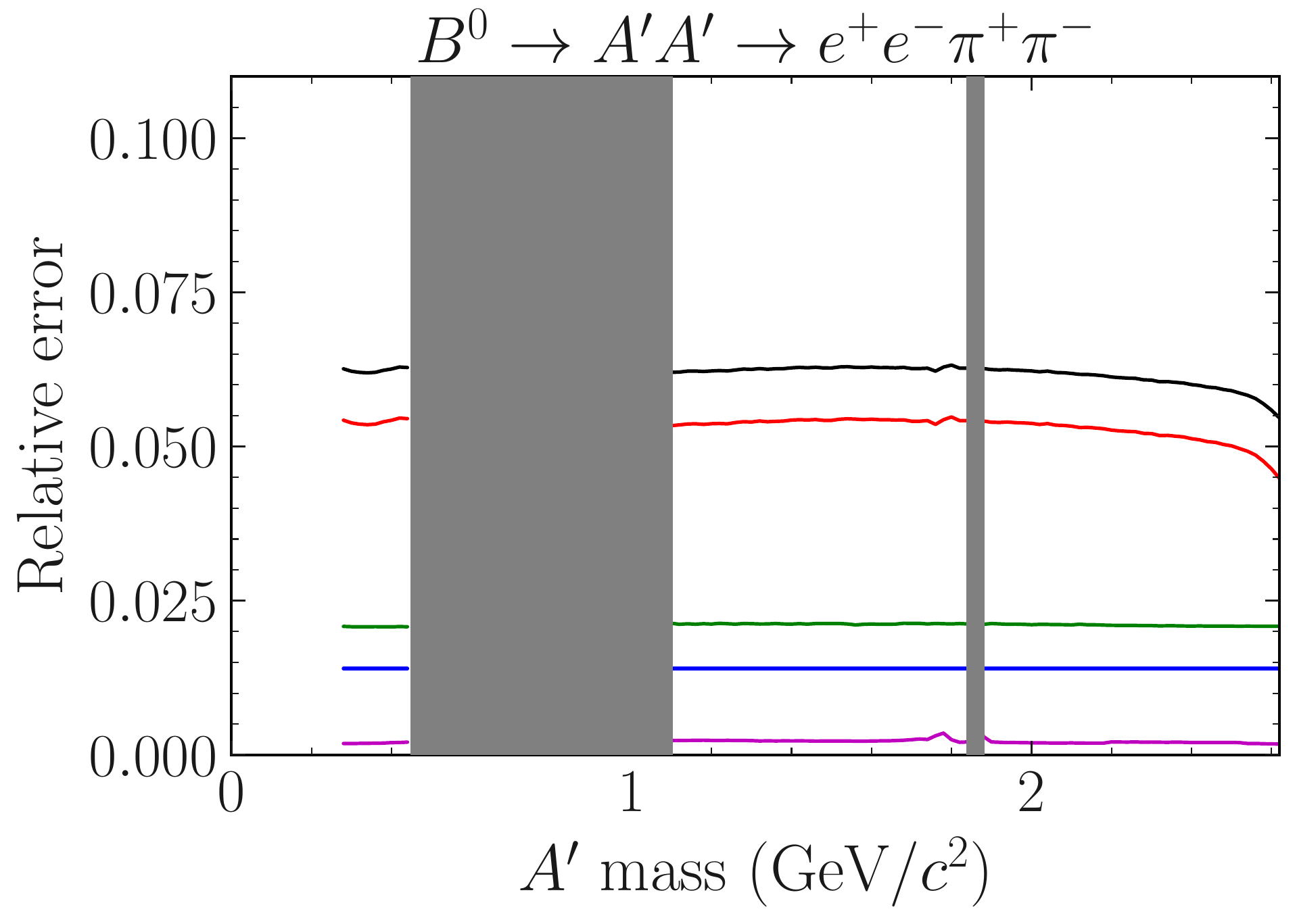}\quad
\includegraphics[width=.46\textwidth]{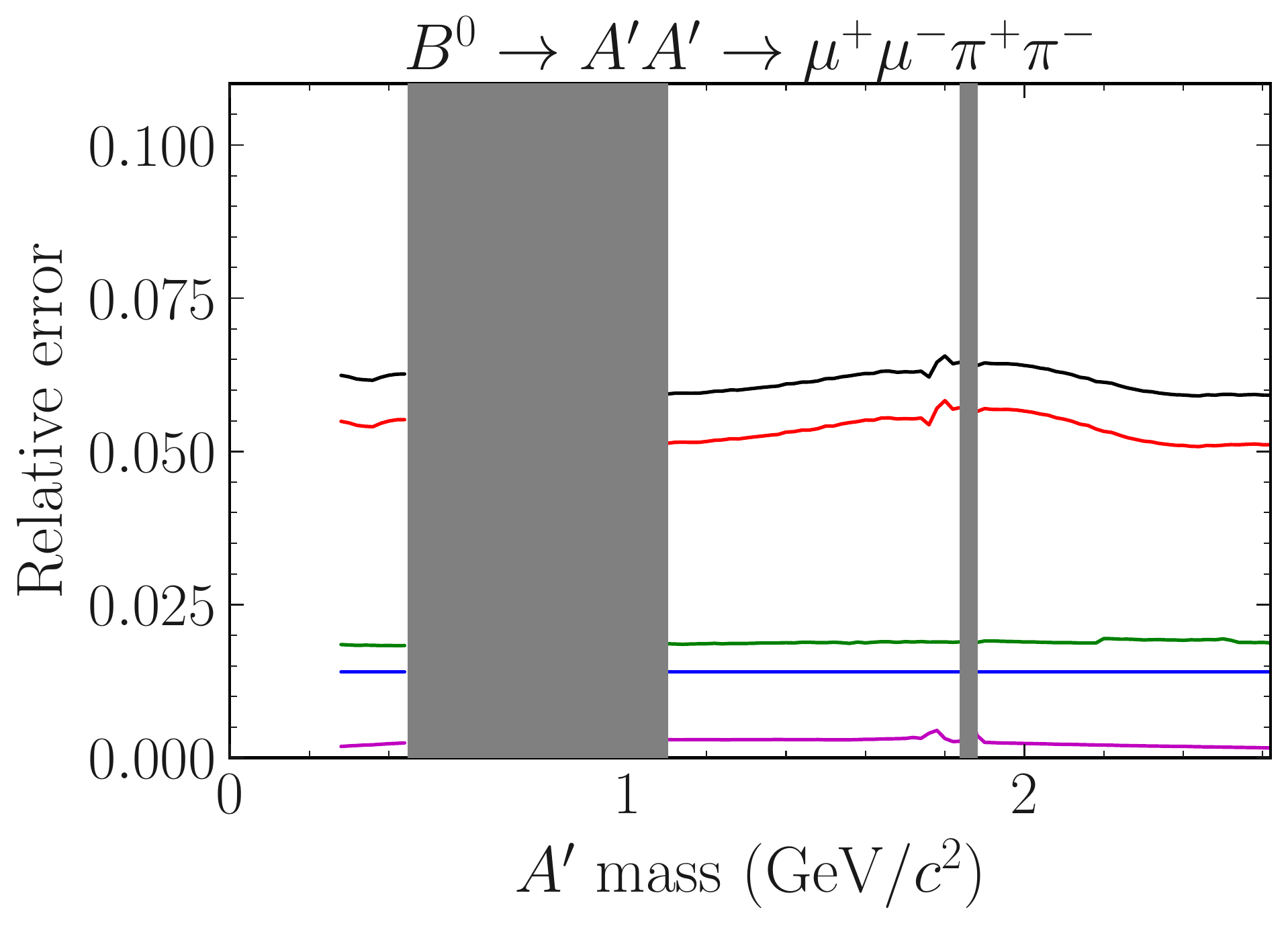}

  \caption{\label{Fig:systematic}Relative uncertainty of signal reconstruction efficiency for each $\Aprime$ mass and final state.}
\end{figure*}

We determine the branching fraction of $B^0 \to \Aprime \Aprime$ as
\begin{equation} \label{eq:br}
  \mathcal{B}(B^0 \to \Aprime \Aprime) = \frac{N_{\textrm{obs}} - N_{\textrm{bkg}}}
    {\epsilon \times 2 \times N_{B\overline{B}} \times \mathcal{B}_0},
\end{equation}
where $\mathcal{B}_0$ is the branching fraction of $\Upsilon(4S)\to B^0 \overline{B}^0$, of which the current 
world-average value is $0.486 \pm 0.006$ \cite{PDG},
$N_{\textrm{obs}}$ is the yield, $N_{\textrm{bkg}}$ is the number of expected background events determined from MC, 
$\epsilon$ is the signal reconstruction efficiency considering branching fraction of $\Aprime$ subdecays, and 
$N_{B\overline{B}} = (772 \pm 11) \times 10^6$ is the number of $B\overline{B}$ pairs which are collected by the Belle detector.

The most important source of systematic uncertainties is the signal reconstruction efficiency, which is obtained by MC. 
The sources of uncertainty include the statistical error in the signal MC, track reconstruction efficiency, particle identification (PID) efficiency, and uncertainties in the MVA method used to suppress continuum background for $\ell^+ \ell^- \pi^+ \pi^-$ final states.  The uncertainties for $N_{B\overline{B}}$ and $\mathcal{B}_0$ also contribute to systematics.
The uncertainties due to background estimation are very small compared to other systematic uncertainties.

Track reconstruction efficiency is studied using the decay chain $D^{*+} \to D^0 \pi^+$, $D^0 \to K_S^0 \pi^+ \pi^-$, and 
$K_S^0 \to \pi^+ \pi^-$ where we tag all the charged tracks in the chain but one from $K_S^0$ decays (`test track') then try to find the test track.  We compare the tracking efficiency difference
of the test track for both data and MC.  The error is 1.4\%, independent of the dark photon mass and final state.

The PID correction is applied to each daughter electron, muon, and pion.
The lepton (pion) identification correction is studied using the difference between MC and data for the process $\gamma \gamma \to e^+ e^- / \mu^+ \mu^-$  ($D^{*+} \to D^0 \pi^+_{\textrm{slow}} \to K^- \pi^+ \pi^+_{\textrm{slow}}$), 
and the errors are approximately $2\%~ (1\%)$ per lepton (pion), with the resulting correction factor being about $90\%$. 
The exact correction factor and uncertainty depend on $m_{\Aprime}$ through different kinematics.

The MVA correction factor and uncertainty are studied using the control mode, $B^0 \to J/\psi {K}^{*0} \to e(\mu)^+ e(\mu)^- \pi^- K^+$.  
We apply MVA training results for the continuum suppression of $\ell^+ \ell^- \pi^+ \pi^-$ modes for each assumed value of $m_{\Aprime}$ to  
$B^0 \to J/\psi K^{*0}$ MC and data.  We then calculate the double ratio
$({N_{\textrm{data,A}} / N_{\textrm{data,B}}})/({N_{\textrm{MC,A}} / N_{\textrm{MC,B}}})$, 
where $N_{\textrm{data(MC),B}}$ and $N_{\textrm{data(MC),A}}$ are the number of signal candidates in data(MC) before and after MVA training application, respectively.
The systematic uncertainty due to MVA training is taken from the uncertainties in the double ratio, and these uncertainties are approximately $2\%$ at all values of $m_{\Aprime}$.

After multiplying all correction factors, signal efficiencies are mostly $5-20$\%.
The efficiencies increase as the $A'$ mass approaches $0$ or $m_{B^0} /2$, in which case both $e^\pm$ ($\mu^\pm$) from the $A'$ decays are more likely to exceed the energy threshold for ECL (KLM) detection.
The summary of signal-efficiency-related systematic uncertainties is shown in Fig.~\ref{Fig:systematic}, and the total systematic uncertainties are $7.5$--$10\%$ for $e^+ e^- e^+ e^-$ and $\mu^+ \mu^- \mu^+ \mu^-$ final states and $5$--$7.5\%$ for $e^+ e^- \mu^+ \mu^-$, $e^+ e^- \pi^+ \pi^-$, and $\mu^+ \mu^- \pi^+ \pi^-$ final states.

\section{Results}

\begin{figure}[]
\centering
\includegraphics[width=0.80\textwidth]{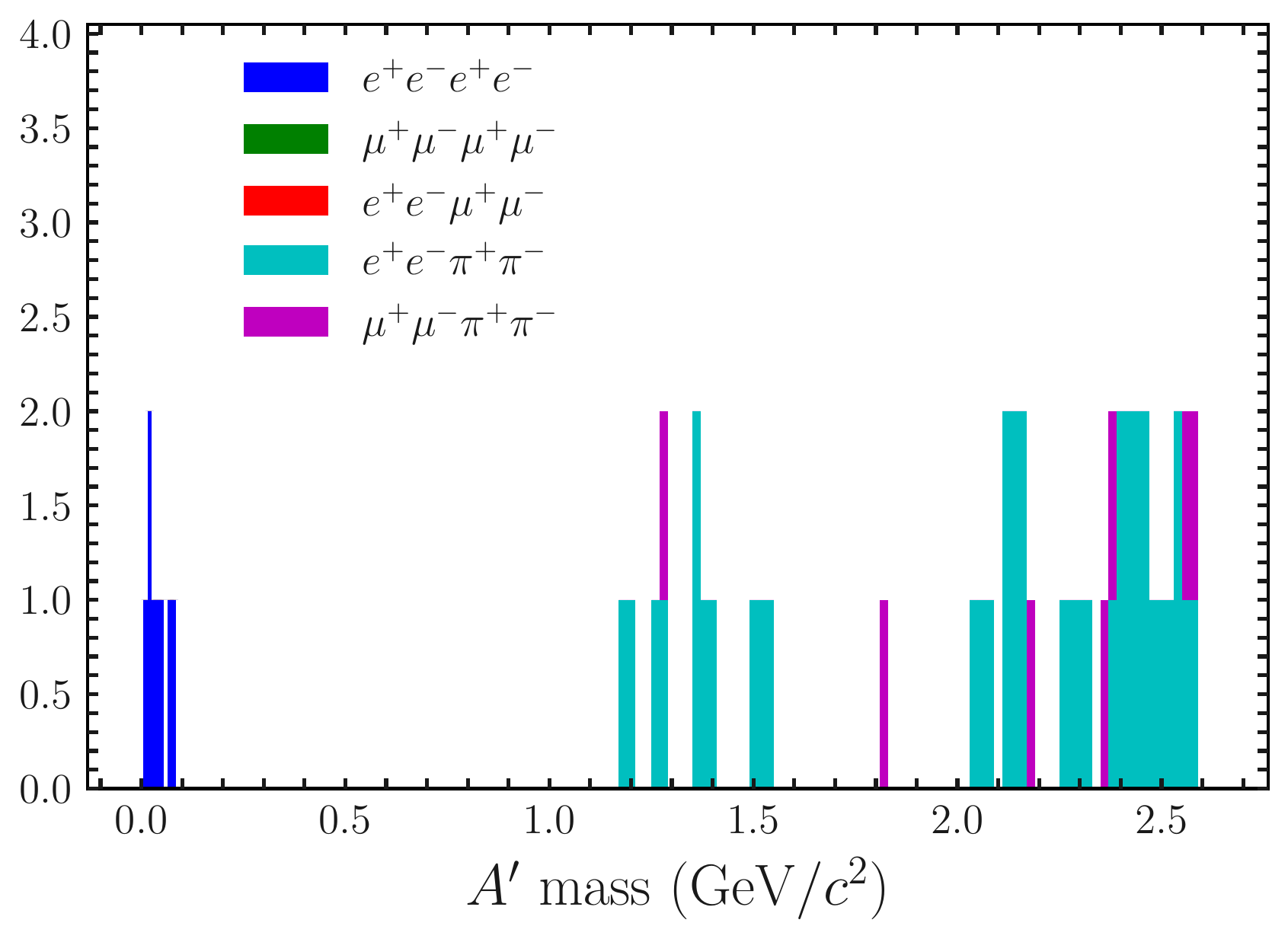}
  \caption{The number of $B^0 \to \Aprime \Aprime$ candidate events for each final state.}
\label{Fig:yield}
\end{figure}

Figure~\ref{Fig:yield} shows the number of $B^0 \to \Aprime \Aprime$ candidate events.
There are no events observed in any bin in the $\eemm$ and $\mmmm$ mode, while we find $N_\mathrm{obs} \leq 2$ events for $\eeee$, $\eepp$, and $\mmpp$ modes.
The yields are consistent with the expected number of background events and we set the upper limits at 90\% C.L. 

\begin{figure}[]
\centering
\includegraphics[width=0.80\textwidth]{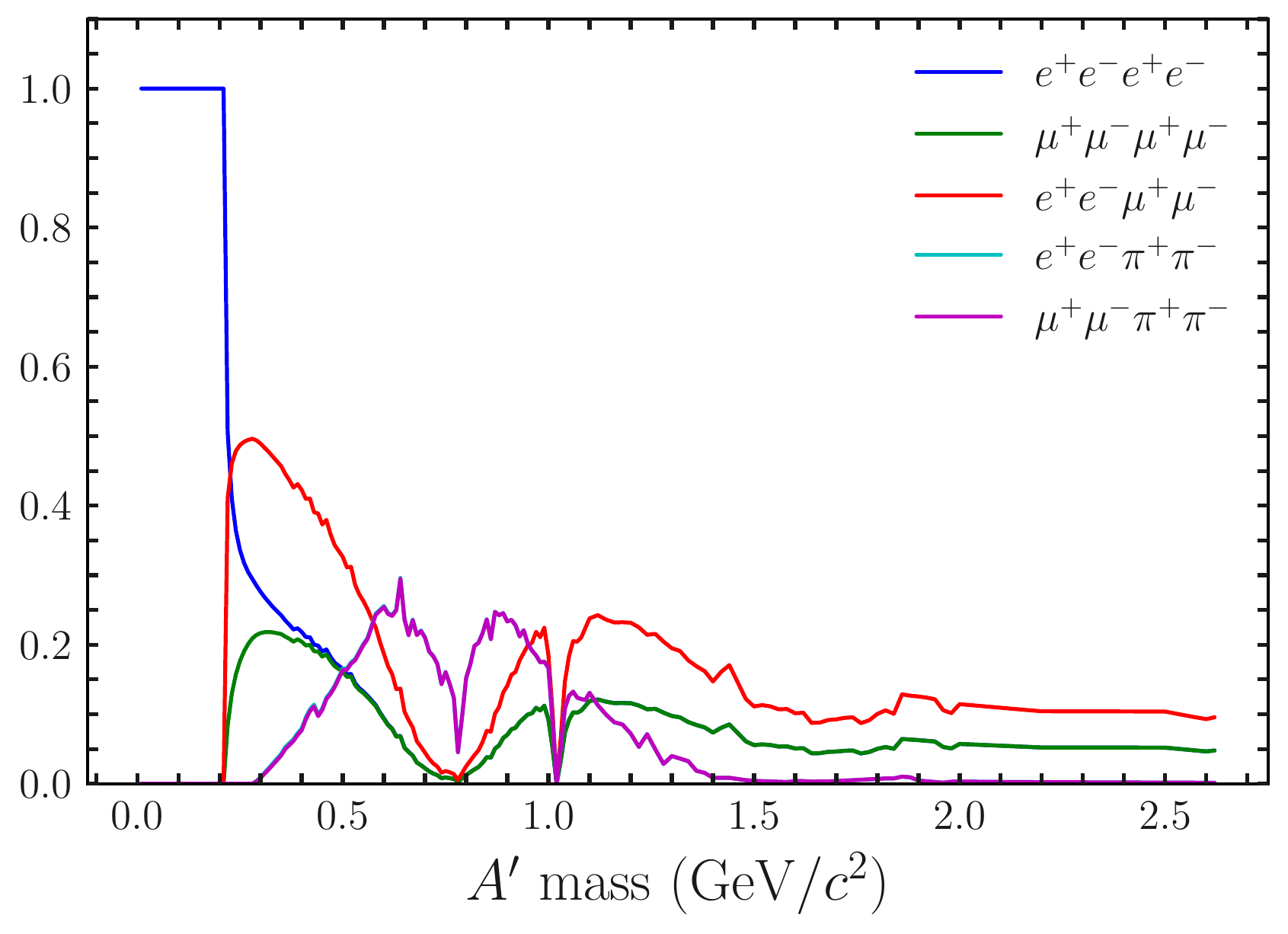}
  \caption{$\mathcal{B}(B^0 \to \Aprime \Aprime \to f)/\mathcal{B}(B^0 \to \Aprime \Aprime)$ distributions for each final state and dark photon mass.  
           $e^+ e^- \pi^+ \pi^-$ and $\mu^+ \mu^- \pi^+ \pi^-$ distributions are almost the same for the whole region. 
           $e^+ e^- e^+ e^-$ and $\mu^+ \mu^- \mu^+ \mu^-$ distributions are the same and $e^+ e^- \mu^+ \mu^-$ distribution is 
           twice that of four-electron or four-muon final states in the region $m_{\Aprime} > 0.5~\GeV/c^2$.
           }
\label{Fig:partialBF}
\end{figure}

For the limits of $\BFBAA$, we combine the number of expected background events, signal candidates in data, and signal reconstruction efficiencies of the five final states.
The combined numbers of expected background events and signal candidates in data are calculated by simply adding the results for the individual final states. 
For the signal efficiencies, we first obtain the ratio  $F_{f} \equiv \mathcal{B}(B^0 \to \Aprime \Aprime \to f)/\mathcal{B}(B^0 \to \Aprime \Aprime)$, where $f$ is each final state, using Eq.~(\ref{eq:darkbr}).
In case of $e^+ e^- \mu^+ \mu^-$, for example, $F_{e^+ e^- \mu^+ \mu^-}$ is $2 \times \mathcal{B}(\Aprime \to e^+ e^-) \times \mathcal{B}(\Aprime \to \mu^+ \mu^-)$. 
The graph of $F_{f}$ is presented in Fig.~\ref{Fig:partialBF}.
With this ratio $F_f$, the combined efficiency is determined as $\sum_f \epsilon_f F_f$ where $\epsilon_f$ is the signal efficiency of the final state $f$. 
The upper limits are calculated using the POLE program \cite{PhysRevD.67.012002}, which is based on the Feldman-Cousins unified approach \cite{PhysRevD.57.3873}.  
We report the limits on the products of branching fractions $\prodee$ and $\prodmm$, as well as the limits on $\BFBAA$. For $\BFBAA$, we use Eq.~(\ref{eq:darkbr}) to combine the five final states. 
The upper limits of $\BFBAA$ are obtained in the mass range 
$0.01~\GeV/c^2 \leq m_{\Aprime} \leq 1.10~\GeV/c^2$ with 10~$\MeV/c^2$ bin and $1.10~\GeV/c^2 \leq m_{\Aprime} \leq 2.62~\GeV/c^2$ with $20~\MeV/c^2$ bin.

\begin{figure}[]
\centering
\includegraphics[width=0.80\textwidth]{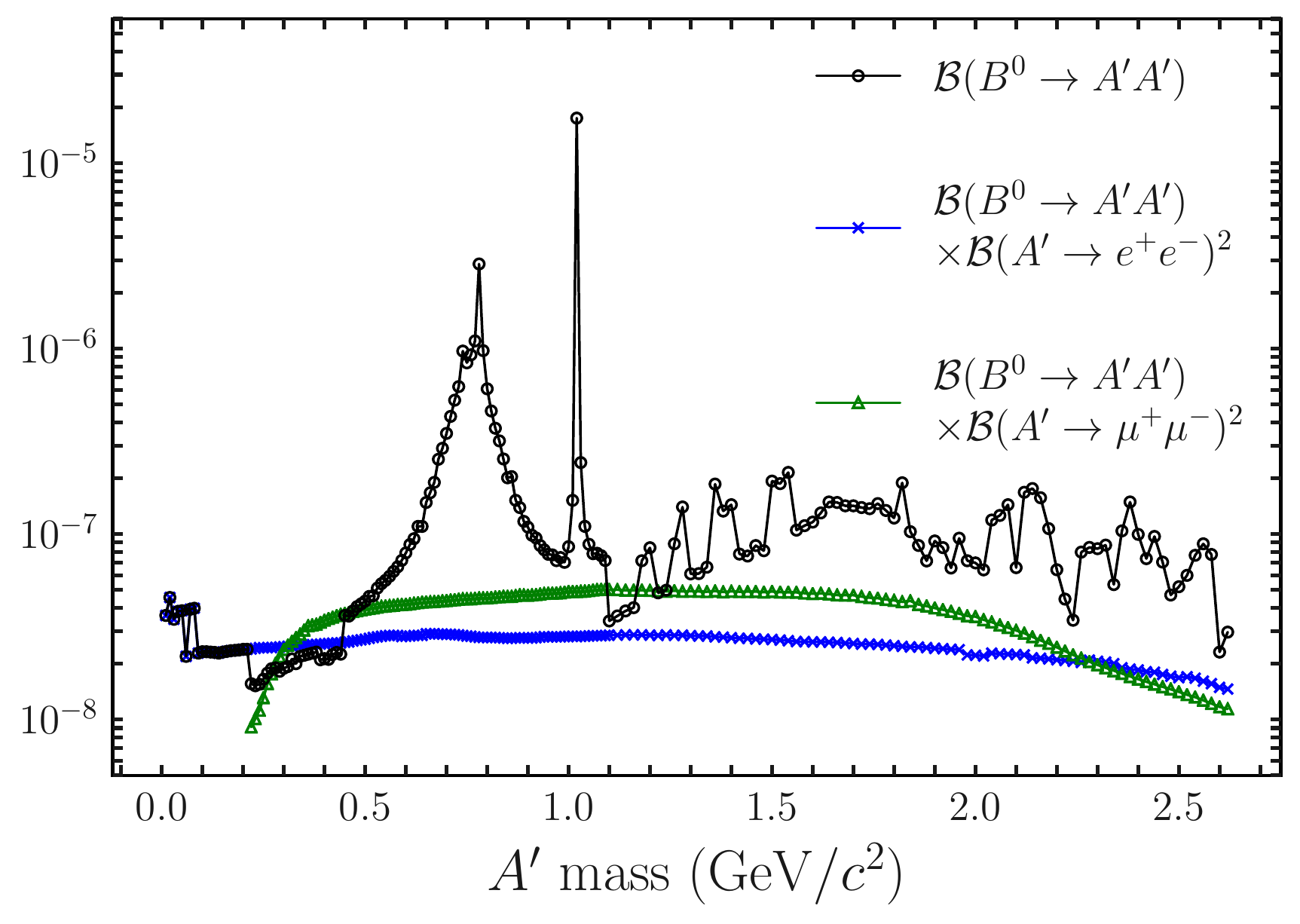}
  \caption{Upper limits of $B^0 \to \Aprime \Aprime$ branching fraction at 90\% C.L.}
\label{Fig:upperlimit}
\end{figure}

The obtained limits are shown in Fig.~\ref{Fig:upperlimit} as functions of $m_{\Aprime}$. 
The limits on the products of branching fractions are $\mathcal{O}(10^{-8})$ for both modes and in all $m_{\Aprime}$ bins.
For $\BFBAA$, the upper limits are $\mathcal{O}(10^{-8})$--$\mathcal{O}(10^{-5})$.
Due to the light meson veto in the $\ell^+ \ell^- \pi^+ \pi^-$ final states and the large fraction of $\Aprime \to \textrm{hadrons}$ in the veto region from Eq.~(\ref{eq:darkbr}), 
the upper limits near the masses of $\rho^0$ and $\phi$ mesons are less restrictive than others. 
Table~\ref{Tab:ul} lists the signal efficiency, the expected number of backgrounds and number of observed events ($N_\mathrm{obs}$) for some of $m_{\Aprime}$. 

\begin{landscape}
\begin{table}[H]
\centering
\scriptsize
\caption{\label{Tab:ul}
\scriptsize
Signal efficnency, expected the number of backgrounds, yields for each $B^0$ final state and upper limits of $B^0 \to A' A'$ branching fraction with 90\% confidence interval.
The table presents a part of the results for dark photons with (i) 20~$\MeV/c^2$ interval in $m_{A'} < 2m_{\mu}$ region, (ii) 100~$\MeV/c^2$ interval on the other region. }
\begin{tabular}{c @{\hskip 0.05in} ccc @{\hskip 0.05in} ccc @{\hskip 0.05in} ccc @{\hskip 0.05in} ccc @{\hskip 0.05in} ccc @{\hskip 0.05in} c}
\hline\hline
  \multicolumn{1}{c}{$m_{A'}$}                  & 
  \multicolumn{3}{c}{$e^+ e^- e^+ e^-$}         & 
  \multicolumn{3}{c}{$e^+ e^- \mu^+ \mu^-$}     & 
  \multicolumn{3}{c}{$\mu^+ \mu^- \mu^+ \mu^-$} & 
  \multicolumn{3}{c}{$e^+ e^- \pi^+ \pi^-$}     & 
  \multicolumn{3}{c}{$\mu^+ \mu^- \pi^+ \pi^-$} & 
  \multicolumn{1}{c}{90\% U.L.}                 \\
  $(\textrm{GeV}/c^2)$                                  & 
  Eff. (\%) & $N_{\textrm{bkg}}^{\textrm{exp}}$ & Yield & 
  Eff. (\%) & $N_{\textrm{bkg}}^{\textrm{exp}}$ & Yield & 
  Eff. (\%) & $N_{\textrm{bkg}}^{\textrm{exp}}$ & Yield & 
  Eff. (\%) & $N_{\textrm{bkg}}^{\textrm{exp}}$ & Yield & 
  Eff. (\%) & $N_{\textrm{bkg}}^{\textrm{exp}}$ & Yield & 
  ($10^{-8}$)                                           \\ 
\hline
  0.02  & 14.82 & 0.83$\pm$0.37 & 2  &     - &    -          & -  &     - &    -          & -  &     - &    -          & -  &     - &    -          & -  & 4.55 \\ 
  0.04  & 14.62 & 0.00$\pm$0.17 & 1  &     - &    -          & -  &     - &    -          & -  &     - &    -          & -  &     - &    -          & -  & 3.85 \\ 
  0.06  & 14.40 & 0.00$\pm$0.17 & 0  &     - &    -          & -  &     - &    -          & -  &     - &    -          & -  &     - &    -          & -  & 2.21 \\ 
  0.08  & 14.07 & 0.00$\pm$0.17 & 1  &     - &    -          & -  &     - &    -          & -  &     - &    -          & -  &     - &    -          & -  & 4.00 \\ 
  0.10  & 13.63 & 0.00$\pm$0.17 & 0  &     - &    -          & -  &     - &    -          & -  &     - &    -          & -  &     - &    -          & -  & 2.34 \\ 
  0.12  & 13.66 & 0.00$\pm$0.17 & 0  &     - &    -          & -  &     - &    -          & -  &     - &    -          & -  &     - &    -          & -  & 2.33 \\ 
  0.14  & 13.85 & 0.00$\pm$0.17 & 0  &     - &    -          & -  &     - &    -          & -  &     - &    -          & -  &     - &    -          & -  & 2.30 \\ 
  0.16  & 13.57 & 0.00$\pm$0.17 & 0  &     - &    -          & -  &     - &    -          & -  &     - &    -          & -  &     - &    -          & -  & 2.35 \\ 
  0.18  & 13.37 & 0.00$\pm$0.17 & 0  &     - &    -          & -  &     - &    -          & -  &     - &    -          & -  &     - &    -          & -  & 2.38 \\ 
  0.20  & 13.25 & 0.00$\pm$0.17 & 0  &     - &    -          & -  &     - &    -          & -  &     - &    -          & -  &     - &    -          & -  & 2.41 \\ 
  0.30  & 12.78 & 0.00$\pm$0.17 & 0  & 15.01 & 0.00$\pm$0.17 & 0  & 13.22 & 0.00$\pm$0.17 & 0  & 21.16 & 0.00$\pm$0.49 & 0  & 19.85 & 0.50$\pm$0.50 & 0  & 1.91 \\ 
  0.40  & 12.35 & 0.00$\pm$0.17 & 0  & 12.44 & 0.00$\pm$0.17 & 0  &  9.18 & 0.00$\pm$0.17 & 0  & 19.25 & 0.50$\pm$0.50 & 0  & 15.30 & 0.00$\pm$0.49 & 0  & 2.15 \\ 
  0.50  & 11.67 & 0.00$\pm$0.17 & 0  & 11.39 & 0.00$\pm$0.17 & 0  &  7.98 & 0.00$\pm$0.17 & 0  &     - &    -          & -  &     - &    -          & -  & 4.39 \\ 
  0.60  & 11.07 & 0.10$\pm$0.19 & 0  & 10.71 & 0.00$\pm$0.17 & 0  &  7.53 & 0.00$\pm$0.17 & 0  &     - &    -          & -  &     - &    -          & -  & 7.99 \\ 
  0.70  & 10.96 & 0.00$\pm$0.17 & 0  & 10.46 & 0.00$\pm$0.17 & 0  &  7.18 & 0.00$\pm$0.17 & 0  &     - &    -          & -  &     - &    -          & -  & 35.2 \\ 
  0.80  & 11.39 & 0.00$\pm$0.17 & 0  & 10.54 & 0.00$\pm$0.17 & 0  &  6.97 & 0.00$\pm$0.17 & 0  &     - &    -          & -  &     - &    -          & -  & 61.3 \\ 
  0.90  & 11.47 & 0.00$\pm$0.17 & 0  & 10.45 & 0.00$\pm$0.17 & 0  &  6.73 & 0.00$\pm$0.17 & 0  &     - &    -          & -  &     - &    -          & -  & 11.0 \\ 
  1.00  & 11.26 & 0.00$\pm$0.17 & 0  & 10.20 & 0.00$\pm$0.17 & 0  &  6.42 & 0.00$\pm$0.17 & 0  &     - &    -          & -  &     - &    -          & -  & 8.63 \\ 
  1.10  & 11.10 & 0.00$\pm$0.17 & 0  &  9.91 & 0.00$\pm$0.17 & 0  &  6.27 & 0.00$\pm$0.17 & 0  & 14.73 & 0.30$\pm$0.52 & 0  &  9.87 & 0.50$\pm$0.50 & 0  & 3.30 \\ 
  1.20  & 11.07 & 0.00$\pm$0.17 & 0  &  9.88 & 0.00$\pm$0.17 & 0  &  6.34 & 0.00$\pm$0.17 & 0  & 14.72 & 0.25$\pm$0.51 & 1  &  9.87 & 0.00$\pm$0.49 & 0  & 8.29 \\ 
  1.30  & 11.22 & 0.00$\pm$0.17 & 0  & 10.10 & 0.00$\pm$0.17 & 0  &  6.40 & 0.00$\pm$0.17 & 0  & 15.08 & 0.30$\pm$0.52 & 0  &  9.81 & 0.00$\pm$0.49 & 0  & 5.95 \\ 
  1.40  & 11.48 & 0.00$\pm$0.17 & 0  & 10.18 & 0.10$\pm$0.19 & 0  &  6.44 & 0.00$\pm$0.17 & 0  & 15.57 & 1.00$\pm$0.55 & 1  &  9.76 & 0.00$\pm$0.49 & 0  & 14.5 \\ 
  1.50  & 11.75 & 0.00$\pm$0.17 & 0  & 10.37 & 0.00$\pm$0.17 & 0  &  6.45 & 0.00$\pm$0.17 & 0  & 15.63 & 1.10$\pm$0.56 & 1  &  9.94 & 0.00$\pm$0.49 & 0  & 19.3 \\ 
  1.60  & 12.13 & 0.00$\pm$0.17 & 0  & 10.57 & 0.00$\pm$0.17 & 0  &  6.62 & 0.00$\pm$0.17 & 0  & 15.84 & 0.85$\pm$0.53 & 0  &  9.44 & 0.00$\pm$0.49 & 0  & 11.8 \\ 
  1.70  & 12.34 & 0.00$\pm$0.17 & 0  & 10.86 & 0.00$\pm$0.17 & 0  &  6.78 & 0.00$\pm$0.17 & 0  & 13.74 & 0.40$\pm$0.53 & 0  &  8.68 & 0.00$\pm$0.49 & 0  & 13.8 \\ 
  1.80  & 12.69 & 0.00$\pm$0.17 & 0  & 11.42 & 0.00$\pm$0.17 & 0  &  7.24 & 0.00$\pm$0.17 & 0  & 13.49 & 0.40$\pm$0.53 & 0  &  8.58 & 0.00$\pm$0.49 & 0  & 11.9 \\
  1.90  & 13.06 & 0.10$\pm$0.19 & 0  & 12.02 & 0.00$\pm$0.17 & 0  &  7.94 & 0.00$\pm$0.17 & 0  & 17.73 & 0.50$\pm$0.54 & 0  & 12.91 & 0.00$\pm$0.49 & 0  & 8.86 \\
  2.00  & 13.43 & 0.25$\pm$0.22 & 0  & 13.08 & 0.00$\pm$0.17 & 0  &  8.84 & 0.00$\pm$0.17 & 0  & 19.94 & 1.32$\pm$0.57 & 0  & 14.62 & 0.10$\pm$0.50 & 0  & 6.80 \\
  2.10  & 13.90 & 0.15$\pm$0.20 & 0  & 14.39 & 0.00$\pm$0.17 & 0  & 10.52 & 0.00$\pm$0.17 & 0  & 20.10 & 0.75$\pm$0.56 & 0  & 16.28 & 0.27$\pm$0.51 & 0  & 7.61 \\
  2.20  & 14.50 & 0.20$\pm$0.22 & 0  & 16.20 & 0.10$\pm$0.19 & 0  & 12.87 & 0.00$\pm$0.17 & 0  & 17.77 & 1.30$\pm$0.61 & 0  & 17.76 & 0.00$\pm$0.49 & 0  & 6.25 \\
  2.30  & 15.32 & 0.00$\pm$0.17 & 0  & 18.47 & 0.10$\pm$0.19 & 0  & 16.01 & 0.00$\pm$0.17 & 0  & 18.05 & 2.04$\pm$0.63 & 1  & 19.74 & 0.20$\pm$0.51 & 0  & 8.38 \\
  2.40  & 16.47 & 0.20$\pm$0.22 & 0  & 20.79 & 0.00$\pm$0.17 & 0  & 19.21 & 0.00$\pm$0.17 & 0  & 19.01 & 2.05$\pm$0.66 & 2  & 20.87 & 0.72$\pm$0.52 & 0  & 10.2 \\
  2.50  & 18.15 & 0.20$\pm$0.22 & 0  & 23.24 & 0.00$\pm$0.17 & 0  & 22.40 & 0.00$\pm$0.17 & 0  & 18.73 & 2.40$\pm$0.66 & 1  & 23.08 & 0.50$\pm$0.50 & 0  & 5.20 \\
  2.60  & 21.05 & 0.00$\pm$0.17 & 0  & 26.34 & 0.10$\pm$0.19 & 0  & 26.85 & 0.00$\pm$0.17 & 0  & 22.52 & 2.25$\pm$0.79 & 0  & 25.34 & 1.52$\pm$0.74 & 0  & 2.31 \\
\hline\hline
\end{tabular}
\end{table}

\end{landscape}

\begin{figure}[]
\centering
\includegraphics[width=0.80\textwidth]{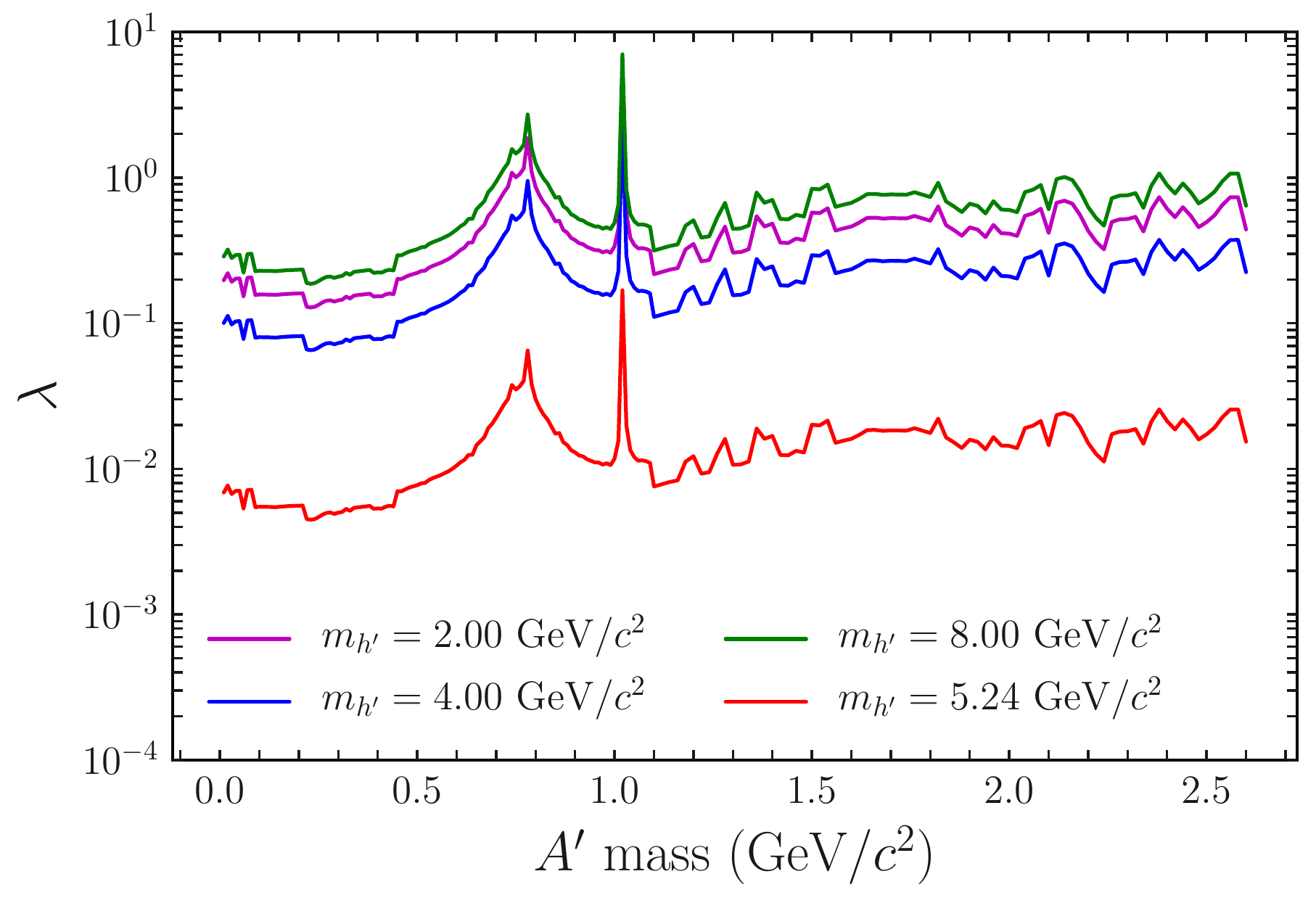}
  \caption{90\% upper limits of the Higgs portal coupling ($\lambda$) versus the dark photon mass for a $2.00$, $4.00$, $5.24$, $8.00~\GeV/c^2$ dark Higgs.}
\label{Fig:darklimit:combined_A}
\end{figure}

\begin{figure}[]
\centering
\includegraphics[width=0.80\textwidth]{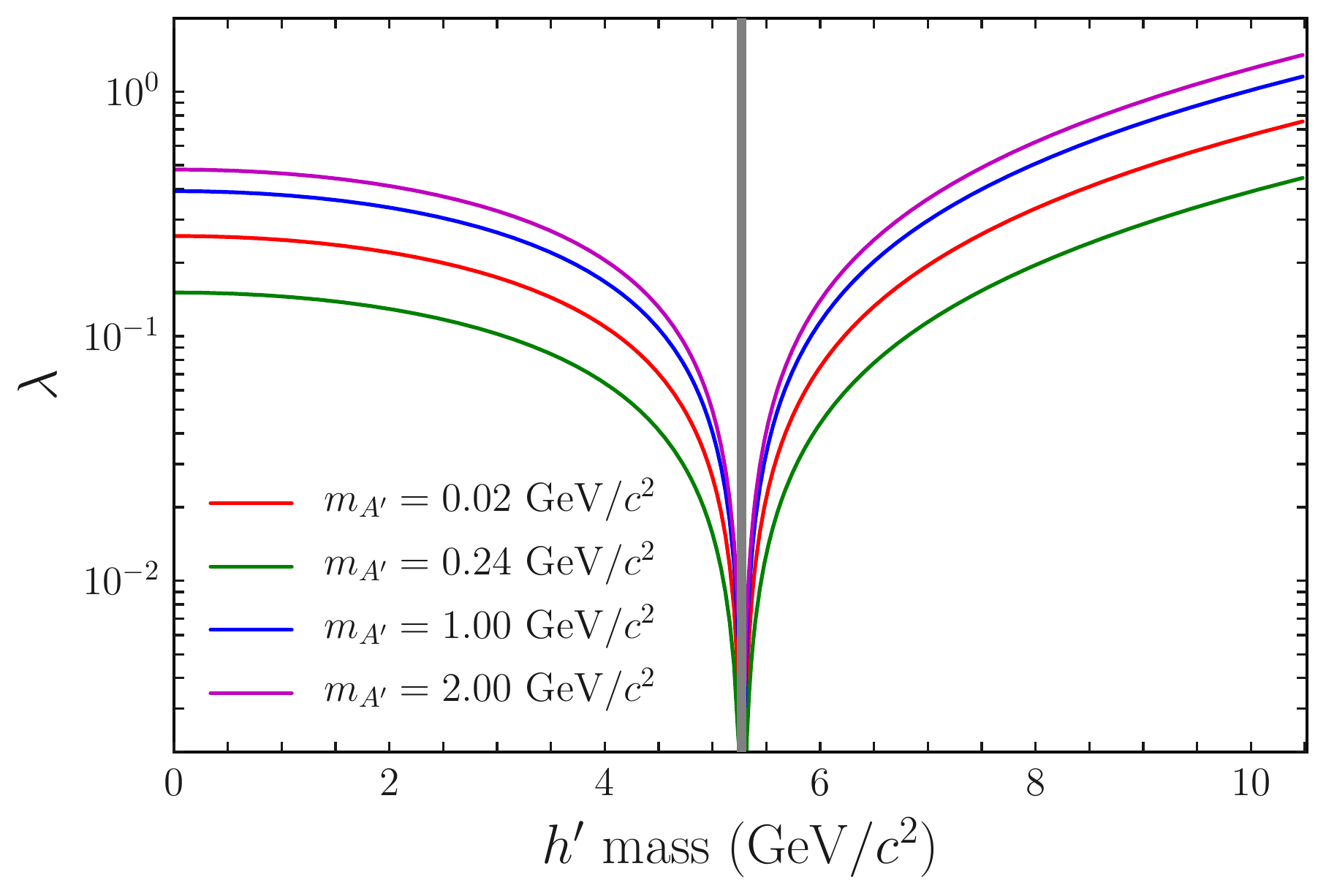}
  \caption{90\% upper limits of the Higgs portal coupling ($\lambda$) versus the dark Higgs mass for the $0.02$, $0.24$, $1.00$, $2.00~\GeV/c^2$ dark photon.}
\label{Fig:darklimit:combined_h}
\end{figure}

The $B^0 \to \Aprime \Aprime$ branching fraction with off-shell $H$--$h'$ mixing, for all but the $m_{h'} \sim m_{B^0}$ region, is calculated as~\cite{PhysRevD.83.054005}
\footnote{B.~Batell, private communication on the numerical factor of Eq.~(\ref{eq:thbr}) of Ref.~\cite{PhysRevD.83.054005}, when we apply $B^0$-meson-related variables instead of $B_s$-meson and the measured Higgs mass.},
\begin{equation} \label{eq:thbr}
\begin{aligned}
  \mathcal{B}(B^0 \to \Aprime \Aprime) \simeq 7 \times 10^{-7} \times \lambda^2 \times 
  V_{\Aprime \Aprime}^{1/2} \times \frac{V_{\Aprime \Aprime} + 12 m_{\Aprime}^4 / m_{B^0}^4}{(1 - m_{h'}^2 / m_{B^0}^2)^2}
\end{aligned}
\end{equation}
where $\lambda$ is the Higgs portal coupling 
with a new scalar field $H'$ from $\mathcal{L}_{\mathrm{Higgs}} = -\lambda(H^{\dagger} H)(H'^{\dagger} H')$ 
and $V_{\Aprime \Aprime} = 1 - {4 m_{\Aprime}^2}/{m_{B^0}^2}$.
From Eq.~(\ref{eq:thbr}) and the limits on $\mathcal{B}(B^0 \to \Aprime \Aprime)$, we determine the 90\% C.L. upper limits on $\lambda$ versus $m_{\Aprime}$ (Fig.~\ref{Fig:darklimit:combined_A}) and $m_{h'}$ (Fig.~\ref{Fig:darklimit:combined_h}).  
In the region where $m_{h'} \simeq m_{B^0}$, the upper limit on $\lambda$ gets as low as $\mathcal{O}(10^{-2})$.  Otherwise, the upper limits are $\mathcal{O}(10^{-1})$--$\mathcal{O}(1)$.

\section{Conclusions}

In summary, we have searched for $B^0 \to \Aprime \Aprime$ decays for the first time using the full data set of $772 \times 10^6$ $B\overline{B}$ events of Belle.  
We restrict our study to the case where $\Aprime$ decays promptly to $e^+ e^-$, $\mu^+ \mu^-$, or hadronic final states, and consider five final states of $B^0$ which are $e^+ e^- e^+ e^-$, 
$e^+ e^- \mu^+ \mu^-$, $\mu^+ \mu^- \mu^+ \mu^-$, $e^+ e^- \pi^+ \pi^-$, and 
$\mu^+ \mu^- \pi^+ \pi^-$.
From the branching fraction of $\Aprime$, the five $B^0$ 
final states are merged to determine the branching fraction of $B^0 \to \Aprime \Aprime$.
We find no significant signal in any assumed $\Aprime$ mass and decay mode, so we determine upper limits on $\prodee$, 
$\prodmm$ and $\BFBAA$, each at 90\% C.L.  The limits on the products of branching fractions are of the order of $\mathcal{O}(10^{-8})$, while the limits on $\BFBAA$ are $\mathcal{O}(10^{-8})$--$\mathcal{O}(10^{-5})$. 
We also set 90\% C.L. upper limits on the Higgs portal coupling $\lambda$ for each assumed value of $m_{\Aprime}$ and $m_{h'}$.
The upper limits on $\lambda$ are of the order of $10^{-2}$--$10^{-1}$ at $m_{h'} \simeq m_{B^0} \pm 40~\MeV/c^2$ and $10^{-1}$--$1$ at $m_{h'} \simeq m_{B^0} \pm 3~\GeV/c^2$.
With minor modifications our analysis can be used to set limits on the other new physics models which include prompt $B^0 \to X X$ and $X \to \ell^+ \ell^- / \pi^+ \pi^-$ decays. 
We expect to have much more stringent results from the Belle II experiment~\cite{Abe:2010gxa, Kou:2018nap}, with nearly two orders of magnitude increase in statistics, in the future.

\acknowledgments

We thank B. Batell and M. Pospelov for the discussion of the $B^0 \to \Aprime \Aprime$ branching fraction which helped to finalize this paper. 
We thank the KEKB group for the excellent operation of the
accelerator; the KEK cryogenics group for the efficient
operation of the solenoid; and the KEK computer group, and the Pacific Northwest National
Laboratory (PNNL) Environmental Molecular Sciences Laboratory (EMSL)
computing group for strong computing support; and the National
Institute of Informatics, and Science Information NETwork 5 (SINET5) for
valuable network support.  We acknowledge support from
the Ministry of Education, Culture, Sports, Science, and
Technology (MEXT) of Japan, the Japan Society for the
Promotion of Science (JSPS), and the Tau-Lepton Physics
Research Center of Nagoya University;
the Australian Research Council including grants
DP180102629, 
DP170102389, 
DP170102204, 
DP150103061, 
FT130100303; 
Austrian Federal Ministry of Education, Science and Research (FWF) and
FWF Austrian Science Fund No.~P~31361-N36;
the National Natural Science Foundation of China under Contracts
No.~11435013,  
No.~11475187,  
No.~11521505,  
No.~11575017,  
No.~11675166,  
No.~11705209;  
Key Research Program of Frontier Sciences, Chinese Academy of Sciences (CAS), Grant No.~QYZDJ-SSW-SLH011; 
the  CAS Center for Excellence in Particle Physics (CCEPP); 
the Shanghai Pujiang Program under Grant No.~18PJ1401000;  
the Ministry of Education, Youth and Sports of the Czech
Republic under Contract No.~LTT17020;
Horizon 2020 ERC Advanced Grant No.~884719 and ERC Starting Grant No.~947006 ``InterLeptons'' (European Union);
the Carl Zeiss Foundation, the Deutsche Forschungsgemeinschaft, the
Excellence Cluster Universe, and the VolkswagenStiftung;
the Department of Atomic Energy (Project Identification No. RTI 4002) and the Department of Science and Technology of India;
the Istituto Nazionale di Fisica Nucleare of Italy;
National Research Foundation (NRF) of Korea Grant
Nos.~2016R1\-D1A1B\-01010135, 2016R1\-D1A1B\-02012900, 2018R1\-A2B\-3003643,
2018R1\-A6A1A\-06024970, 2018R1\-D1A1B\-07047294, 2018R1\-A4A\-1025334,
2019K1\-A3A7A\-09033840, 2019K1\-A3A7A\-09034999, 2019R1\-I1A3A\-01058933;
Radiation Science Research Institute, Foreign Large-size Research Facility Application Supporting project, the Global Science Experimental Data Hub Center of the Korea Institute of Science and Technology Information and KREONET/GLORIAD;
the Polish Ministry of Science and Higher Education and
the National Science Center;
the Ministry of Science and Higher Education of the Russian Federation, Agreement 14.W03.31.0026, 
and the HSE University Basic Research Program, Moscow; 
University of Tabuk research grants
S-1440-0321, S-0256-1438, and S-0280-1439 (Saudi Arabia);
the Slovenian Research Agency Grant Nos. J1-9124 and P1-0135;
Ikerbasque, Basque Foundation for Science, Spain;
the Swiss National Science Foundation;
the Ministry of Education and the Ministry of Science and Technology of Taiwan;
and the United States Department of Energy and the National Science Foundation.

\end{document}



\section{Tables of signal efficiency, the number of expected background events, and the number of observed events at Belle}

In this section, tables of signal efficnency ($\epsilon$) in signal MC, 
the number of expected background events ($N_{\rm bkg}$) in Belle background MC, 
and the number of observed events ($N_{\rm obs}$) in the Belle data are introduced.
The dark photon masses are in the range of ($m_{A^{\prime}}$) in $0.01 - 2.62~\mathrm{GeV}/c^2$ with 
$10~\mathrm{MeV}/c^2$ ($0.01 - 1.10~\mathrm{GeV}/c^2$) and 
$20~\mathrm{MeV}/c^2$ ($1.10 - 2.62~\mathrm{GeV}/c^2$) intervals
for $e^+ e^- e^+ e^-$, $e^+ e^- \mu^+ \mu^-$, $\mu^+ \mu^- \mu^+ \mu^-$, $e^+ e^- \pi^+ \pi^-$,
$\mu^+ \mu^- \pi^+ \pi^-$ final states and $B^0 \to A^{\prime} A^{\prime}$ combined results.

For the combined signal efficiencies,
we first obtain the ratio  $F_{f} \equiv \mathcal{B}(B^0 \to A^{\prime} A^{\prime} \to f)/\mathcal{B}(B^0 \to A^{\prime} A^{\prime})$, where $f$ is each final state, using the dark photon branching fraction.
The dark photon branching fraction below the $\tau^+ \tau^-$ threshold is obtained as
\begin{equation} \label{eq:darkbr}
\begin{aligned}
  \mathcal{B}(A^{\prime} \to \ell^+ \ell^- / \pi^+ \pi^-) = \frac{\Gamma_{A^{\prime} \to \ell^+ \ell^- / \pi^+ \pi^-      }}{\Gamma_{A^{\prime} \to e^+ e^-} + \Gamma_{A^{\prime} \to \mu^+ \mu^-} + \Gamma_{A^{\prime} \to \textrm{hadrons}}},
\end{aligned}
\end{equation}
where $\ell = e~\textrm{or}~\mu$.
Following Ref.~\cite{PhysRevD.79.115008}, we write down the partial widths to $\ell^+ \ell^-$ and hadrons as
\begin{equation} \label{eq:llwidth}
\begin{aligned}
  \Gamma_{A^{\prime} \to \ell^+ \ell^-} =& \frac{1}{3} \alpha \varepsilon_{\textrm{mixing}}^2 m_{A^{\prime}} \sqrt{1 - {4 m_\ell^2}/{m_{A^{\prime}}^2}}(1 + 2 m_\ell^2 / m_{A^{\prime}}^2),\\
  \Gamma_{A^{\prime} \to \textrm{hadrons}} =& \Gamma_{A^{\prime} \to \mu^+ \mu^-} \times R(s = m_{A^{\prime}}^2),
\end{aligned}
\end{equation}
with the square of the total center-of-mass (CM) frame energy $s$, 
the kinetic mixing parameter $\varepsilon_{\textrm{mix}}$, and
$R(s) = \sigma_{e^+ e^- \to \textrm{hadrons}} / \sigma_{e^+ e^- \to \mu^+ \mu^-}$ which is determined by various experiments \cite{PDG}.  
The branching fraction of $A^{\prime} \to \pi^+ \pi^-$ is then obtained as ~\cite{PhysRevLett.108.211801}:
\begin{equation}
\begin{aligned}
\mathcal{B}(A^{\prime} \to \pi^+ \pi^-) = \mathcal{B}(A^{\prime} \to \textrm{hadrons}) \times \sigma(e^+ e^- \to \pi^+ \pi^-) / \sigma(e^+ e^- \to \textrm{hadrons}).
\end{aligned}
\end{equation}

From the $A^{\prime}$ branching fraction, we calculate $F_{f}$ for each final state.  In case of $e^+ e^- \mu^+ \mu^-$, for example, $F_{e^+ e^- \mu^+ \mu^-}$ is $2 \times \mathcal{B}(A^{\prime} \to  e^+ e^-) \times \mathcal{B}(A^{\prime} \to \mu^+ \mu^-)$.
With this ratio $F_f$, the combined efficiency is determined as $\sum_f \epsilon_f F_f$ where $\epsilon_f$ is the signal efficiency of the final state $f$.

\begin{longtable}{cccc | cccc}
\caption{\label{sub:eeee}
  Signal efficnency ($\epsilon$), the number of expected background events ($N_{\rm bkg}$), 
  and the number of observed events ($N_{\rm obs}$)
  for $B^0 \to A^{\prime} A^{\prime} \to e^+ e^- e^+ e^-$ mode.
  }\\
\hline
\hline
              $m_{A^{\prime}}$ & $\epsilon$ & $N_{\rm bkg}$ & $N_{\rm obs}$ &             $m_{A^{\prime}}$ & $\epsilon$ & $N_{\rm bkg}$ & $N_{\rm obs}$ \\
  ($\mathrm{GeV}/c^2$) &       (\%) &               &               & ($\mathrm{GeV}/c^2$) &       (\%) &               &               \\
\hline
  0.01 & $14.56\pm1.34$ & $0.33\pm0.24$ & 1 & 0.94 & $11.27\pm1.04$ & $0.00\pm0.17$ & 0 \\ 
  0.02 & $14.82\pm1.36$ & $0.83\pm0.37$ & 2 & 0.95 & $11.31\pm1.05$ & $0.00\pm0.17$ & 0 \\ 
  0.03 & $14.72\pm1.35$ & $0.50\pm0.29$ & 1 & 0.96 & $11.34\pm1.05$ & $0.00\pm0.17$ & 0 \\ 
  0.04 & $14.62\pm1.33$ & $0.00\pm0.17$ & 1 & 0.97 & $11.33\pm1.05$ & $0.00\pm0.17$ & 0 \\ 
  0.05 & $14.51\pm1.32$ & $0.00\pm0.17$ & 1 & 0.98 & $11.32\pm1.05$ & $0.00\pm0.17$ & 0 \\ 
  0.06 & $14.40\pm1.31$ & $0.00\pm0.17$ & 0 & 0.99 & $11.29\pm1.05$ & $0.00\pm0.17$ & 0 \\ 
  0.07 & $14.24\pm1.30$ & $0.00\pm0.17$ & 1 & 1.00 & $11.26\pm1.05$ & $0.00\pm0.17$ & 0 \\ 
  0.08 & $14.08\pm1.28$ & $0.00\pm0.17$ & 1 & 1.01 & $11.26\pm1.05$ & $0.00\pm0.17$ & 0 \\ 
  0.09 & $13.86\pm1.26$ & $0.00\pm0.17$ & 0 & 1.02 & $11.27\pm1.05$ & $0.00\pm0.17$ & 0 \\ 
  0.10 & $13.64\pm1.24$ & $0.00\pm0.17$ & 0 & 1.03 & $11.25\pm1.05$ & $0.00\pm0.17$ & 0 \\ 
  0.11 & $13.65\pm1.24$ & $0.00\pm0.17$ & 0 & 1.04 & $11.24\pm1.05$ & $0.00\pm0.17$ & 0 \\ 
  0.12 & $13.66\pm1.24$ & $0.00\pm0.17$ & 0 & 1.05 & $11.20\pm1.05$ & $0.00\pm0.17$ & 0 \\ 
  0.13 & $13.76\pm1.25$ & $0.00\pm0.17$ & 0 & 1.06 & $11.17\pm1.05$ & $0.00\pm0.17$ & 0 \\ 
  0.14 & $13.85\pm1.26$ & $0.00\pm0.17$ & 0 & 1.07 & $11.18\pm1.05$ & $0.00\pm0.17$ & 0 \\ 
  0.15 & $13.71\pm1.25$ & $0.00\pm0.17$ & 0 & 1.08 & $11.18\pm1.05$ & $0.00\pm0.17$ & 0 \\ 
  0.16 & $13.57\pm1.24$ & $0.00\pm0.17$ & 0 & 1.09 & $11.14\pm1.04$ & $0.00\pm0.17$ & 0 \\ 
  0.17 & $13.47\pm1.23$ & $0.00\pm0.17$ & 0 & 1.10 & $11.10\pm1.04$ & $0.00\pm0.17$ & 0 \\ 
  0.18 & $13.37\pm1.22$ & $0.00\pm0.17$ & 0 & 1.12 & $11.11\pm1.04$ & $0.00\pm0.17$ & 0 \\ 
  0.19 & $13.31\pm1.21$ & $0.00\pm0.17$ & 0 & 1.14 & $11.06\pm1.04$ & $0.00\pm0.17$ & 0 \\ 
  0.20 & $13.25\pm1.21$ & $0.00\pm0.17$ & 0 & 1.16 & $11.07\pm1.04$ & $0.00\pm0.17$ & 0 \\ 
  0.21 & $13.22\pm1.20$ & $0.00\pm0.17$ & 0 & 1.18 & $11.06\pm1.05$ & $0.00\pm0.17$ & 0 \\ 
  0.22 & $13.19\pm1.20$ & $0.00\pm0.17$ & 0 & 1.20 & $11.07\pm1.05$ & $0.00\pm0.17$ & 0 \\ 
  0.23 & $13.08\pm1.19$ & $0.00\pm0.17$ & 0 & 1.22 & $11.11\pm1.06$ & $0.00\pm0.17$ & 0 \\ 
  0.24 & $12.97\pm1.18$ & $0.00\pm0.17$ & 0 & 1.24 & $11.05\pm1.05$ & $0.00\pm0.17$ & 0 \\ 
  0.25 & $12.97\pm1.18$ & $0.00\pm0.17$ & 0 & 1.26 & $11.13\pm1.06$ & $0.00\pm0.17$ & 0 \\ 
  0.26 & $12.97\pm1.18$ & $0.00\pm0.17$ & 0 & 1.28 & $11.14\pm1.07$ & $0.00\pm0.17$ & 0 \\ 
  0.27 & $12.94\pm1.17$ & $0.00\pm0.17$ & 0 & 1.30 & $11.22\pm1.08$ & $0.00\pm0.17$ & 0 \\ 
  0.28 & $12.90\pm1.17$ & $0.00\pm0.17$ & 0 & 1.32 & $11.26\pm1.08$ & $0.00\pm0.17$ & 0 \\ 
  0.29 & $12.84\pm1.16$ & $0.00\pm0.17$ & 0 & 1.34 & $11.27\pm1.09$ & $0.00\pm0.17$ & 0 \\ 
  0.30 & $12.78\pm1.16$ & $0.00\pm0.17$ & 0 & 1.36 & $11.36\pm1.09$ & $0.00\pm0.17$ & 0 \\ 
  0.31 & $12.73\pm1.15$ & $0.00\pm0.17$ & 0 & 1.38 & $11.42\pm1.10$ & $0.00\pm0.17$ & 0 \\ 
  0.32 & $12.69\pm1.15$ & $0.00\pm0.17$ & 0 & 1.40 & $11.48\pm1.11$ & $0.00\pm0.17$ & 0 \\ 
  0.33 & $12.64\pm1.15$ & $0.00\pm0.17$ & 0 & 1.42 & $11.58\pm1.12$ & $0.00\pm0.17$ & 0 \\ 
  0.34 & $12.58\pm1.14$ & $0.00\pm0.17$ & 0 & 1.44 & $11.57\pm1.12$ & $0.00\pm0.17$ & 0 \\ 
  0.35 & $12.54\pm1.14$ & $0.00\pm0.17$ & 0 & 1.46 & $11.66\pm1.13$ & $0.00\pm0.17$ & 0 \\ 
  0.36 & $12.50\pm1.13$ & $0.00\pm0.17$ & 0 & 1.48 & $11.73\pm1.13$ & $0.00\pm0.17$ & 0 \\ 
  0.37 & $12.47\pm1.13$ & $0.00\pm0.17$ & 0 & 1.50 & $11.75\pm1.14$ & $0.00\pm0.17$ & 0 \\ 
  0.38 & $12.45\pm1.13$ & $0.00\pm0.17$ & 0 & 1.52 & $11.86\pm1.15$ & $0.00\pm0.17$ & 0 \\ 
  0.39 & $12.40\pm1.12$ & $0.00\pm0.17$ & 0 & 1.54 & $11.93\pm1.15$ & $0.00\pm0.17$ & 0 \\ 
  0.40 & $12.35\pm1.12$ & $0.00\pm0.17$ & 0 & 1.56 & $12.07\pm1.17$ & $0.00\pm0.17$ & 0 \\ 
  0.41 & $12.29\pm1.11$ & $0.00\pm0.17$ & 0 & 1.58 & $12.05\pm1.16$ & $0.00\pm0.17$ & 0 \\ 
  0.42 & $12.24\pm1.11$ & $0.00\pm0.17$ & 0 & 1.60 & $12.13\pm1.17$ & $0.00\pm0.17$ & 0 \\ 
  0.43 & $12.22\pm1.11$ & $0.00\pm0.17$ & 0 & 1.62 & $12.11\pm1.17$ & $0.00\pm0.17$ & 0 \\ 
  0.44 & $12.20\pm1.11$ & $0.00\pm0.17$ & 0 & 1.64 & $12.18\pm1.18$ & $0.00\pm0.17$ & 0 \\ 
  0.45 & $12.12\pm1.10$ & $0.00\pm0.17$ & 0 & 1.66 & $12.23\pm1.18$ & $0.00\pm0.17$ & 0 \\ 
  0.46 & $12.05\pm1.09$ & $0.00\pm0.17$ & 0 & 1.68 & $12.31\pm1.19$ & $0.00\pm0.17$ & 0 \\ 
  0.47 & $11.94\pm1.08$ & $0.00\pm0.17$ & 0 & 1.70 & $12.34\pm1.19$ & $0.00\pm0.17$ & 0 \\ 
  0.48 & $11.83\pm1.08$ & $0.00\pm0.17$ & 0 & 1.72 & $12.46\pm1.21$ & $0.00\pm0.17$ & 0 \\ 
  0.49 & $11.75\pm1.07$ & $0.00\pm0.17$ & 0 & 1.74 & $12.47\pm1.21$ & $0.00\pm0.17$ & 0 \\ 
  0.50 & $11.67\pm1.06$ & $0.00\pm0.17$ & 0 & 1.76 & $12.51\pm1.21$ & $0.00\pm0.17$ & 0 \\ 
  0.51 & $11.55\pm1.05$ & $0.00\pm0.17$ & 0 & 1.78 & $12.59\pm1.21$ & $0.00\pm0.17$ & 0 \\ 
  0.52 & $11.42\pm1.04$ & $0.00\pm0.17$ & 0 & 1.80 & $12.69\pm1.22$ & $0.00\pm0.17$ & 0 \\ 
  0.53 & $11.34\pm1.03$ & $0.00\pm0.17$ & 0 & 1.82 & $12.79\pm1.23$ & $0.00\pm0.17$ & 0 \\ 
  0.54 & $11.26\pm1.03$ & $0.00\pm0.17$ & 0 & 1.84 & $12.90\pm1.24$ & $0.00\pm0.17$ & 0 \\ 
  0.55 & $11.19\pm1.02$ & $0.00\pm0.17$ & 0 & 1.86 & $12.93\pm1.24$ & $0.00\pm0.17$ & 0 \\ 
  0.56 & $11.11\pm1.01$ & $0.00\pm0.17$ & 0 & 1.88 & $13.02\pm1.25$ & $0.00\pm0.17$ & 0 \\ 
  0.57 & $11.12\pm1.02$ & $0.00\pm0.17$ & 0 & 1.90 & $13.06\pm1.25$ & $0.10\pm0.19$ & 0 \\ 
  0.58 & $11.12\pm1.02$ & $0.00\pm0.17$ & 0 & 1.92 & $13.12\pm1.26$ & $0.10\pm0.19$ & 0 \\ 
  0.59 & $11.10\pm1.02$ & $0.10\pm0.19$ & 0 & 1.94 & $13.25\pm1.27$ & $0.10\pm0.19$ & 0 \\ 
  0.60 & $11.07\pm1.01$ & $0.10\pm0.19$ & 0 & 1.96 & $13.27\pm1.27$ & $0.10\pm0.19$ & 0 \\ 
  0.61 & $11.00\pm1.01$ & $0.10\pm0.19$ & 0 & 1.98 & $13.40\pm1.28$ & $0.25\pm0.22$ & 0 \\ 
  0.62 & $10.94\pm1.00$ & $0.10\pm0.19$ & 0 & 2.00 & $13.43\pm1.28$ & $0.25\pm0.22$ & 0 \\ 
  0.63 & $10.93\pm1.00$ & $0.10\pm0.19$ & 0 & 2.02 & $13.55\pm1.29$ & $0.25\pm0.22$ & 0 \\ 
  0.64 & $10.92\pm1.00$ & $0.10\pm0.19$ & 0 & 2.04 & $13.63\pm1.30$ & $0.15\pm0.20$ & 0 \\ 
  0.65 & $10.91\pm1.00$ & $0.00\pm0.17$ & 0 & 2.06 & $13.74\pm1.30$ & $0.15\pm0.20$ & 0 \\ 
  0.66 & $10.90\pm1.00$ & $0.00\pm0.17$ & 0 & 2.08 & $13.83\pm1.31$ & $0.15\pm0.20$ & 0 \\ 
  0.67 & $10.91\pm1.00$ & $0.00\pm0.17$ & 0 & 2.10 & $13.90\pm1.31$ & $0.15\pm0.20$ & 0 \\ 
  0.68 & $10.92\pm1.00$ & $0.00\pm0.17$ & 0 & 2.12 & $13.95\pm1.31$ & $0.15\pm0.20$ & 0 \\ 
  0.69 & $10.94\pm1.00$ & $0.00\pm0.17$ & 0 & 2.14 & $14.14\pm1.33$ & $0.20\pm0.22$ & 0 \\ 
  0.70 & $10.96\pm1.01$ & $0.00\pm0.17$ & 0 & 2.16 & $14.21\pm1.33$ & $0.20\pm0.22$ & 0 \\ 
  0.71 & $10.99\pm1.01$ & $0.00\pm0.17$ & 0 & 2.18 & $14.31\pm1.34$ & $0.20\pm0.22$ & 0 \\ 
  0.72 & $11.02\pm1.02$ & $0.00\pm0.17$ & 0 & 2.20 & $14.51\pm1.36$ & $0.20\pm0.22$ & 0 \\ 
  0.73 & $11.07\pm1.02$ & $0.00\pm0.17$ & 0 & 2.22 & $14.63\pm1.36$ & $0.20\pm0.22$ & 0 \\ 
  0.74 & $11.12\pm1.02$ & $0.00\pm0.17$ & 0 & 2.24 & $14.79\pm1.38$ & $0.20\pm0.22$ & 0 \\ 
  0.75 & $11.19\pm1.03$ & $0.00\pm0.17$ & 0 & 2.26 & $14.98\pm1.38$ & $0.20\pm0.22$ & 0 \\ 
  0.76 & $11.26\pm1.04$ & $0.00\pm0.17$ & 0 & 2.28 & $15.09\pm1.39$ & $0.10\pm0.19$ & 0 \\ 
  0.77 & $11.33\pm1.04$ & $0.00\pm0.17$ & 0 & 2.30 & $15.32\pm1.41$ & $0.00\pm0.17$ & 0 \\ 
  0.78 & $11.39\pm1.05$ & $0.00\pm0.17$ & 0 & 2.32 & $15.52\pm1.42$ & $0.00\pm0.17$ & 0 \\ 
  0.79 & $11.39\pm1.05$ & $0.00\pm0.17$ & 0 & 2.34 & $15.77\pm1.44$ & $0.10\pm0.19$ & 0 \\ 
  0.80 & $11.40\pm1.05$ & $0.00\pm0.17$ & 0 & 2.36 & $15.99\pm1.45$ & $0.20\pm0.22$ & 0 \\ 
  0.81 & $11.41\pm1.05$ & $0.00\pm0.17$ & 0 & 2.38 & $16.25\pm1.47$ & $0.20\pm0.22$ & 0 \\ 
  0.82 & $11.42\pm1.05$ & $0.00\pm0.17$ & 0 & 2.40 & $16.47\pm1.48$ & $0.20\pm0.22$ & 0 \\ 
  0.83 & $11.45\pm1.06$ & $0.00\pm0.17$ & 0 & 2.42 & $16.79\pm1.50$ & $0.20\pm0.22$ & 0 \\ 
  0.84 & $11.49\pm1.06$ & $0.00\pm0.17$ & 0 & 2.44 & $16.88\pm1.50$ & $0.20\pm0.22$ & 0 \\ 
  0.85 & $11.52\pm1.06$ & $0.00\pm0.17$ & 0 & 2.46 & $17.41\pm1.54$ & $0.20\pm0.22$ & 0 \\ 
  0.86 & $11.55\pm1.06$ & $0.00\pm0.17$ & 0 & 2.48 & $17.76\pm1.56$ & $0.20\pm0.22$ & 0 \\ 
  0.87 & $11.50\pm1.06$ & $0.00\pm0.17$ & 0 & 2.50 & $18.15\pm1.58$ & $0.20\pm0.22$ & 0 \\ 
  0.88 & $11.45\pm1.06$ & $0.00\pm0.17$ & 0 & 2.52 & $18.53\pm1.60$ & $0.00\pm0.17$ & 0 \\ 
  0.89 & $11.46\pm1.06$ & $0.00\pm0.17$ & 0 & 2.54 & $18.82\pm1.61$ & $0.00\pm0.17$ & 0 \\ 
  0.90 & $11.47\pm1.06$ & $0.00\pm0.17$ & 0 & 2.56 & $19.53\pm1.64$ & $0.00\pm0.17$ & 0 \\ 
  0.91 & $11.47\pm1.06$ & $0.00\pm0.17$ & 0 & 2.58 & $20.21\pm1.65$ & $0.00\pm0.17$ & 0 \\ 
  0.92 & $11.47\pm1.06$ & $0.00\pm0.17$ & 0 & 2.60 & $21.05\pm1.67$ & $0.00\pm0.17$ & 0 \\ 
  0.93 & $11.37\pm1.05$ & $0.00\pm0.17$ & 0 & 2.62 & $21.53\pm1.64$ & $0.00\pm0.17$ & 0 \\ 
\hline
\hline

\end{longtable}

\begin{longtable}{cccc | cccc}
\caption{\label{sub:eemm}
  Signal efficnency ($\epsilon$), the number of expected background events ($N_{\rm bkg}$), 
  and the number of observed events ($N_{\rm obs}$)
  for $B^0 \to A^{\prime} A^{\prime} \to e^+ e^- \mu^+ \mu^-$ mode.
  }\\
\hline
\hline
              $m_{A^{\prime}}$ & $\epsilon$ & $N_{\rm bkg}$ & $N_{\rm obs}$ &             $m_{A^{\prime}}$ & $\epsilon$ & $N_{\rm bkg}$ & $N_{\rm obs}$ \\
  ($\mathrm{GeV}/c^2$) &       (\%) &               &               & ($\mathrm{GeV}/c^2$) &       (\%) &               &               \\
\hline
  0.22 & $23.93\pm1.68$ & $0.00\pm0.17$ & 0 & 1.05 & $ 9.89\pm0.68$ & $0.00\pm0.17$ & 0 \\ 
  0.23 & $22.79\pm1.59$ & $0.00\pm0.17$ & 0 & 1.06 & $ 9.99\pm0.68$ & $0.00\pm0.17$ & 0 \\ 
  0.24 & $21.65\pm1.51$ & $0.00\pm0.17$ & 0 & 1.07 & $ 9.99\pm0.69$ & $0.00\pm0.17$ & 0 \\ 
  0.25 & $20.06\pm1.39$ & $0.00\pm0.17$ & 0 & 1.08 & $ 9.98\pm0.69$ & $0.00\pm0.17$ & 0 \\ 
  0.26 & $18.46\pm1.28$ & $0.00\pm0.17$ & 0 & 1.09 & $ 9.95\pm0.68$ & $0.00\pm0.17$ & 0 \\ 
  0.27 & $17.35\pm1.21$ & $0.00\pm0.17$ & 0 & 1.10 & $ 9.91\pm0.68$ & $0.00\pm0.17$ & 0 \\ 
  0.28 & $16.25\pm1.13$ & $0.00\pm0.17$ & 0 & 1.12 & $ 9.95\pm0.69$ & $0.00\pm0.17$ & 0 \\ 
  0.29 & $15.63\pm1.09$ & $0.00\pm0.17$ & 0 & 1.14 & $ 9.94\pm0.69$ & $0.00\pm0.17$ & 0 \\ 
  0.30 & $15.01\pm1.04$ & $0.00\pm0.17$ & 0 & 1.16 & $ 9.92\pm0.69$ & $0.00\pm0.17$ & 0 \\ 
  0.31 & $14.63\pm1.02$ & $0.00\pm0.17$ & 0 & 1.18 & $ 9.93\pm0.69$ & $0.00\pm0.17$ & 0 \\ 
  0.32 & $14.26\pm0.99$ & $0.00\pm0.17$ & 0 & 1.20 & $ 9.88\pm0.69$ & $0.00\pm0.17$ & 0 \\ 
  0.33 & $13.92\pm0.97$ & $0.00\pm0.17$ & 0 & 1.22 & $10.03\pm0.70$ & $0.00\pm0.17$ & 0 \\ 
  0.34 & $13.58\pm0.94$ & $0.00\pm0.17$ & 0 & 1.24 & $ 9.96\pm0.70$ & $0.00\pm0.17$ & 0 \\ 
  0.35 & $13.33\pm0.92$ & $0.00\pm0.17$ & 0 & 1.26 & $10.01\pm0.70$ & $0.00\pm0.17$ & 0 \\ 
  0.36 & $13.07\pm0.90$ & $0.00\pm0.17$ & 0 & 1.28 & $10.05\pm0.71$ & $0.00\pm0.17$ & 0 \\ 
  0.37 & $12.88\pm0.89$ & $0.00\pm0.17$ & 0 & 1.30 & $10.11\pm0.71$ & $0.00\pm0.17$ & 0 \\ 
  0.38 & $12.70\pm0.88$ & $0.00\pm0.17$ & 0 & 1.32 & $10.08\pm0.71$ & $0.00\pm0.17$ & 0 \\ 
  0.39 & $12.57\pm0.87$ & $0.00\pm0.17$ & 0 & 1.34 & $10.09\pm0.72$ & $0.10\pm0.19$ & 0 \\ 
  0.40 & $12.44\pm0.86$ & $0.00\pm0.17$ & 0 & 1.36 & $10.09\pm0.72$ & $0.10\pm0.19$ & 0 \\ 
  0.41 & $12.30\pm0.85$ & $0.00\pm0.17$ & 0 & 1.38 & $10.12\pm0.72$ & $0.20\pm0.22$ & 0 \\ 
  0.42 & $12.15\pm0.84$ & $0.00\pm0.17$ & 0 & 1.40 & $10.18\pm0.73$ & $0.10\pm0.19$ & 0 \\ 
  0.43 & $12.05\pm0.83$ & $0.00\pm0.17$ & 0 & 1.42 & $10.24\pm0.73$ & $0.10\pm0.19$ & 0 \\ 
  0.44 & $11.95\pm0.82$ & $0.00\pm0.17$ & 0 & 1.44 & $10.27\pm0.74$ & $0.00\pm0.17$ & 0 \\ 
  0.45 & $11.85\pm0.82$ & $0.00\pm0.17$ & 0 & 1.46 & $10.31\pm0.74$ & $0.00\pm0.17$ & 0 \\ 
  0.46 & $11.75\pm0.81$ & $0.00\pm0.17$ & 0 & 1.48 & $10.12\pm0.73$ & $0.00\pm0.17$ & 0 \\ 
  0.47 & $11.67\pm0.80$ & $0.00\pm0.17$ & 0 & 1.50 & $10.37\pm0.75$ & $0.00\pm0.17$ & 0 \\ 
  0.48 & $11.60\pm0.80$ & $0.00\pm0.17$ & 0 & 1.52 & $10.37\pm0.75$ & $0.00\pm0.17$ & 0 \\ 
  0.49 & $11.50\pm0.79$ & $0.00\pm0.17$ & 0 & 1.54 & $10.44\pm0.76$ & $0.00\pm0.17$ & 0 \\ 
  0.50 & $11.39\pm0.78$ & $0.00\pm0.17$ & 0 & 1.56 & $10.47\pm0.76$ & $0.00\pm0.17$ & 0 \\ 
  0.51 & $11.25\pm0.77$ & $0.00\pm0.17$ & 0 & 1.58 & $10.53\pm0.77$ & $0.00\pm0.17$ & 0 \\ 
  0.52 & $11.12\pm0.76$ & $0.00\pm0.17$ & 0 & 1.60 & $10.57\pm0.77$ & $0.00\pm0.17$ & 0 \\ 
  0.53 & $11.05\pm0.76$ & $0.00\pm0.17$ & 0 & 1.62 & $10.66\pm0.78$ & $0.00\pm0.17$ & 0 \\ 
  0.54 & $10.98\pm0.75$ & $0.00\pm0.17$ & 0 & 1.64 & $10.74\pm0.79$ & $0.00\pm0.17$ & 0 \\ 
  0.55 & $10.92\pm0.75$ & $0.00\pm0.17$ & 0 & 1.66 & $10.74\pm0.79$ & $0.00\pm0.17$ & 0 \\ 
  0.56 & $10.87\pm0.74$ & $0.00\pm0.17$ & 0 & 1.68 & $10.88\pm0.81$ & $0.00\pm0.17$ & 0 \\ 
  0.57 & $10.80\pm0.74$ & $0.00\pm0.17$ & 0 & 1.70 & $10.86\pm0.81$ & $0.00\pm0.17$ & 0 \\ 
  0.58 & $10.74\pm0.73$ & $0.00\pm0.17$ & 0 & 1.72 & $11.04\pm0.82$ & $0.00\pm0.17$ & 0 \\ 
  0.59 & $10.73\pm0.73$ & $0.00\pm0.17$ & 0 & 1.74 & $11.13\pm0.83$ & $0.00\pm0.17$ & 0 \\ 
  0.60 & $10.71\pm0.73$ & $0.00\pm0.17$ & 0 & 1.76 & $11.13\pm0.83$ & $0.00\pm0.17$ & 0 \\ 
  0.61 & $10.70\pm0.73$ & $0.00\pm0.17$ & 0 & 1.78 & $11.33\pm0.85$ & $0.00\pm0.17$ & 0 \\ 
  0.62 & $10.68\pm0.73$ & $0.00\pm0.17$ & 0 & 1.80 & $11.42\pm0.85$ & $0.00\pm0.17$ & 0 \\ 
  0.63 & $10.64\pm0.73$ & $0.00\pm0.17$ & 0 & 1.82 & $11.58\pm0.87$ & $0.00\pm0.17$ & 0 \\ 
  0.64 & $10.59\pm0.72$ & $0.00\pm0.17$ & 0 & 1.84 & $11.62\pm0.87$ & $0.00\pm0.17$ & 0 \\ 
  0.65 & $10.58\pm0.72$ & $0.00\pm0.17$ & 0 & 1.86 & $11.87\pm0.89$ & $0.00\pm0.17$ & 0 \\ 
  0.66 & $10.57\pm0.72$ & $0.00\pm0.17$ & 0 & 1.88 & $11.93\pm0.89$ & $0.00\pm0.17$ & 0 \\ 
  0.67 & $10.51\pm0.72$ & $0.00\pm0.17$ & 0 & 1.90 & $12.02\pm0.90$ & $0.00\pm0.17$ & 0 \\ 
  0.68 & $10.45\pm0.71$ & $0.00\pm0.17$ & 0 & 1.92 & $12.29\pm0.92$ & $0.00\pm0.17$ & 0 \\ 
  0.69 & $10.46\pm0.71$ & $0.00\pm0.17$ & 0 & 1.94 & $12.43\pm0.93$ & $0.00\pm0.17$ & 0 \\ 
  0.70 & $10.46\pm0.71$ & $0.00\pm0.17$ & 0 & 1.96 & $12.67\pm0.95$ & $0.00\pm0.17$ & 0 \\ 
  0.71 & $10.48\pm0.71$ & $0.00\pm0.17$ & 0 & 1.98 & $12.81\pm0.96$ & $0.00\pm0.17$ & 0 \\ 
  0.72 & $10.49\pm0.71$ & $0.00\pm0.17$ & 0 & 2.00 & $13.08\pm0.97$ & $0.00\pm0.17$ & 0 \\ 
  0.73 & $10.49\pm0.71$ & $0.00\pm0.17$ & 0 & 2.02 & $13.20\pm0.98$ & $0.10\pm0.19$ & 0 \\ 
  0.74 & $10.48\pm0.71$ & $0.00\pm0.17$ & 0 & 2.04 & $13.48\pm1.00$ & $0.10\pm0.19$ & 0 \\ 
  0.75 & $10.53\pm0.71$ & $0.00\pm0.17$ & 0 & 2.06 & $13.76\pm1.02$ & $0.10\pm0.19$ & 0 \\ 
  0.76 & $10.59\pm0.72$ & $0.00\pm0.17$ & 0 & 2.08 & $14.01\pm1.03$ & $0.00\pm0.17$ & 0 \\ 
  0.77 & $10.57\pm0.72$ & $0.00\pm0.17$ & 0 & 2.10 & $14.39\pm1.05$ & $0.00\pm0.17$ & 0 \\ 
  0.78 & $10.55\pm0.72$ & $0.00\pm0.17$ & 0 & 2.12 & $14.55\pm1.06$ & $0.00\pm0.17$ & 0 \\ 
  0.79 & $10.55\pm0.72$ & $0.00\pm0.17$ & 0 & 2.14 & $15.07\pm1.09$ & $0.00\pm0.17$ & 0 \\ 
  0.80 & $10.54\pm0.72$ & $0.00\pm0.17$ & 0 & 2.16 & $15.44\pm1.12$ & $0.00\pm0.17$ & 0 \\ 
  0.81 & $10.53\pm0.71$ & $0.00\pm0.17$ & 0 & 2.18 & $15.79\pm1.13$ & $0.10\pm0.19$ & 0 \\ 
  0.82 & $10.51\pm0.71$ & $0.00\pm0.17$ & 0 & 2.20 & $16.20\pm1.16$ & $0.10\pm0.19$ & 0 \\ 
  0.83 & $10.53\pm0.71$ & $0.00\pm0.17$ & 0 & 2.22 & $16.68\pm1.18$ & $0.10\pm0.19$ & 0 \\ 
  0.84 & $10.55\pm0.72$ & $0.00\pm0.17$ & 0 & 2.24 & $17.13\pm1.21$ & $0.10\pm0.19$ & 0 \\ 
  0.85 & $10.53\pm0.71$ & $0.00\pm0.17$ & 0 & 2.26 & $17.60\pm1.23$ & $0.00\pm0.17$ & 0 \\ 
  0.86 & $10.51\pm0.71$ & $0.00\pm0.17$ & 0 & 2.28 & $18.10\pm1.26$ & $0.10\pm0.19$ & 0 \\ 
  0.87 & $10.52\pm0.71$ & $0.00\pm0.17$ & 0 & 2.30 & $18.47\pm1.28$ & $0.10\pm0.19$ & 0 \\ 
  0.88 & $10.54\pm0.71$ & $0.00\pm0.17$ & 0 & 2.32 & $18.91\pm1.30$ & $0.10\pm0.19$ & 0 \\ 
  0.89 & $10.49\pm0.71$ & $0.00\pm0.17$ & 0 & 2.34 & $19.36\pm1.33$ & $0.10\pm0.19$ & 0 \\ 
  0.90 & $10.45\pm0.71$ & $0.00\pm0.17$ & 0 & 2.36 & $19.90\pm1.36$ & $0.00\pm0.17$ & 0 \\ 
  0.91 & $10.44\pm0.71$ & $0.00\pm0.17$ & 0 & 2.38 & $20.28\pm1.38$ & $0.00\pm0.17$ & 0 \\ 
  0.92 & $10.42\pm0.71$ & $0.10\pm0.19$ & 0 & 2.40 & $20.79\pm1.41$ & $0.00\pm0.17$ & 0 \\ 
  0.93 & $10.39\pm0.70$ & $0.10\pm0.19$ & 0 & 2.42 & $21.24\pm1.44$ & $0.00\pm0.17$ & 0 \\ 
  0.94 & $10.35\pm0.70$ & $0.10\pm0.19$ & 0 & 2.44 & $21.73\pm1.46$ & $0.00\pm0.17$ & 0 \\ 
  0.95 & $10.33\pm0.70$ & $0.10\pm0.19$ & 0 & 2.46 & $22.17\pm1.49$ & $0.00\pm0.17$ & 0 \\ 
  0.96 & $10.32\pm0.70$ & $0.00\pm0.17$ & 0 & 2.48 & $22.77\pm1.52$ & $0.00\pm0.17$ & 0 \\ 
  0.97 & $10.26\pm0.70$ & $0.00\pm0.17$ & 0 & 2.50 & $23.24\pm1.55$ & $0.00\pm0.17$ & 0 \\ 
  0.98 & $10.20\pm0.69$ & $0.00\pm0.17$ & 0 & 2.52 & $23.82\pm1.58$ & $0.00\pm0.17$ & 0 \\ 
  0.99 & $10.20\pm0.69$ & $0.00\pm0.17$ & 0 & 2.54 & $24.41\pm1.61$ & $0.00\pm0.17$ & 0 \\ 
  1.00 & $10.20\pm0.69$ & $0.00\pm0.17$ & 0 & 2.56 & $25.01\pm1.64$ & $0.00\pm0.17$ & 0 \\ 
  1.01 & $10.16\pm0.69$ & $0.00\pm0.17$ & 0 & 2.58 & $25.69\pm1.66$ & $0.10\pm0.19$ & 0 \\ 
  1.02 & $10.13\pm0.69$ & $0.00\pm0.17$ & 0 & 2.60 & $26.34\pm1.68$ & $0.10\pm0.19$ & 0 \\ 
  1.03 & $ 9.96\pm0.68$ & $0.00\pm0.17$ & 0 & 2.62 & $26.97\pm1.69$ & $0.10\pm0.19$ & 0 \\ 
  1.04 & $ 9.80\pm0.67$ & $0.00\pm0.17$ & 0 &      &                &               &   \\ 
\hline
\hline

\end{longtable}

\begin{longtable}{cccc | cccc}
\caption{\label{sub:mmmm}
  Signal efficnency ($\epsilon$), the number of expected background events ($N_{\rm bkg}$), 
  and the number of observed events ($N_{\rm obs}$)
  for $B^0 \to A^{\prime} A^{\prime} \to \mu^+ \mu^- \mu^+ \mu^-$ mode.
  }\\
\hline
\hline
              $m_{A^{\prime}}$ & $\epsilon$ & $N_{\rm bkg}$ & $N_{\rm obs}$ &             $m_{A^{\prime}}$ & $\epsilon$ & $N_{\rm bkg}$ & $N_{\rm obs}$ \\
  ($\mathrm{GeV}/c^2$) &       (\%) &               &               & ($\mathrm{GeV}/c^2$) &       (\%) &               &               \\
\hline
  0.22 & $34.57\pm2.86$ & $0.00\pm0.17$ & 0 & 1.05 & $ 6.35\pm0.52$ & $0.00\pm0.17$ & 0 \\ 
  0.23 & $31.31\pm2.54$ & $0.00\pm0.17$ & 0 & 1.06 & $ 6.31\pm0.52$ & $0.00\pm0.17$ & 0 \\ 
  0.24 & $28.05\pm2.23$ & $0.00\pm0.17$ & 0 & 1.07 & $ 6.28\pm0.51$ & $0.00\pm0.17$ & 0 \\ 
  0.25 & $24.10\pm1.91$ & $0.00\pm0.17$ & 0 & 1.08 & $ 6.26\pm0.51$ & $0.00\pm0.17$ & 0 \\ 
  0.26 & $20.16\pm1.60$ & $0.00\pm0.17$ & 0 & 1.09 & $ 6.26\pm0.52$ & $0.00\pm0.17$ & 0 \\ 
  0.27 & $17.91\pm1.42$ & $0.00\pm0.17$ & 0 & 1.10 & $ 6.27\pm0.52$ & $0.00\pm0.17$ & 0 \\ 
  0.28 & $15.67\pm1.25$ & $0.00\pm0.17$ & 0 & 1.12 & $ 6.28\pm0.52$ & $0.00\pm0.17$ & 0 \\ 
  0.29 & $14.45\pm1.15$ & $0.00\pm0.17$ & 0 & 1.14 & $ 6.32\pm0.53$ & $0.00\pm0.17$ & 0 \\ 
  0.30 & $13.22\pm1.05$ & $0.00\pm0.17$ & 0 & 1.16 & $ 6.32\pm0.53$ & $0.00\pm0.17$ & 0 \\ 
  0.31 & $12.56\pm1.00$ & $0.00\pm0.17$ & 0 & 1.18 & $ 6.32\pm0.53$ & $0.00\pm0.17$ & 0 \\ 
  0.32 & $11.90\pm0.95$ & $0.00\pm0.17$ & 0 & 1.20 & $ 6.34\pm0.54$ & $0.00\pm0.17$ & 0 \\ 
  0.33 & $11.41\pm0.91$ & $0.00\pm0.17$ & 0 & 1.22 & $ 6.39\pm0.54$ & $0.00\pm0.17$ & 0 \\ 
  0.34 & $10.93\pm0.87$ & $0.00\pm0.17$ & 0 & 1.24 & $ 6.41\pm0.55$ & $0.00\pm0.17$ & 0 \\ 
  0.35 & $10.38\pm0.82$ & $0.00\pm0.17$ & 0 & 1.26 & $ 6.36\pm0.54$ & $0.00\pm0.17$ & 0 \\ 
  0.36 & $ 9.83\pm0.78$ & $0.00\pm0.17$ & 0 & 1.28 & $ 6.41\pm0.55$ & $0.00\pm0.17$ & 0 \\ 
  0.37 & $ 9.74\pm0.77$ & $0.00\pm0.17$ & 0 & 1.30 & $ 6.40\pm0.55$ & $0.00\pm0.17$ & 0 \\ 
  0.38 & $ 9.65\pm0.77$ & $0.00\pm0.17$ & 0 & 1.32 & $ 6.41\pm0.56$ & $0.00\pm0.17$ & 0 \\ 
  0.39 & $ 9.41\pm0.75$ & $0.00\pm0.17$ & 0 & 1.34 & $ 6.43\pm0.56$ & $0.00\pm0.17$ & 0 \\ 
  0.40 & $ 9.18\pm0.73$ & $0.00\pm0.17$ & 0 & 1.36 & $ 6.41\pm0.56$ & $0.00\pm0.17$ & 0 \\ 
  0.41 & $ 9.00\pm0.71$ & $0.00\pm0.17$ & 0 & 1.38 & $ 6.48\pm0.57$ & $0.00\pm0.17$ & 0 \\ 
  0.42 & $ 8.83\pm0.70$ & $0.00\pm0.17$ & 0 & 1.40 & $ 6.44\pm0.57$ & $0.00\pm0.17$ & 0 \\ 
  0.43 & $ 8.67\pm0.69$ & $0.00\pm0.17$ & 0 & 1.42 & $ 6.46\pm0.57$ & $0.00\pm0.17$ & 0 \\ 
  0.44 & $ 8.52\pm0.68$ & $0.00\pm0.17$ & 0 & 1.44 & $ 6.46\pm0.58$ & $0.00\pm0.17$ & 0 \\ 
  0.45 & $ 8.43\pm0.67$ & $0.00\pm0.17$ & 0 & 1.46 & $ 6.45\pm0.58$ & $0.00\pm0.17$ & 0 \\ 
  0.46 & $ 8.34\pm0.66$ & $0.00\pm0.17$ & 0 & 1.48 & $ 6.50\pm0.59$ & $0.00\pm0.17$ & 0 \\ 
  0.47 & $ 8.25\pm0.65$ & $0.00\pm0.17$ & 0 & 1.50 & $ 6.45\pm0.59$ & $0.00\pm0.17$ & 0 \\ 
  0.48 & $ 8.15\pm0.65$ & $0.00\pm0.17$ & 0 & 1.52 & $ 6.50\pm0.59$ & $0.00\pm0.17$ & 0 \\ 
  0.49 & $ 8.06\pm0.64$ & $0.00\pm0.17$ & 0 & 1.54 & $ 6.50\pm0.60$ & $0.00\pm0.17$ & 0 \\ 
  0.50 & $ 7.98\pm0.63$ & $0.00\pm0.17$ & 0 & 1.56 & $ 6.52\pm0.60$ & $0.00\pm0.17$ & 0 \\ 
  0.51 & $ 7.93\pm0.63$ & $0.00\pm0.17$ & 0 & 1.58 & $ 6.58\pm0.61$ & $0.00\pm0.17$ & 0 \\ 
  0.52 & $ 7.88\pm0.62$ & $0.00\pm0.17$ & 0 & 1.60 & $ 6.62\pm0.62$ & $0.00\pm0.17$ & 0 \\ 
  0.53 & $ 7.81\pm0.62$ & $0.00\pm0.17$ & 0 & 1.62 & $ 6.59\pm0.62$ & $0.00\pm0.17$ & 0 \\ 
  0.54 & $ 7.74\pm0.61$ & $0.00\pm0.17$ & 0 & 1.64 & $ 6.66\pm0.63$ & $0.00\pm0.17$ & 0 \\ 
  0.55 & $ 7.70\pm0.61$ & $0.00\pm0.17$ & 0 & 1.66 & $ 6.74\pm0.64$ & $0.00\pm0.17$ & 0 \\ 
  0.56 & $ 7.66\pm0.60$ & $0.00\pm0.17$ & 0 & 1.68 & $ 6.79\pm0.65$ & $0.00\pm0.17$ & 0 \\ 
  0.57 & $ 7.61\pm0.60$ & $0.00\pm0.17$ & 0 & 1.70 & $ 6.78\pm0.66$ & $0.00\pm0.17$ & 0 \\ 
  0.58 & $ 7.57\pm0.60$ & $0.00\pm0.17$ & 0 & 1.72 & $ 6.89\pm0.67$ & $0.00\pm0.17$ & 0 \\ 
  0.59 & $ 7.55\pm0.60$ & $0.00\pm0.17$ & 0 & 1.74 & $ 6.99\pm0.68$ & $0.00\pm0.17$ & 0 \\ 
  0.60 & $ 7.53\pm0.59$ & $0.00\pm0.17$ & 0 & 1.76 & $ 7.03\pm0.69$ & $0.00\pm0.17$ & 0 \\ 
  0.61 & $ 7.49\pm0.59$ & $0.00\pm0.17$ & 0 & 1.78 & $ 7.14\pm0.70$ & $0.00\pm0.17$ & 0 \\ 
  0.62 & $ 7.44\pm0.59$ & $0.00\pm0.17$ & 0 & 1.80 & $ 7.24\pm0.71$ & $0.00\pm0.17$ & 0 \\ 
  0.63 & $ 7.39\pm0.58$ & $0.00\pm0.17$ & 0 & 1.82 & $ 7.37\pm0.73$ & $0.00\pm0.17$ & 0 \\ 
  0.64 & $ 7.34\pm0.58$ & $0.00\pm0.17$ & 0 & 1.84 & $ 7.29\pm0.72$ & $0.00\pm0.17$ & 0 \\ 
  0.65 & $ 7.32\pm0.58$ & $0.00\pm0.17$ & 0 & 1.86 & $ 7.60\pm0.76$ & $0.00\pm0.17$ & 0 \\ 
  0.66 & $ 7.31\pm0.58$ & $0.00\pm0.17$ & 0 & 1.88 & $ 7.76\pm0.77$ & $0.00\pm0.17$ & 0 \\ 
  0.67 & $ 7.28\pm0.57$ & $0.00\pm0.17$ & 0 & 1.90 & $ 7.94\pm0.79$ & $0.00\pm0.17$ & 0 \\ 
  0.68 & $ 7.25\pm0.57$ & $0.00\pm0.17$ & 0 & 1.92 & $ 8.12\pm0.81$ & $0.00\pm0.17$ & 0 \\ 
  0.69 & $ 7.22\pm0.57$ & $0.00\pm0.17$ & 0 & 1.94 & $ 8.29\pm0.83$ & $0.00\pm0.17$ & 0 \\ 
  0.70 & $ 7.18\pm0.57$ & $0.00\pm0.17$ & 0 & 1.96 & $ 8.51\pm0.85$ & $0.00\pm0.17$ & 0 \\ 
  0.71 & $ 7.15\pm0.56$ & $0.00\pm0.17$ & 0 & 1.98 & $ 8.76\pm0.87$ & $0.00\pm0.17$ & 0 \\ 
  0.72 & $ 7.12\pm0.56$ & $0.00\pm0.17$ & 0 & 2.00 & $ 8.84\pm0.88$ & $0.00\pm0.17$ & 0 \\ 
  0.73 & $ 7.08\pm0.56$ & $0.00\pm0.17$ & 0 & 2.02 & $ 9.20\pm0.91$ & $0.00\pm0.17$ & 0 \\ 
  0.74 & $ 7.04\pm0.55$ & $0.00\pm0.17$ & 0 & 2.04 & $ 9.49\pm0.94$ & $0.00\pm0.17$ & 0 \\ 
  0.75 & $ 7.04\pm0.55$ & $0.00\pm0.17$ & 0 & 2.06 & $ 9.82\pm0.97$ & $0.00\pm0.17$ & 0 \\ 
  0.76 & $ 7.04\pm0.55$ & $0.00\pm0.17$ & 0 & 2.08 & $10.13\pm1.00$ & $0.00\pm0.17$ & 0 \\ 
  0.77 & $ 7.00\pm0.55$ & $0.00\pm0.17$ & 0 & 2.10 & $10.52\pm1.03$ & $0.00\pm0.17$ & 0 \\ 
  0.78 & $ 6.96\pm0.55$ & $0.00\pm0.17$ & 0 & 2.12 & $10.88\pm1.05$ & $0.00\pm0.17$ & 0 \\ 
  0.79 & $ 6.96\pm0.55$ & $0.00\pm0.17$ & 0 & 2.14 & $11.37\pm1.09$ & $0.00\pm0.17$ & 0 \\ 
  0.80 & $ 6.97\pm0.55$ & $0.00\pm0.17$ & 0 & 2.16 & $11.82\pm1.13$ & $0.00\pm0.17$ & 0 \\ 
  0.81 & $ 6.93\pm0.55$ & $0.00\pm0.17$ & 0 & 2.18 & $12.30\pm1.16$ & $0.00\pm0.17$ & 0 \\ 
  0.82 & $ 6.89\pm0.54$ & $0.00\pm0.17$ & 0 & 2.20 & $12.88\pm1.21$ & $0.00\pm0.17$ & 0 \\ 
  0.83 & $ 6.86\pm0.54$ & $0.00\pm0.17$ & 0 & 2.22 & $13.53\pm1.26$ & $0.00\pm0.17$ & 0 \\ 
  0.84 & $ 6.82\pm0.54$ & $0.00\pm0.17$ & 0 & 2.24 & $14.16\pm1.31$ & $0.00\pm0.17$ & 0 \\ 
  0.85 & $ 6.83\pm0.54$ & $0.00\pm0.17$ & 0 & 2.26 & $14.80\pm1.35$ & $0.00\pm0.17$ & 0 \\ 
  0.86 & $ 6.83\pm0.54$ & $0.00\pm0.17$ & 0 & 2.28 & $15.38\pm1.39$ & $0.00\pm0.17$ & 0 \\ 
  0.87 & $ 6.76\pm0.54$ & $0.00\pm0.17$ & 0 & 2.30 & $16.01\pm1.44$ & $0.00\pm0.17$ & 0 \\ 
  0.88 & $ 6.69\pm0.53$ & $0.00\pm0.17$ & 0 & 2.32 & $16.61\pm1.48$ & $0.00\pm0.17$ & 0 \\ 
  0.89 & $ 6.71\pm0.53$ & $0.00\pm0.17$ & 0 & 2.34 & $17.25\pm1.53$ & $0.00\pm0.17$ & 0 \\ 
  0.90 & $ 6.73\pm0.54$ & $0.00\pm0.17$ & 0 & 2.36 & $17.84\pm1.58$ & $0.00\pm0.17$ & 0 \\ 
  0.91 & $ 6.71\pm0.53$ & $0.00\pm0.17$ & 0 & 2.38 & $18.54\pm1.63$ & $0.00\pm0.17$ & 0 \\ 
  0.92 & $ 6.69\pm0.53$ & $0.00\pm0.17$ & 0 & 2.40 & $19.21\pm1.68$ & $0.00\pm0.17$ & 0 \\ 
  0.93 & $ 6.63\pm0.53$ & $0.00\pm0.17$ & 0 & 2.42 & $19.76\pm1.73$ & $0.00\pm0.17$ & 0 \\ 
  0.94 & $ 6.58\pm0.52$ & $0.00\pm0.17$ & 0 & 2.44 & $20.43\pm1.78$ & $0.00\pm0.17$ & 0 \\ 
  0.95 & $ 6.57\pm0.52$ & $0.00\pm0.17$ & 0 & 2.46 & $21.05\pm1.84$ & $0.00\pm0.17$ & 0 \\ 
  0.96 & $ 6.56\pm0.53$ & $0.00\pm0.17$ & 0 & 2.48 & $21.69\pm1.89$ & $0.00\pm0.17$ & 0 \\ 
  0.97 & $ 6.55\pm0.53$ & $0.00\pm0.17$ & 0 & 2.50 & $22.40\pm1.95$ & $0.00\pm0.17$ & 0 \\ 
  0.98 & $ 6.53\pm0.52$ & $0.00\pm0.17$ & 0 & 2.52 & $23.15\pm2.02$ & $0.00\pm0.17$ & 0 \\ 
  0.99 & $ 6.48\pm0.52$ & $0.00\pm0.17$ & 0 & 2.54 & $23.90\pm2.09$ & $0.00\pm0.17$ & 0 \\ 
  1.00 & $ 6.42\pm0.52$ & $0.00\pm0.17$ & 0 & 2.56 & $24.81\pm2.17$ & $0.00\pm0.17$ & 0 \\ 
  1.01 & $ 6.42\pm0.52$ & $0.00\pm0.17$ & 0 & 2.58 & $25.70\pm2.24$ & $0.00\pm0.17$ & 0 \\ 
  1.02 & $ 6.43\pm0.52$ & $0.00\pm0.17$ & 0 & 2.60 & $26.85\pm2.34$ & $0.00\pm0.17$ & 0 \\ 
  1.03 & $ 6.41\pm0.52$ & $0.00\pm0.17$ & 0 & 2.62 & $27.62\pm2.40$ & $0.00\pm0.17$ & 0 \\ 
  1.04 & $ 6.40\pm0.52$ & $0.00\pm0.17$ & 0 &      &                &               &   \\ 
\hline
\hline

\end{longtable}

\begin{longtable}{cccc | cccc}
\caption{\label{sub:eepp}
  Signal efficnency ($\epsilon$), the number of expected background events ($N_{\rm bkg}$), 
  and the number of observed events ($N_{\rm obs}$)
  for $B^0 \to A^{\prime} A^{\prime} \to e^+ e^- \pi^+ \pi^-$ mode.
  }\\
\hline
\hline
              $m_{A^{\prime}}$ & $\epsilon$ & $N_{\rm bkg}$ & $N_{\rm obs}$ &             $m_{A^{\prime}}$ & $\epsilon$ & $N_{\rm bkg}$ & $N_{\rm obs}$ \\
  ($\mathrm{GeV}/c^2$) &       (\%) &               &               & ($\mathrm{GeV}/c^2$) &       (\%) &               &               \\
\hline
  0.28 & $20.84\pm1.25$ & $0.00\pm0.49$ & 0 & 1.70 & $13.74\pm0.82$ & $0.40\pm0.53$ & 0 \\ 
  0.29 & $21.00\pm1.25$ & $0.50\pm0.50$ & 0 & 1.72 & $12.70\pm0.76$ & $0.30\pm0.52$ & 0 \\ 
  0.30 & $21.16\pm1.26$ & $0.00\pm0.49$ & 0 & 1.74 & $13.48\pm0.81$ & $0.20\pm0.51$ & 0 \\ 
  0.31 & $21.08\pm1.25$ & $0.00\pm0.49$ & 0 & 1.76 & $ 9.27\pm0.55$ & $0.30\pm0.52$ & 0 \\ 
  0.32 & $21.00\pm1.24$ & $0.00\pm0.49$ & 0 & 1.78 & $ 7.13\pm0.43$ & $0.40\pm0.53$ & 0 \\ 
  0.33 & $20.97\pm1.24$ & $0.00\pm0.49$ & 0 & 1.80 & $13.49\pm0.82$ & $0.40\pm0.53$ & 0 \\ 
  0.34 & $20.94\pm1.24$ & $0.00\pm0.49$ & 0 & 1.82 & $18.39\pm1.10$ & $0.60\pm0.55$ & 0 \\ 
  0.35 & $20.82\pm1.23$ & $0.00\pm0.49$ & 0 & 1.84 & $17.80\pm1.07$ & $0.90\pm0.54$ & 0 \\ 
  0.36 & $20.70\pm1.23$ & $0.00\pm0.49$ & 0 & 1.88 & $10.01\pm0.60$ & $1.30\pm0.73$ & 0 \\ 
  0.37 & $20.52\pm1.22$ & $0.00\pm0.49$ & 0 & 1.90 & $17.73\pm1.06$ & $0.50\pm0.54$ & 0 \\ 
  0.38 & $20.33\pm1.21$ & $0.00\pm0.49$ & 0 & 1.92 & $18.53\pm1.11$ & $0.85\pm0.57$ & 0 \\ 
  0.39 & $19.79\pm1.18$ & $0.50\pm0.50$ & 0 & 1.94 & $19.28\pm1.15$ & $1.05\pm0.58$ & 0 \\ 
  0.40 & $19.25\pm1.15$ & $0.50\pm0.50$ & 0 & 1.96 & $19.23\pm1.15$ & $0.97\pm0.58$ & 0 \\ 
  0.41 & $19.10\pm1.15$ & $0.50\pm0.50$ & 0 & 1.98 & $19.51\pm1.16$ & $1.52\pm0.59$ & 0 \\ 
  0.42 & $18.96\pm1.14$ & $0.00\pm0.49$ & 0 & 2.00 & $19.94\pm1.19$ & $1.32\pm0.57$ & 0 \\ 
  0.43 & $18.52\pm1.11$ & $0.00\pm0.49$ & 0 & 2.02 & $19.80\pm1.17$ & $1.39\pm0.58$ & 0 \\ 
  0.44 & $18.08\pm1.09$ & $0.50\pm0.50$ & 0 & 2.04 & $20.18\pm1.20$ & $1.57\pm0.59$ & 1 \\ 
  1.10 & $14.73\pm0.87$ & $0.30\pm0.52$ & 0 & 2.06 & $20.27\pm1.20$ & $1.32\pm0.60$ & 1 \\ 
  1.12 & $14.70\pm0.87$ & $0.24\pm0.51$ & 0 & 2.08 & $20.39\pm1.21$ & $0.95\pm0.58$ & 1 \\ 
  1.14 & $14.77\pm0.88$ & $0.64\pm0.51$ & 0 & 2.10 & $20.10\pm1.19$ & $0.75\pm0.56$ & 0 \\ 
  1.16 & $14.69\pm0.87$ & $0.62\pm0.51$ & 0 & 2.12 & $20.50\pm1.21$ & $1.80\pm0.62$ & 2 \\ 
  1.18 & $14.40\pm0.86$ & $0.85\pm0.53$ & 1 & 2.14 & $20.58\pm1.21$ & $1.60\pm0.60$ & 2 \\ 
  1.20 & $14.72\pm0.88$ & $0.25\pm0.51$ & 1 & 2.16 & $20.61\pm1.21$ & $2.00\pm0.63$ & 2 \\ 
  1.22 & $14.82\pm0.88$ & $0.20\pm0.50$ & 0 & 2.18 & $20.68\pm1.21$ & $1.60\pm0.63$ & 0 \\ 
  1.24 & $14.58\pm0.87$ & $0.25\pm0.51$ & 0 & 2.20 & $17.77\pm1.04$ & $1.30\pm0.61$ & 0 \\ 
  1.26 & $14.83\pm0.88$ & $0.75\pm0.52$ & 1 & 2.22 & $18.07\pm1.05$ & $2.10\pm0.78$ & 0 \\ 
  1.28 & $14.80\pm0.89$ & $0.60\pm0.51$ & 1 & 2.24 & $18.20\pm1.06$ & $2.12\pm0.78$ & 0 \\ 
  1.30 & $15.08\pm0.90$ & $0.30\pm0.52$ & 0 & 2.26 & $17.84\pm1.04$ & $2.72\pm0.93$ & 1 \\ 
  1.32 & $15.64\pm0.94$ & $0.30\pm0.52$ & 0 & 2.28 & $18.45\pm1.07$ & $2.74\pm0.93$ & 1 \\ 
  1.34 & $15.49\pm0.93$ & $0.20\pm0.51$ & 0 & 2.30 & $18.05\pm1.05$ & $2.04\pm0.63$ & 1 \\ 
  1.36 & $15.50\pm0.93$ & $0.00\pm0.49$ & 2 & 2.32 & $18.34\pm1.06$ & $1.72\pm0.61$ & 1 \\ 
  1.38 & $15.48\pm0.93$ & $0.60\pm0.51$ & 1 & 2.34 & $18.49\pm1.07$ & $0.77\pm0.56$ & 0 \\ 
  1.40 & $15.57\pm0.93$ & $1.00\pm0.55$ & 1 & 2.36 & $18.18\pm1.05$ & $1.15\pm0.59$ & 0 \\ 
  1.42 & $15.52\pm0.93$ & $0.40\pm0.53$ & 0 & 2.38 & $18.40\pm1.06$ & $1.15\pm0.59$ & 1 \\ 
  1.44 & $15.64\pm0.94$ & $0.60\pm0.51$ & 0 & 2.40 & $19.01\pm1.09$ & $2.05\pm0.66$ & 2 \\ 
  1.46 & $15.84\pm0.95$ & $0.70\pm0.52$ & 0 & 2.42 & $18.79\pm1.07$ & $2.50\pm0.70$ & 2 \\ 
  1.48 & $15.80\pm0.95$ & $1.50\pm0.74$ & 0 & 2.44 & $18.69\pm1.06$ & $2.30\pm0.68$ & 2 \\ 
  1.50 & $15.63\pm0.94$ & $1.10\pm0.56$ & 1 & 2.46 & $18.69\pm1.06$ & $3.10\pm0.71$ & 2 \\ 
  1.52 & $15.75\pm0.95$ & $1.15\pm0.56$ & 1 & 2.48 & $18.79\pm1.06$ & $3.40\pm0.86$ & 1 \\ 
  1.54 & $15.69\pm0.94$ & $0.60\pm0.54$ & 1 & 2.50 & $18.73\pm1.05$ & $2.40\pm0.66$ & 1 \\ 
  1.56 & $16.25\pm0.98$ & $1.15\pm0.55$ & 0 & 2.52 & $19.12\pm1.07$ & $2.05\pm0.63$ & 1 \\ 
  1.58 & $16.15\pm0.97$ & $0.90\pm0.53$ & 0 & 2.54 & $21.35\pm1.18$ & $3.75\pm1.08$ & 2 \\ 
  1.60 & $15.84\pm0.95$ & $0.85\pm0.53$ & 0 & 2.56 & $21.30\pm1.17$ & $3.05\pm0.95$ & 1 \\ 
  1.62 & $15.61\pm0.94$ & $0.45\pm0.53$ & 0 & 2.58 & $21.96\pm1.18$ & $3.35\pm0.97$ & 1 \\ 
  1.64 & $15.60\pm0.94$ & $0.55\pm0.54$ & 0 & 2.60 & $22.52\pm1.19$ & $2.25\pm0.79$ & 0 \\ 
  1.66 & $15.11\pm0.91$ & $0.55\pm0.54$ & 0 & 2.62 & $22.94\pm1.18$ & $0.80\pm0.56$ & 0 \\ 
  1.68 & $14.50\pm0.87$ & $0.50\pm0.54$ & 0 &      &                &               &   \\ 
\hline
\hline

\end{longtable}

\begin{longtable}{cccc | cccc}
\caption{\label{sub:mmpp}
  Signal efficnency ($\epsilon$), the number of expected background events ($N_{\rm bkg}$), 
  and the number of observed events ($N_{\rm obs}$)
  for $B^0 \to A^{\prime} A^{\prime} \to \mu^+ \mu^- \pi^+ \pi^-$ mode.
  }\\
\hline
\hline
              $m_{A^{\prime}}$ & $\epsilon$ & $N_{\rm bkg}$ & $N_{\rm obs}$ &             $m_{A^{\prime}}$ & $\epsilon$ & $N_{\rm bkg}$ & $N_{\rm obs}$ \\
  ($\mathrm{GeV}/c^2$) &       (\%) &               &               & ($\mathrm{GeV}/c^2$) &       (\%) &               &               \\
\hline
  0.28 & $20.99\pm1.25$ & $0.00\pm0.49$ & 0 & 1.70 & $ 8.68\pm0.52$ & $0.00\pm0.49$ & 0 \\ 
  0.29 & $20.42\pm1.22$ & $0.00\pm0.49$ & 0 & 1.72 & $ 8.02\pm0.48$ & $0.00\pm0.49$ & 0 \\ 
  0.30 & $19.85\pm1.18$ & $0.50\pm0.50$ & 0 & 1.74 & $ 8.67\pm0.52$ & $0.00\pm0.49$ & 0 \\ 
  0.31 & $19.41\pm1.15$ & $0.50\pm0.50$ & 0 & 1.76 & $ 5.68\pm0.34$ & $0.10\pm0.50$ & 0 \\ 
  0.32 & $18.97\pm1.12$ & $0.00\pm0.49$ & 0 & 1.78 & $ 4.48\pm0.28$ & $0.10\pm0.50$ & 0 \\ 
  0.33 & $18.45\pm1.09$ & $0.50\pm0.50$ & 0 & 1.80 & $ 8.58\pm0.54$ & $0.00\pm0.49$ & 0 \\ 
  0.34 & $17.93\pm1.06$ & $0.00\pm0.49$ & 0 & 1.82 & $11.76\pm0.72$ & $0.20\pm0.51$ & 1 \\ 
  0.35 & $17.67\pm1.04$ & $0.00\pm0.49$ & 0 & 1.84 & $11.40\pm0.71$ & $0.10\pm0.50$ & 0 \\ 
  0.36 & $17.42\pm1.02$ & $0.00\pm0.49$ & 0 & 1.88 & $ 6.46\pm0.40$ & $0.10\pm0.50$ & 0 \\ 
  0.37 & $16.87\pm1.00$ & $0.00\pm0.49$ & 0 & 1.90 & $12.91\pm0.80$ & $0.00\pm0.49$ & 0 \\ 
  0.38 & $16.33\pm0.97$ & $0.00\pm0.49$ & 0 & 1.92 & $13.31\pm0.82$ & $0.00\pm0.49$ & 0 \\ 
  0.39 & $15.81\pm0.94$ & $0.00\pm0.49$ & 0 & 1.94 & $13.75\pm0.85$ & $0.10\pm0.50$ & 0 \\ 
  0.40 & $15.30\pm0.91$ & $0.00\pm0.49$ & 0 & 1.96 & $14.01\pm0.86$ & $0.00\pm0.49$ & 0 \\ 
  0.41 & $14.98\pm0.90$ & $0.00\pm0.49$ & 0 & 1.98 & $14.22\pm0.88$ & $0.20\pm0.51$ & 0 \\ 
  0.42 & $14.66\pm0.88$ & $0.00\pm0.49$ & 0 & 2.00 & $14.62\pm0.90$ & $0.10\pm0.50$ & 0 \\ 
  0.43 & $14.31\pm0.86$ & $0.00\pm0.49$ & 0 & 2.02 & $14.84\pm0.91$ & $0.20\pm0.51$ & 0 \\ 
  0.44 & $13.95\pm0.84$ & $0.00\pm0.49$ & 0 & 2.04 & $15.19\pm0.92$ & $0.10\pm0.50$ & 0 \\ 
  1.10 & $ 9.87\pm0.56$ & $0.50\pm0.50$ & 0 & 2.06 & $15.53\pm0.94$ & $0.10\pm0.50$ & 0 \\ 
  1.12 & $ 9.88\pm0.56$ & $0.50\pm0.50$ & 0 & 2.08 & $15.96\pm0.96$ & $0.10\pm0.50$ & 0 \\ 
  1.14 & $ 9.88\pm0.56$ & $0.10\pm0.50$ & 0 & 2.10 & $16.28\pm0.98$ & $0.27\pm0.51$ & 0 \\ 
  1.16 & $ 9.82\pm0.56$ & $0.00\pm0.49$ & 0 & 2.12 & $16.46\pm0.98$ & $0.22\pm0.51$ & 0 \\ 
  1.18 & $ 9.77\pm0.55$ & $0.00\pm0.49$ & 0 & 2.14 & $16.84\pm1.00$ & $0.20\pm0.51$ & 0 \\ 
  1.20 & $ 9.87\pm0.56$ & $0.00\pm0.49$ & 0 & 2.16 & $17.15\pm1.01$ & $0.10\pm0.50$ & 0 \\ 
  1.22 & $ 9.76\pm0.56$ & $0.50\pm0.50$ & 0 & 2.18 & $17.52\pm1.03$ & $0.10\pm0.50$ & 1 \\ 
  1.24 & $ 9.96\pm0.57$ & $0.00\pm0.49$ & 0 & 2.20 & $17.76\pm1.04$ & $0.00\pm0.49$ & 0 \\ 
  1.26 & $ 9.93\pm0.57$ & $0.10\pm0.50$ & 0 & 2.22 & $18.08\pm1.05$ & $0.10\pm0.50$ & 0 \\ 
  1.28 & $ 9.73\pm0.56$ & $0.50\pm0.50$ & 1 & 2.24 & $18.66\pm1.08$ & $0.60\pm0.51$ & 0 \\ 
  1.30 & $ 9.81\pm0.56$ & $0.00\pm0.49$ & 0 & 2.26 & $18.95\pm1.09$ & $0.30\pm0.52$ & 0 \\ 
  1.32 & $ 9.86\pm0.57$ & $0.10\pm0.50$ & 0 & 2.28 & $19.26\pm1.10$ & $0.00\pm0.49$ & 0 \\ 
  1.34 & $ 9.77\pm0.56$ & $0.10\pm0.50$ & 0 & 2.30 & $19.74\pm1.12$ & $0.20\pm0.51$ & 0 \\ 
  1.36 & $ 9.76\pm0.56$ & $0.50\pm0.50$ & 0 & 2.32 & $19.99\pm1.14$ & $0.30\pm0.52$ & 0 \\ 
  1.38 & $ 9.58\pm0.55$ & $0.00\pm0.49$ & 0 & 2.34 & $20.31\pm1.15$ & $0.10\pm0.50$ & 0 \\ 
  1.40 & $ 9.76\pm0.57$ & $0.00\pm0.49$ & 0 & 2.36 & $20.66\pm1.17$ & $0.00\pm0.49$ & 1 \\ 
  1.42 & $ 9.71\pm0.57$ & $0.10\pm0.50$ & 0 & 2.38 & $20.98\pm1.18$ & $0.12\pm0.50$ & 1 \\ 
  1.44 & $ 9.75\pm0.57$ & $0.10\pm0.50$ & 0 & 2.40 & $20.87\pm1.17$ & $0.72\pm0.52$ & 0 \\ 
  1.46 & $ 9.76\pm0.57$ & $0.10\pm0.50$ & 0 & 2.42 & $21.57\pm1.21$ & $1.30\pm0.73$ & 0 \\ 
  1.48 & $ 9.65\pm0.57$ & $0.00\pm0.49$ & 0 & 2.44 & $22.02\pm1.24$ & $0.40\pm0.53$ & 0 \\ 
  1.50 & $ 9.94\pm0.59$ & $0.00\pm0.49$ & 0 & 2.46 & $22.41\pm1.26$ & $0.70\pm0.52$ & 0 \\ 
  1.52 & $ 9.83\pm0.58$ & $0.00\pm0.49$ & 0 & 2.48 & $22.53\pm1.27$ & $0.50\pm0.50$ & 0 \\ 
  1.54 & $ 9.85\pm0.59$ & $0.00\pm0.49$ & 0 & 2.50 & $23.08\pm1.30$ & $0.50\pm0.50$ & 0 \\ 
  1.56 & $ 9.92\pm0.59$ & $0.00\pm0.49$ & 0 & 2.52 & $23.45\pm1.32$ & $0.62\pm0.51$ & 0 \\ 
  1.58 & $ 9.65\pm0.58$ & $0.00\pm0.49$ & 0 & 2.54 & $23.91\pm1.35$ & $0.32\pm0.52$ & 0 \\ 
  1.60 & $ 9.44\pm0.57$ & $0.00\pm0.49$ & 0 & 2.56 & $24.17\pm1.36$ & $0.42\pm0.53$ & 1 \\ 
  1.62 & $ 9.41\pm0.56$ & $0.00\pm0.49$ & 0 & 2.58 & $24.50\pm1.38$ & $0.52\pm0.54$ & 1 \\ 
  1.64 & $ 9.07\pm0.55$ & $0.00\pm0.49$ & 0 & 2.60 & $25.34\pm1.43$ & $1.52\pm0.74$ & 0 \\ 
  1.66 & $ 8.93\pm0.54$ & $0.00\pm0.49$ & 0 & 2.62 & $25.74\pm1.45$ & $1.52\pm0.87$ & 0 \\ 
  1.68 & $ 8.86\pm0.53$ & $0.00\pm0.49$ & 0 &      &                &               &   \\ 
\hline
\hline

\end{longtable}

\begin{longtable}{cccc | cccc}
\caption{\label{sub:AA}
  Signal efficnency ($\epsilon$), the number of expected background events ($N_{\rm bkg}$), 
  and the number of observed events ($N_{\rm obs}$)
  for a combined $B^0 \to A^{\prime} A^{\prime}$.
  }\\
\hline
\hline
              $m_{A^{\prime}}$ & $\epsilon$ & $N_{\rm bkg}$ & $N_{\rm obs}$ &             $m_{A^{\prime}}$ & $\epsilon$ & $N_{\rm bkg}$ & $N_{\rm obs}$ \\
  ($\mathrm{GeV}/c^2$) &       (\%) &               &               & ($\mathrm{GeV}/c^2$) &       (\%) &               &               \\
\hline
  0.01 & $14.56\pm1.34$ & $0.33\pm0.24$ & 1 & 0.94 & $ 3.63\pm0.08$ & $0.10\pm0.30$ & 0 \\ 
  0.02 & $14.82\pm1.36$ & $0.83\pm0.37$ & 2 & 0.95 & $ 3.84\pm0.09$ & $0.10\pm0.30$ & 0 \\ 
  0.03 & $14.72\pm1.35$ & $0.50\pm0.29$ & 1 & 0.96 & $ 3.91\pm0.09$ & $0.00\pm0.29$ & 0 \\ 
  0.04 & $14.62\pm1.33$ & $0.00\pm0.17$ & 1 & 0.97 & $ 4.19\pm0.10$ & $0.00\pm0.29$ & 0 \\ 
  0.05 & $14.51\pm1.32$ & $0.00\pm0.17$ & 1 & 0.98 & $ 4.03\pm0.10$ & $0.00\pm0.29$ & 0 \\ 
  0.06 & $14.40\pm1.31$ & $0.00\pm0.17$ & 0 & 0.99 & $ 4.28\pm0.11$ & $0.00\pm0.29$ & 0 \\ 
  0.07 & $14.24\pm1.30$ & $0.00\pm0.17$ & 1 & 1.00 & $ 3.53\pm0.09$ & $0.00\pm0.29$ & 0 \\ 
  0.08 & $14.08\pm1.28$ & $0.00\pm0.17$ & 1 & 1.01 & $ 1.98\pm0.05$ & $0.00\pm0.29$ & 0 \\ 
  0.09 & $13.86\pm1.26$ & $0.00\pm0.17$ & 0 & 1.02 & $ 0.02\pm0.00$ & $0.00\pm0.29$ & 0 \\ 
  0.10 & $13.64\pm1.24$ & $0.00\pm0.17$ & 0 & 1.03 & $ 1.24\pm0.03$ & $0.00\pm0.29$ & 0 \\ 
  0.11 & $13.65\pm1.24$ & $0.00\pm0.17$ & 0 & 1.04 & $ 2.73\pm0.07$ & $0.00\pm0.29$ & 0 \\ 
  0.12 & $13.66\pm1.24$ & $0.00\pm0.17$ & 0 & 1.05 & $ 3.42\pm0.09$ & $0.00\pm0.29$ & 0 \\ 
  0.13 & $13.76\pm1.25$ & $0.00\pm0.17$ & 0 & 1.06 & $ 3.84\pm0.11$ & $0.00\pm0.29$ & 0 \\ 
  0.14 & $13.85\pm1.26$ & $0.00\pm0.17$ & 0 & 1.07 & $ 3.83\pm0.11$ & $0.00\pm0.29$ & 0 \\ 
  0.15 & $13.71\pm1.25$ & $0.00\pm0.17$ & 0 & 1.08 & $ 3.94\pm0.12$ & $0.00\pm0.29$ & 0 \\ 
  0.16 & $13.57\pm1.24$ & $0.00\pm0.17$ & 0 & 1.09 & $ 4.19\pm0.13$ & $0.00\pm0.29$ & 0 \\ 
  0.17 & $13.47\pm1.23$ & $0.00\pm0.17$ & 0 & 1.10 & $ 7.64\pm0.25$ & $0.80\pm0.78$ & 0 \\ 
  0.18 & $13.37\pm1.22$ & $0.00\pm0.17$ & 0 & 1.12 & $ 7.30\pm0.25$ & $0.74\pm0.77$ & 0 \\ 
  0.19 & $13.31\pm1.21$ & $0.00\pm0.17$ & 0 & 1.14 & $ 6.85\pm0.24$ & $0.74\pm0.77$ & 0 \\ 
  0.20 & $13.25\pm1.21$ & $0.00\pm0.17$ & 0 & 1.16 & $ 6.49\pm0.23$ & $0.62\pm0.76$ & 0 \\ 
  0.21 & $13.22\pm1.20$ & $0.00\pm0.17$ & 0 & 1.18 & $ 6.38\pm0.23$ & $0.85\pm0.78$ & 1 \\ 
  0.22 & $19.39\pm1.06$ & $0.00\pm0.29$ & 0 & 1.20 & $ 6.08\pm0.23$ & $0.25\pm0.76$ & 1 \\ 
  0.23 & $19.91\pm1.00$ & $0.00\pm0.29$ & 0 & 1.22 & $ 5.53\pm0.22$ & $0.70\pm0.77$ & 0 \\ 
  0.24 & $19.50\pm0.95$ & $0.00\pm0.29$ & 0 & 1.24 & $ 5.75\pm0.21$ & $0.25\pm0.76$ & 0 \\ 
  0.25 & $18.39\pm0.88$ & $0.00\pm0.29$ & 0 & 1.26 & $ 5.27\pm0.21$ & $0.85\pm0.78$ & 1 \\ 
  0.26 & $17.04\pm0.80$ & $0.00\pm0.29$ & 0 & 1.28 & $ 4.54\pm0.19$ & $1.10\pm0.77$ & 2 \\ 
  0.27 & $16.12\pm0.76$ & $0.00\pm0.29$ & 0 & 1.30 & $ 4.68\pm0.19$ & $0.30\pm0.77$ & 0 \\ 
  0.28 & $15.14\pm0.71$ & $0.00\pm0.75$ & 0 & 1.32 & $ 4.53\pm0.19$ & $0.40\pm0.77$ & 0 \\ 
  0.29 & $14.62\pm0.68$ & $0.50\pm0.76$ & 0 & 1.34 & $ 4.18\pm0.17$ & $0.40\pm0.77$ & 0 \\ 
  0.30 & $14.10\pm0.64$ & $0.50\pm0.76$ & 0 & 1.36 & $ 3.68\pm0.16$ & $0.60\pm0.76$ & 2 \\ 
  0.31 & $13.84\pm0.62$ & $0.50\pm0.76$ & 0 & 1.38 & $ 3.48\pm0.15$ & $0.80\pm0.77$ & 1 \\ 
  0.32 & $13.58\pm0.60$ & $0.00\pm0.75$ & 0 & 1.40 & $ 3.03\pm0.14$ & $1.10\pm0.79$ & 1 \\ 
  0.33 & $13.37\pm0.58$ & $0.50\pm0.76$ & 0 & 1.42 & $ 3.32\pm0.15$ & $0.60\pm0.79$ & 0 \\ 
  0.34 & $13.16\pm0.57$ & $0.00\pm0.75$ & 0 & 1.44 & $ 3.51\pm0.16$ & $0.70\pm0.77$ & 0 \\ 
  0.35 & $12.97\pm0.55$ & $0.00\pm0.75$ & 0 & 1.46 & $ 3.00\pm0.14$ & $0.80\pm0.77$ & 0 \\ 
  0.36 & $12.80\pm0.53$ & $0.00\pm0.75$ & 0 & 1.48 & $ 2.48\pm0.12$ & $1.50\pm0.93$ & 0 \\ 
  0.37 & $12.65\pm0.52$ & $0.00\pm0.75$ & 0 & 1.50 & $ 2.27\pm0.11$ & $1.10\pm0.79$ & 1 \\ 
  0.38 & $12.45\pm0.50$ & $0.00\pm0.75$ & 0 & 1.52 & $ 2.31\pm0.11$ & $1.15\pm0.80$ & 1 \\ 
  0.39 & $12.69\pm0.51$ & $0.50\pm0.76$ & 0 & 1.54 & $ 2.27\pm0.11$ & $0.60\pm0.78$ & 1 \\ 
  0.40 & $12.54\pm0.49$ & $0.50\pm0.76$ & 0 & 1.56 & $ 2.18\pm0.11$ & $1.15\pm0.79$ & 0 \\ 
  0.41 & $12.63\pm0.48$ & $0.50\pm0.76$ & 0 & 1.58 & $ 2.20\pm0.11$ & $0.90\pm0.78$ & 0 \\ 
  0.42 & $12.89\pm0.48$ & $0.00\pm0.75$ & 0 & 1.60 & $ 2.12\pm0.10$ & $0.85\pm0.78$ & 0 \\ 
  0.43 & $12.50\pm0.46$ & $0.00\pm0.75$ & 0 & 1.62 & $ 2.14\pm0.10$ & $0.45\pm0.78$ & 0 \\ 
  0.44 & $11.85\pm0.44$ & $0.50\pm0.76$ & 0 & 1.64 & $ 1.84\pm0.09$ & $0.55\pm0.78$ & 0 \\ 
  0.45 & $ 8.27\pm0.29$ & $0.00\pm0.29$ & 0 & 1.66 & $ 1.85\pm0.09$ & $0.55\pm0.78$ & 0 \\ 
  0.46 & $ 8.34\pm0.29$ & $0.00\pm0.29$ & 0 & 1.68 & $ 1.95\pm0.10$ & $0.50\pm0.78$ & 0 \\ 
  0.47 & $ 7.83\pm0.26$ & $0.00\pm0.29$ & 0 & 1.70 & $ 1.98\pm0.10$ & $0.40\pm0.77$ & 0 \\ 
  0.48 & $ 7.42\pm0.24$ & $0.00\pm0.29$ & 0 & 1.72 & $ 2.06\pm0.10$ & $0.30\pm0.77$ & 0 \\ 
  0.49 & $ 7.18\pm0.23$ & $0.00\pm0.29$ & 0 & 1.74 & $ 2.11\pm0.10$ & $0.20\pm0.76$ & 0 \\ 
  0.50 & $ 6.93\pm0.21$ & $0.00\pm0.29$ & 0 & 1.76 & $ 1.90\pm0.09$ & $0.40\pm0.77$ & 0 \\ 
  0.51 & $ 6.54\pm0.20$ & $0.00\pm0.29$ & 0 & 1.78 & $ 2.00\pm0.10$ & $0.50\pm0.78$ & 0 \\ 
  0.52 & $ 6.49\pm0.19$ & $0.00\pm0.29$ & 0 & 1.80 & $ 2.30\pm0.11$ & $0.40\pm0.77$ & 0 \\ 
  0.53 & $ 5.90\pm0.17$ & $0.00\pm0.29$ & 0 & 1.82 & $ 2.52\pm0.12$ & $0.80\pm0.80$ & 1 \\ 
  0.54 & $ 5.58\pm0.15$ & $0.00\pm0.29$ & 0 & 1.84 & $ 2.41\pm0.11$ & $1.00\pm0.79$ & 0 \\ 
  0.55 & $ 5.36\pm0.14$ & $0.00\pm0.29$ & 0 & 1.86 & $ 2.87\pm0.14$ & $0.80\pm0.78$ & 0 \\ 
  0.56 & $ 5.09\pm0.13$ & $0.00\pm0.29$ & 0 & 1.88 & $ 2.98\pm0.14$ & $1.40\pm0.93$ & 0 \\ 
  0.57 & $ 4.79\pm0.11$ & $0.00\pm0.29$ & 0 & 1.90 & $ 2.96\pm0.15$ & $0.60\pm0.79$ & 0 \\ 
  0.58 & $ 4.52\pm0.10$ & $0.00\pm0.29$ & 0 & 1.92 & $ 2.95\pm0.15$ & $0.95\pm0.81$ & 0 \\ 
  0.59 & $ 4.10\pm0.08$ & $0.10\pm0.30$ & 0 & 1.94 & $ 2.91\pm0.15$ & $1.25\pm0.82$ & 0 \\ 
  0.60 & $ 3.74\pm0.07$ & $0.10\pm0.30$ & 0 & 1.96 & $ 2.54\pm0.13$ & $1.07\pm0.81$ & 0 \\ 
  0.61 & $ 3.37\pm0.06$ & $0.10\pm0.30$ & 0 & 1.98 & $ 2.53\pm0.13$ & $1.97\pm0.84$ & 0 \\ 
  0.62 & $ 3.14\pm0.06$ & $0.10\pm0.30$ & 0 & 2.00 & $ 2.88\pm0.14$ & $1.67\pm0.83$ & 0 \\ 
  0.63 & $ 2.69\pm0.04$ & $0.10\pm0.30$ & 0 & 2.02 & $ 2.89\pm0.14$ & $1.94\pm0.84$ & 0 \\ 
  0.64 & $ 2.70\pm0.04$ & $0.10\pm0.30$ & 0 & 2.04 & $ 2.91\pm0.14$ & $1.92\pm0.84$ & 1 \\ 
  0.65 & $ 2.04\pm0.03$ & $0.00\pm0.29$ & 0 & 2.06 & $ 2.94\pm0.14$ & $1.67\pm0.85$ & 1 \\ 
  0.66 & $ 1.80\pm0.03$ & $0.00\pm0.29$ & 0 & 2.08 & $ 2.96\pm0.15$ & $1.20\pm0.82$ & 1 \\ 
  0.67 & $ 1.59\pm0.02$ & $0.00\pm0.29$ & 0 & 2.10 & $ 3.00\pm0.15$ & $1.17\pm0.82$ & 0 \\ 
  0.68 & $ 1.19\pm0.01$ & $0.00\pm0.29$ & 0 & 2.12 & $ 3.01\pm0.15$ & $2.17\pm0.86$ & 2 \\ 
  0.69 & $ 1.04\pm0.01$ & $0.00\pm0.29$ & 0 & 2.14 & $ 3.07\pm0.15$ & $2.00\pm0.85$ & 2 \\ 
  0.70 & $ 0.86\pm0.01$ & $0.00\pm0.29$ & 0 & 2.16 & $ 3.11\pm0.15$ & $2.30\pm0.87$ & 2 \\ 
  0.71 & $ 0.70\pm0.01$ & $0.00\pm0.29$ & 0 & 2.18 & $ 3.15\pm0.15$ & $2.00\pm0.87$ & 1 \\ 
  0.72 & $ 0.57\pm0.00$ & $0.00\pm0.29$ & 0 & 2.20 & $ 3.20\pm0.15$ & $1.60\pm0.85$ & 0 \\ 
  0.73 & $ 0.48\pm0.00$ & $0.00\pm0.29$ & 0 & 2.22 & $ 3.29\pm0.16$ & $2.50\pm0.98$ & 0 \\ 
  0.74 & $ 0.31\pm0.00$ & $0.00\pm0.29$ & 0 & 2.24 & $ 3.38\pm0.16$ & $3.02\pm0.99$ & 0 \\ 
  0.75 & $ 0.36\pm0.00$ & $0.00\pm0.29$ & 0 & 2.26 & $ 3.47\pm0.16$ & $3.22\pm1.11$ & 1 \\ 
  0.76 & $ 0.33\pm0.00$ & $0.00\pm0.29$ & 0 & 2.28 & $ 3.56\pm0.17$ & $2.94\pm1.10$ & 1 \\ 
  0.77 & $ 0.28\pm0.00$ & $0.00\pm0.29$ & 0 & 2.30 & $ 3.64\pm0.17$ & $2.34\pm0.87$ & 1 \\ 
  0.78 & $ 0.11\pm0.00$ & $0.00\pm0.29$ & 0 & 2.32 & $ 3.73\pm0.17$ & $2.12\pm0.85$ & 1 \\ 
  0.79 & $ 0.31\pm0.00$ & $0.00\pm0.29$ & 0 & 2.34 & $ 3.82\pm0.18$ & $1.07\pm0.81$ & 0 \\ 
  0.80 & $ 0.50\pm0.00$ & $0.00\pm0.29$ & 0 & 2.36 & $ 3.92\pm0.18$ & $1.35\pm0.83$ & 1 \\ 
  0.81 & $ 0.65\pm0.01$ & $0.00\pm0.29$ & 0 & 2.38 & $ 4.01\pm0.18$ & $1.47\pm0.84$ & 2 \\ 
  0.82 & $ 0.81\pm0.01$ & $0.00\pm0.29$ & 0 & 2.40 & $ 4.11\pm0.19$ & $2.97\pm0.90$ & 2 \\ 
  0.83 & $ 0.95\pm0.01$ & $0.00\pm0.29$ & 0 & 2.42 & $ 4.20\pm0.19$ & $4.00\pm1.06$ & 2 \\ 
  0.84 & $ 1.19\pm0.01$ & $0.00\pm0.29$ & 0 & 2.44 & $ 4.28\pm0.19$ & $2.90\pm0.92$ & 2 \\ 
  0.85 & $ 1.50\pm0.02$ & $0.00\pm0.29$ & 0 & 2.46 & $ 4.38\pm0.20$ & $4.00\pm0.94$ & 2 \\ 
  0.86 & $ 1.48\pm0.02$ & $0.00\pm0.29$ & 0 & 2.48 & $ 4.49\pm0.20$ & $4.10\pm1.05$ & 1 \\ 
  0.87 & $ 1.98\pm0.03$ & $0.00\pm0.29$ & 0 & 2.50 & $ 4.59\pm0.21$ & $3.10\pm0.89$ & 1 \\ 
  0.88 & $ 2.17\pm0.03$ & $0.00\pm0.29$ & 0 & 2.52 & $ 4.61\pm0.20$ & $2.67\pm0.86$ & 1 \\ 
  0.89 & $ 2.57\pm0.04$ & $0.00\pm0.29$ & 0 & 2.54 & $ 4.62\pm0.20$ & $4.07\pm1.23$ & 2 \\ 
  0.90 & $ 2.76\pm0.05$ & $0.00\pm0.29$ & 0 & 2.56 & $ 4.65\pm0.20$ & $3.47\pm1.12$ & 2 \\ 
  0.91 & $ 3.06\pm0.06$ & $0.00\pm0.29$ & 0 & 2.58 & $ 4.69\pm0.20$ & $3.97\pm1.15$ & 2 \\ 
  0.92 & $ 3.14\pm0.06$ & $0.10\pm0.30$ & 0 & 2.60 & $ 4.74\pm0.20$ & $3.87\pm1.12$ & 0 \\ 
  0.93 & $ 3.44\pm0.07$ & $0.10\pm0.30$ & 0 & 2.62 & $ 5.00\pm0.21$ & $2.42\pm1.08$ & 0 \\ 
\hline
\hline

\end{longtable}

\section{Upper limits of branching fraction}

The following table presents upper limits at 90\% confidence level for three kinds of branching fraction:
(a) $\mathrm{B.F.}(B^0 \to A^{\prime} A^{\prime}) \equiv \mathcal{B}$,
(b) $\mathrm{B.F.}(B^0 \to A^{\prime} A^{\prime}) \times \mathrm{B.F.}(A^{\prime} \to e^+ e^-)^2 \equiv \mathcal{B}_{ee}$,
(c) $\mathrm{B.F.}(B^0 \to A^{\prime} A^{\prime}) \times \mathrm{B.F.}(A^{\prime} \to \mu^+ \mu^-)^2 \equiv \mathcal{B}_{\mu\mu}$.
The dark photon is scanned in $0.01 - 2.62~\mathrm{GeV}/c^2$ mass range with 
$10~\mathrm{MeV}/c^2$ ($0.01 - 1.10~\mathrm{GeV}/c^2$) and 
$20~\mathrm{MeV}/c^2$ ($1.10 - 2.62~\mathrm{GeV}/c^2$) intervals.
The upper limits are calculated using the POLE program \cite{PhysRevD.67.012002} which is based on the Feldman-Cousins unified approach \cite{PhysRevD.57.3873}.

\begin{longtable}{cccc | cccc}
\caption{\label{sub:AAul}
Upper limits of branching fractions at 90\% confidence level.
}\\
\hline
\hline
              $m_{A^{\prime}}$ & $\mathcal{B}$ & $\mathcal{B}_{ee}$ & $\mathcal{B}_{\mu\mu}$ &             $m_{A^{\prime}}$ & $\mathcal{B}$ & $\mathcal{B}_{ee}$ & $\mathcal{B}_{\mu\mu}$ \\
  ($\mathrm{GeV}/c^2$) &   ($10^{-8}$) &        ($10^{-8}$) &            ($10^{-8}$) & ($\mathrm{GeV}/c^2$) &   ($10^{-8}$) &        ($10^{-8}$) &            ($10^{-8}$) \\
\hline
  0.01 & 3.64 & 3.64 &    - & 0.94 & 8.22 & 2.83 & 4.80 \\ 
  0.02 & 4.55 & 4.55 &    - & 0.95 & 7.78 & 2.82 & 4.81 \\ 
  0.03 & 3.47 & 3.47 &    - & 0.96 & 7.79 & 2.81 & 4.81 \\ 
  0.04 & 3.85 & 3.85 &    - & 0.97 & 7.25 & 2.81 & 4.83 \\ 
  0.05 & 3.88 & 3.88 &    - & 0.98 & 7.54 & 2.82 & 4.84 \\ 
  0.06 & 2.21 & 2.21 &    - & 0.99 & 7.10 & 2.82 & 4.88 \\ 
  0.07 & 3.95 & 3.95 &    - & 1.00 & 8.63 & 2.83 & 4.92 \\ 
  0.08 & 4.00 & 4.00 &    - & 1.01 & 15.4 & 2.83 & 4.94 \\ 
  0.09 & 2.30 & 2.30 &    - & 1.02 & 1763 & 2.83 & 4.94 \\ 
  0.10 & 2.34 & 2.34 &    - & 1.03 & 24.5 & 2.83 & 4.95 \\ 
  0.11 & 2.33 & 2.33 &    - & 1.04 & 11.1 & 2.84 & 4.96 \\ 
  0.12 & 2.33 & 2.33 &    - & 1.05 & 8.89 & 2.84 & 4.99 \\ 
  0.13 & 2.32 & 2.32 &    - & 1.06 & 7.91 & 2.85 & 5.03 \\ 
  0.14 & 2.30 & 2.30 &    - & 1.07 & 7.95 & 2.85 & 5.05 \\ 
  0.15 & 2.32 & 2.32 &    - & 1.08 & 7.72 & 2.85 & 5.07 \\ 
  0.16 & 2.35 & 2.35 &    - & 1.09 & 7.26 & 2.86 & 5.07 \\ 
  0.17 & 2.37 & 2.37 &    - & 1.10 & 3.30 & 2.87 & 5.06 \\ 
  0.18 & 2.38 & 2.38 &    - & 1.12 & 3.51 & 2.87 & 5.06 \\ 
  0.19 & 2.39 & 2.39 &    - & 1.14 & 3.74 & 2.88 & 5.02 \\ 
  0.20 & 2.41 & 2.41 &    - & 1.16 & 4.05 & 2.88 & 5.02 \\ 
  0.21 & 2.41 & 2.41 &    - & 1.18 & 7.21 & 2.88 & 5.02 \\ 
  0.22 & 1.58 & 2.42 & 0.92 & 1.20 & 8.29 & 2.88 & 5.00 \\ 
  0.23 & 1.53 & 2.44 & 1.01 & 1.22 & 4.68 & 2.87 & 4.97 \\ 
  0.24 & 1.57 & 2.46 & 1.13 & 1.24 & 4.89 & 2.88 & 4.95 \\ 
  0.25 & 1.66 & 2.46 & 1.31 & 1.26 & 8.74 & 2.86 & 4.99 \\ 
  0.26 & 1.79 & 2.46 & 1.57 & 1.28 & 14.0 & 2.86 & 4.95 \\ 
  0.27 & 1.89 & 2.46 & 1.76 & 1.30 & 5.95 & 2.84 & 4.96 \\ 
  0.28 & 1.94 & 2.47 & 2.02 & 1.32 & 6.03 & 2.83 & 4.95 \\ 
  0.29 & 1.84 & 2.48 & 2.19 & 1.34 & 6.54 & 2.83 & 4.93 \\ 
  0.30 & 1.91 & 2.49 & 2.39 & 1.36 & 18.6 & 2.81 & 4.95 \\ 
  0.31 & 1.95 & 2.50 & 2.52 & 1.38 & 13.3 & 2.79 & 4.90 \\ 
  0.32 & 2.16 & 2.51 & 2.66 & 1.40 & 14.5 & 2.78 & 4.93 \\ 
  0.33 & 2.01 & 2.52 & 2.77 & 1.42 & 7.92 & 2.75 & 4.91 \\ 
  0.34 & 2.23 & 2.53 & 2.89 & 1.44 & 7.38 & 2.76 & 4.91 \\ 
  0.35 & 2.26 & 2.54 & 3.04 & 1.46 & 8.41 & 2.73 & 4.92 \\ 
  0.36 & 2.29 & 2.55 & 3.21 & 1.48 & 8.19 & 2.72 & 4.90 \\ 
  0.37 & 2.32 & 2.56 & 3.24 & 1.50 & 19.3 & 2.71 & 4.94 \\ 
  0.38 & 2.36 & 2.56 & 3.27 & 1.52 & 18.7 & 2.69 & 4.90 \\ 
  0.39 & 2.12 & 2.57 & 3.36 & 1.54 & 21.1 & 2.67 & 4.90 \\ 
  0.40 & 2.15 & 2.58 & 3.44 & 1.56 & 10.6 & 2.64 & 4.89 \\ 
  0.41 & 2.13 & 2.59 & 3.51 & 1.58 & 11.2 & 2.64 & 4.84 \\ 
  0.42 & 2.28 & 2.60 & 3.58 & 1.60 & 11.8 & 2.63 & 4.81 \\ 
  0.43 & 2.35 & 2.61 & 3.64 & 1.62 & 12.7 & 2.63 & 4.83 \\ 
  0.44 & 2.27 & 2.61 & 3.71 & 1.64 & 14.4 & 2.62 & 4.79 \\ 
  0.45 & 3.69 & 2.63 & 3.75 & 1.66 & 14.3 & 2.61 & 4.73 \\ 
  0.46 & 3.66 & 2.65 & 3.79 & 1.68 & 13.7 & 2.59 & 4.70 \\ 
  0.47 & 3.90 & 2.67 & 3.83 & 1.70 & 13.8 & 2.58 & 4.70 \\ 
  0.48 & 4.11 & 2.69 & 3.88 & 1.72 & 13.6 & 2.56 & 4.62 \\ 
  0.49 & 4.25 & 2.71 & 3.92 & 1.74 & 13.5 & 2.56 & 4.56 \\ 
  0.50 & 4.39 & 2.73 & 3.96 & 1.76 & 14.4 & 2.55 & 4.53 \\ 
  0.51 & 4.65 & 2.76 & 3.99 & 1.78 & 13.4 & 2.53 & 4.48 \\ 
  0.52 & 4.69 & 2.79 & 4.01 & 1.80 & 11.9 & 2.51 & 4.42 \\ 
  0.53 & 5.15 & 2.81 & 4.05 & 1.82 & 18.4 & 2.49 & 4.34 \\ 
  0.54 & 5.44 & 2.83 & 4.09 & 1.84 & 9.98 & 2.47 & 4.39 \\ 
  0.55 & 5.67 & 2.85 & 4.11 & 1.86 & 8.77 & 2.47 & 4.21 \\ 
  0.56 & 5.97 & 2.87 & 4.13 & 1.88 & 6.98 & 2.45 & 4.13 \\ 
  0.57 & 6.34 & 2.87 & 4.15 & 1.90 & 8.86 & 2.38 & 4.03 \\ 
  0.58 & 6.72 & 2.87 & 4.18 & 1.92 & 8.18 & 2.37 & 3.94 \\ 
  0.59 & 7.28 & 2.80 & 4.19 & 1.94 & 7.66 & 2.35 & 3.86 \\ 
  0.60 & 7.99 & 2.81 & 4.20 & 1.96 & 9.18 & 2.34 & 3.76 \\ 
  0.61 & 8.85 & 2.82 & 4.22 & 1.98 & 7.02 & 2.22 & 3.65 \\ 
  0.62 & 9.52 & 2.84 & 4.25 & 2.00 & 6.80 & 2.21 & 3.62 \\ 
  0.63 & 11.1 & 2.84 & 4.28 & 2.02 & 6.23 & 2.19 & 3.48 \\ 
  0.64 & 11.1 & 2.84 & 4.31 & 2.04 & 11.9 & 2.25 & 3.37 \\ 
  0.65 & 14.9 & 2.92 & 4.32 & 2.06 & 12.8 & 2.23 & 3.26 \\ 
  0.66 & 16.9 & 2.92 & 4.32 & 2.08 & 14.4 & 2.22 & 3.16 \\ 
  0.67 & 19.1 & 2.92 & 4.34 & 2.10 & 7.61 & 2.21 & 3.03 \\ 
  0.68 & 25.5 & 2.92 & 4.36 & 2.12 & 17.1 & 2.20 & 2.93 \\ 
  0.69 & 29.2 & 2.91 & 4.38 & 2.14 & 17.7 & 2.14 & 2.80 \\ 
  0.70 & 35.2 & 2.91 & 4.40 & 2.16 & 16.0 & 2.13 & 2.70 \\ 
  0.71 & 43.5 & 2.90 & 4.42 & 2.18 & 10.8 & 2.12 & 2.59 \\ 
  0.72 & 53.2 & 2.89 & 4.44 & 2.20 & 6.25 & 2.09 & 2.48 \\ 
  0.73 & 62.9 & 2.88 & 4.46 & 2.22 & 4.34 & 2.06 & 2.36 \\ 
  0.74 & 97.8 & 2.87 & 4.49 & 2.24 & 3.39 & 2.04 & 2.25 \\ 
  0.75 & 84.6 & 2.85 & 4.49 & 2.26 & 7.99 & 2.01 & 2.15 \\ 
  0.76 & 93.3 & 2.83 & 4.49 & 2.28 & 8.47 & 2.06 & 2.07 \\ 
  0.77 &  111 & 2.81 & 4.52 & 2.30 & 8.38 & 2.08 & 1.99 \\ 
  0.78 &  288 & 2.80 & 4.54 & 2.32 & 8.76 & 2.05 & 1.91 \\ 
  0.79 & 98.4 & 2.80 & 4.54 & 2.34 & 6.14 & 1.97 & 1.84 \\ 
  0.80 & 61.3 & 2.80 & 4.54 & 2.36 & 10.4 & 1.88 & 1.78 \\ 
  0.81 & 46.5 & 2.79 & 4.56 & 2.38 & 14.9 & 1.85 & 1.71 \\ 
  0.82 & 37.6 & 2.79 & 4.59 & 2.40 & 10.2 & 1.83 & 1.65 \\ 
  0.83 & 32.1 & 2.78 & 4.61 & 2.42 & 8.61 & 1.79 & 1.61 \\ 
  0.84 & 25.6 & 2.77 & 4.63 & 2.44 & 10.1 & 1.79 & 1.55 \\ 
  0.85 & 20.3 & 2.77 & 4.63 & 2.46 & 7.06 & 1.73 & 1.51 \\ 
  0.86 & 20.6 & 2.76 & 4.63 & 2.48 & 4.69 & 1.70 & 1.46 \\ 
  0.87 & 15.4 & 2.77 & 4.67 & 2.50 & 5.20 & 1.66 & 1.42 \\ 
  0.88 & 14.0 & 2.78 & 4.72 & 2.52 & 5.96 & 1.71 & 1.37 \\ 
  0.89 & 11.8 & 2.78 & 4.71 & 2.54 & 7.71 & 1.69 & 1.33 \\ 
  0.90 & 11.0 & 2.78 & 4.69 & 2.56 & 8.89 & 1.62 & 1.28 \\ 
  0.91 & 9.94 & 2.78 & 4.71 & 2.58 & 7.76 & 1.57 & 1.23 \\ 
  0.92 & 9.52 & 2.78 & 4.73 & 2.60 & 2.31 & 1.50 & 1.18 \\ 
  0.93 & 8.68 & 2.80 & 4.76 & 2.62 & 2.88 & 1.47 & 1.15 \\ 
\hline
\hline

\end{longtable}